\documentstyle[12pt,utdiss,epsf]{report}

\author{Rusty Shane Towell}
\title{Measurement of the Antiquark Flavor \\Asymmetry in the Nucleon Sea}
\previousdegrees{B.S.}
\address{Rt. 4 Box 155 A\\ Brownwood, TX 76801}
\supervisor[L. Donald Isenhower]{Peter J. Riley}
\committeesize{6}	

\begin{document}
\widowpenalty = 10000
\epsfclipon
%
%


\titlepage			
\copyrightpage			
\signaturepage			
\begin{dedication}		
\begin{center}
To my wife Amy,\\ 
my children Ramsey, Marshall, Cecily, Ensley, and Travis,\\ 
and my parents Delbert and Helen.
\end{center}
	%
	%
\end{dedication}

\acknowledgements

This dissertation would not have been possible without the 
efforts of the FNAL E866/NuSea collaboration.  I have been blessed to 
be a member of this talented, hard working, and fun collaboration. I 
thank each of you for your contribution to the experiment, my education,
and this dissertation.  I hope I have the opportunity to work with
each of you again.

\begin{center}
The FNAL E866/NuSea Collaboration\\
T.C. Awes, M.E. Beddo, C.N. Brown, J.D. Bush, T.A. Carey, \\
T.H. Chang, W.E. Cooper, C.A. Gagliardi, G.T. Garvey, D.F. Geesaman, \\
E.A. Hawker, X.C. He, L.D. Isenhower, S.B. Kaufman, D.M. Kaplan, \\
P.N. Kirk, D.D. Koetke, G. Kyle, D.M. Lee, W.M. Lee, M.J. Leitch, \\
N. Makins, P.L. McGaughey, J.M. Moss, B.A. Mueller, P.M. Nord, \\
B.K. Park, V. Papavassiliou, J.C. Peng, G. Petitt, P.E. Reimer, \\
M.E. Sadler, J. Selden, P.W. Stankus, W.E. Sondheim, T.N. Thompson, \\
R.S. Towell, R.E. Tribble, M.A. Vasiliev, Y.C. Wang, Z.F. Wang, \\
J.C. Webb, J.L. Willis, D.K. Wise, and G.R. Young
\end{center}

Many members of this collaboration have gone above and beyond what 
could reasonably be expected of them.  For this they deserve many special thanks.

Mike Sadler and Donald Isenhower have been teaching me about physics
and life for over twelve years.  Thank you for everything you have taught me
from basic physics to how to exchange money in Russia to how to be a friend
to thermodynamics (I still can't believe you gave me a B) to how to finish 
graduate school.  I know you have more to teach me, I hope you don't stop.

Gerry Garvey, Pat McGaughey, and Mike Leitch have each done any excellent job
as spokesman for E866.  They have not only over seen the success of the 
experiment, but the success of the graduate students.  Thank you for 
the time you have spend working with me.

Carl Gagliardi has been an incredible support to me throughout the 
analysis of the data and the writing of this dissertation.  Thank you
for watching what I did, correcting me when I was wrong, complimenting
me when I was right, and always encouraging me to finish.

Eric, Bill, Ting, and Jason have been wonderful fellow graduate students.  
Thanks for helping to keep the physics fun.  For four summers I worked with 
Derek.  I could not have asked for a better friend.  Thanks for being 
my fellow `science guy'.

Finally, I must thank Paul Reimer.  I can't imagine going through the past
three years without his help and friendship.  He has help me and taught me 
so much.  Thank you for answering all my dumb questions.  Thank you for 
your hard work to help me finish this dissertation.

In addition to the support I received from the members of this collaboration,
I have also been blessed by the support of many friends and family.  

For the past ten years I have enjoyed the constant support of my wife.
Thank you Amy for supporting my decision to go back to graduate school. 
Thank you for your patience, love, and encouragement.  I could not have 
done this without you.  

For more years than anyone else, my parents have blessed my life. Thank you
for teaching me the important things.  Thank you for helping me through 
twenty years of school.  I'm finally done.

	%
	%
\abstract
A precise measurement of the ratio of Drell-Yan yields from an
800 GeV/c proton beam incident on liquid hydrogen and deuterium
targets is reported.  Over 370,000 Drell-Yan muon pairs were recorded.
From these data, the ratio of anti-down ($\bar{d}$) to anti-up 
($\bar{u}$) quark distributions in the proton sea is determined
over a wide range in Bjorken-$x$.  A strong $x$ dependence is 
observed in the ratio $\bar{d}/\bar{u}$.  From this result,
the integral of $(\bar{d}-\bar{u})$ is evaluated for \mbox{$0.015<x<0.35$}
and compared with deep inelastic scattering results.  
The origin of this asymmetry is probably due to non-perturbative 
effects such as those contained in pion cloud models.  This measurement
has instigated new global fits since previous parameterizations
of the proton could not accommodate this large asymmetry.
	%
	%
\tableofcontents	

\listoftables		
\listoffigures		



\chapter{Introduction}

In the beginning God created the heavens and the earth.  Ever since that
time, man has striven to understand the composition of matter.
The history of the study of the building blocks of matter is a series
of discoveries.  With each new discovery the number and size of building blocks 
has decreased while the strength of the bond holding the constituents together 
has increased.
The starting point 
could be considered to be the discovery of molecules.  All compounds 
are composed of weakly bound molecules, of which there are a plethora of types.
Next came the discovery that molecules are composed of atoms. 
It was then discovered that the atoms were made of
a dense nucleus surrounded by orbiting electrons.
As machines were invented that could probe matter
at higher energies and correspondingly smaller sizes, it became known that 
nuclei had internal structure and were composed of protons and neutrons which
are collectively referred to as nucleons. 
This series of discoveries continues as the structure of 
the nucleon is probed.

The discovery of large numbers of new particles, such as mesons, heavier hadrons, 
and antiparticles, implied
there were some smaller, more basic building blocks yet to be discovered.  These
smallest, most tightly bound, and relatively few building blocks are described 
by the Standard Model.  
This model has been extremely successful explaining the composition of all
matter and the forces that bind them\footnote{The Standard Model does not address
the weakest fundamental force, which is gravity.}.
The Standard Model has accomplished this using only six quarks ($u, d, s$,...), 
six leptons, 
their antiparticles ($\bar{u}, \bar{d}, \bar{s}$, ...), and five force carriers.
In fact, most of the matter we interact with every day is
made of only two types, or flavors, of quarks and a single lepton.  

\begin{figure}                                                     
  \begin{center} 
    \mbox{\epsfxsize=3in\epsffile{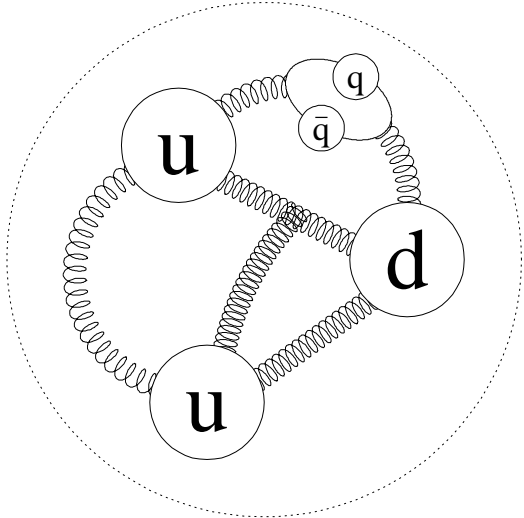}}
    \vspace*{-0.25in}                           
  \end{center}       
  \caption{A simple picture of the proton according to the Parton Model\@.  The 
		proton contains 2 valence up quarks ($u$), one valence down quark
		($d$), gluons (helix), and the sea of any number of quark antiquark pairs 
		($q$ and $\bar{q}$).}             
  \label{fig:proton}
\end{figure}

The bulk of the matter around us is composed of nucleons and electrons.  
Electrons are currently believed to be fundamental particles with no 
discernible internal structure.  Nucleons, on the other hand, 
do have internal structure.
The Parton Model describes the nucleon as a bound state
of three quarks that give the nucleon its electrical charge and other basic properties.  
For example, a nucleon composed of two up quarks and one down quark ($uud$) 
is a proton, while a nucleon composed of one up quark and two down quarks ($udd$)
is a neutron.
These three quarks, called valence quarks, are held together by the strong force which is
mediated, or carried, by gluons ($g$).  In addition to valence quarks and gluons, there is 
a background of virtual quark-antiquark pairs which come and go without affecting the 
basic characteristics of the nucleon, since any effect of the quark is canceled 
by the opposite effect of the antiquark.  These components of the nucleon, 
shown in Fig. \ref{fig:proton}, are collectively referred to as partons.

One source of these quark-antiquark pairs, 
commonly called the sea quarks, is gluon splitting.  
According to a theoretical description of the strong force, 
Quantum Chromodynamics (QCD), a gluon can split
into a quark-antiquark pair and exist for a short time before recombining
into a gluon.  This simple production mechanism implies that the sea of 
antiquarks in a nucleon would be flavor symmetric with respect to the up and down 
antiquarks.  Therefore, it was initially assumed that there were as many anti-up quarks
in a nucleon as there were anti-down quarks.

Over the past decade evidence has accumulated   
that indicates there are more anti-down
quarks in a proton than anti-up quarks.  
The primary motivation for
Experiment 866 at Fermi National Accelerator Laboratory (FNAL E866/NuSea)
was to measure the ratio of anti-down quarks 
to anti-up quarks in the nucleon.  
This experiment has 
measured the antiquark content of the nucleon sea more precisely and 
extensively than any previous measurement.
This dissertation will describe the NuSea experiment, including its motivation, apparatus, procedures,
analysis, and conclusions.



\chapter{Theory and Motivation}
\label{ch:theory}

The structure of the nucleon is difficult to study because of its small size and 
the large binding energy of its constituents.  
Nevertheless, with the aid of particle accelerators, 
much has been learned about the structure of the 
nucleon in recent years.  
Two electromagnetic interactions that can be produced
by modern accelerators and studied at high energies 
are muon induced deep inelastic scattering
\footnote{Deep inelastic scattering induced by leptons, other than muons, will not be discussed 
in this dissertation.}
and the Drell-Yan process. 
These two methods of probing the nucleon have revealed much about the 
nucleon sea.

\section{Muon Induced Deep Inelastic Scattering}
\label{sect:dis}

\begin{figure}                                                     
  \begin{center} 
    \mbox{\epsfxsize=2.3in\epsffile{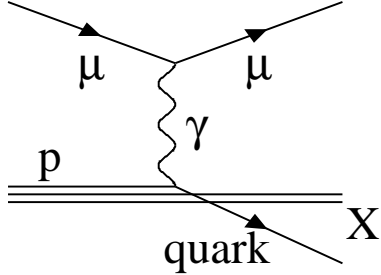}}
    \vspace*{-0.25in}                           
  \end{center}       
  \caption{Muon induced deep inelastic scattering.  In this process the 
		high energy muon scatters off a single quark in the nucleon by 
		exchanging a high energy virtual photon.}             
  \label{fig:deep}
\end{figure}

Deep inelastic scattering (DIS) is the process of inelastically scattering
a high energy lepton off a target nucleon.  
In the case of muon induced DIS, shown in Fig. \ref{fig:deep}, the muon scatters off a single quark
in the nucleon by exchanging a high energy virtual photon.  The differential cross
section for this interaction can be measured experimentally and
expressed in terms of the two structure functions $F_1(x,Q^2)$ and $F_2(x,Q^2)$.
These structure functions are functions of Bjorken-$x$~($x$), which is a dimensionless
scaling variable that at high energies represents the fraction of the nucleon's 
longitudinal momentum carried by a parton, and $Q^2$, which is the four momentum 
squared of the virtual photon exchanged in the process.  

These two structure functions can be expressed as linear combinations of 
parton functions, $f_i(x,Q^2)$, which are commonly called parton 
distribution functions (PDF's).
These PDF's are the probability of finding a parton in the nucleon that carries 
some fraction ($x$) of the nucleon's total longitudinal momentum.  The relationship between
the structure functions and the PDF's is:\footnote{Since the $Q^2$ 
dependence is small, it will be dropped from the PDF notation for the remainder of this dissertation.}
\begin{eqnarray}
F_1(x) = \frac{1}{2}\sum_i e^2_i f_i(x)\\
F_2(x) = \sum_i e^2_i x f_i(x),
\label{eqn:struct}
\end{eqnarray}
where the sum is over all the partons ($i=u,d,s,...,\bar u,\bar d,\bar s, ...,g$) 
and $e_i$ is the electric charge of the $i$-th parton. 
So by measuring the cross section for DIS, the
internal structure of the nucleon can be studied.

\begin{figure}                                                     
  \begin{center} 
    \mbox{\epsffile{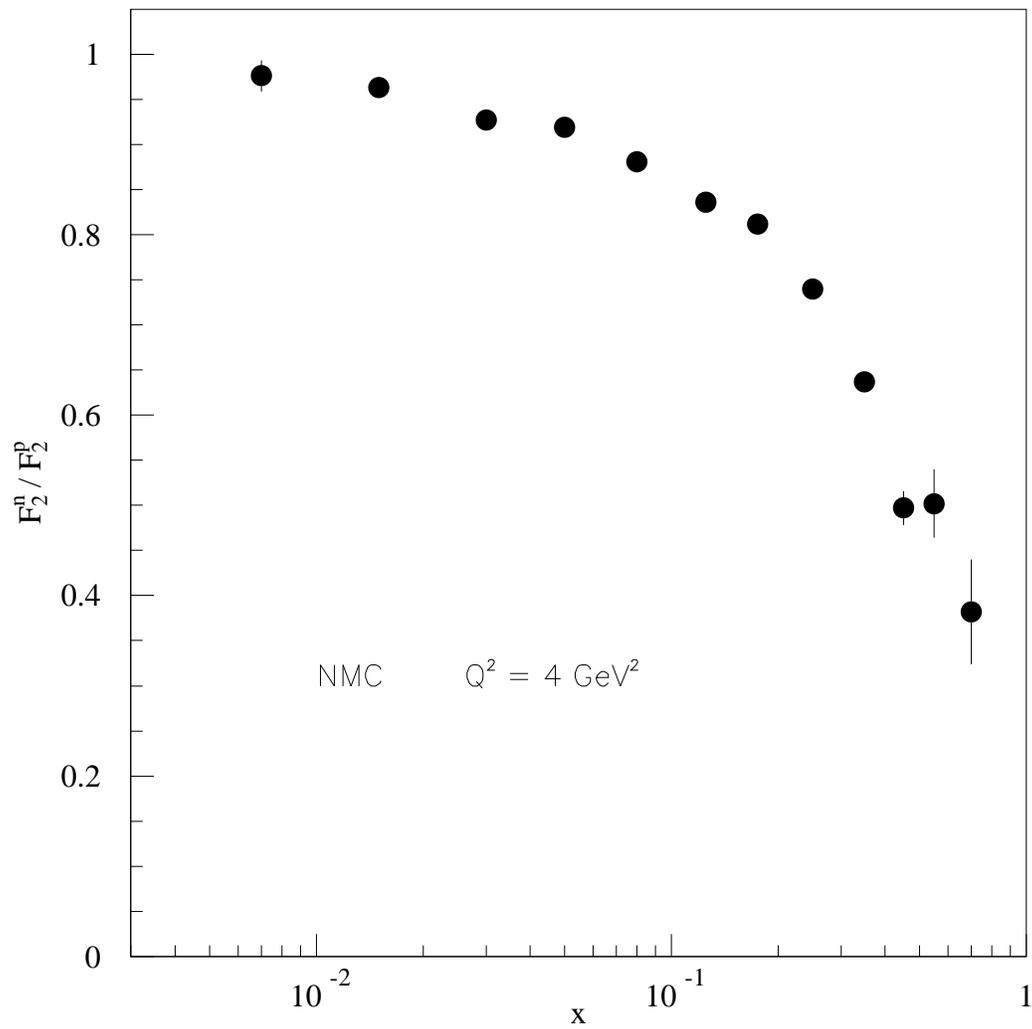}}
  \end{center}       
  \caption{ The NMC measured ratio $F^n_2/F^p_2$ at $Q^2 = 4$~GeV$^2$.
            Only the statistical uncertainty is shown.}             
  \label{fig:nmc1}
\end{figure}

An experiment performed by the New Muon Collaboration (NMC) measured
the cross section ratio for deep inelastic scattering of muons from hydrogen 
and deuterium~\cite{NMC}.  Their measurement\footnote{NMC actually measured 
$F^d_2/F^p_2 \equiv 0.5(1 + F^n_2/F^p_2)$ where their convention of normalizing 
$F^d_2$ per nucleon has been used here.}
of $F^{n}_2/F^{p}_2$ over the 
kinematic range of $0.004<x<0.8$ is shown in Fig. \ref{fig:nmc1}.
These data points are the result of combining both the 90 GeV and 
280 GeV incident muon energy measurements and 
adjusting the data for the small $Q^2$ dependence to a fixed $Q^2$ of 4 GeV$^2$.

\begin{figure}                                                     
  \begin{center} 
    \mbox{\epsfxsize=5.2in\epsffile{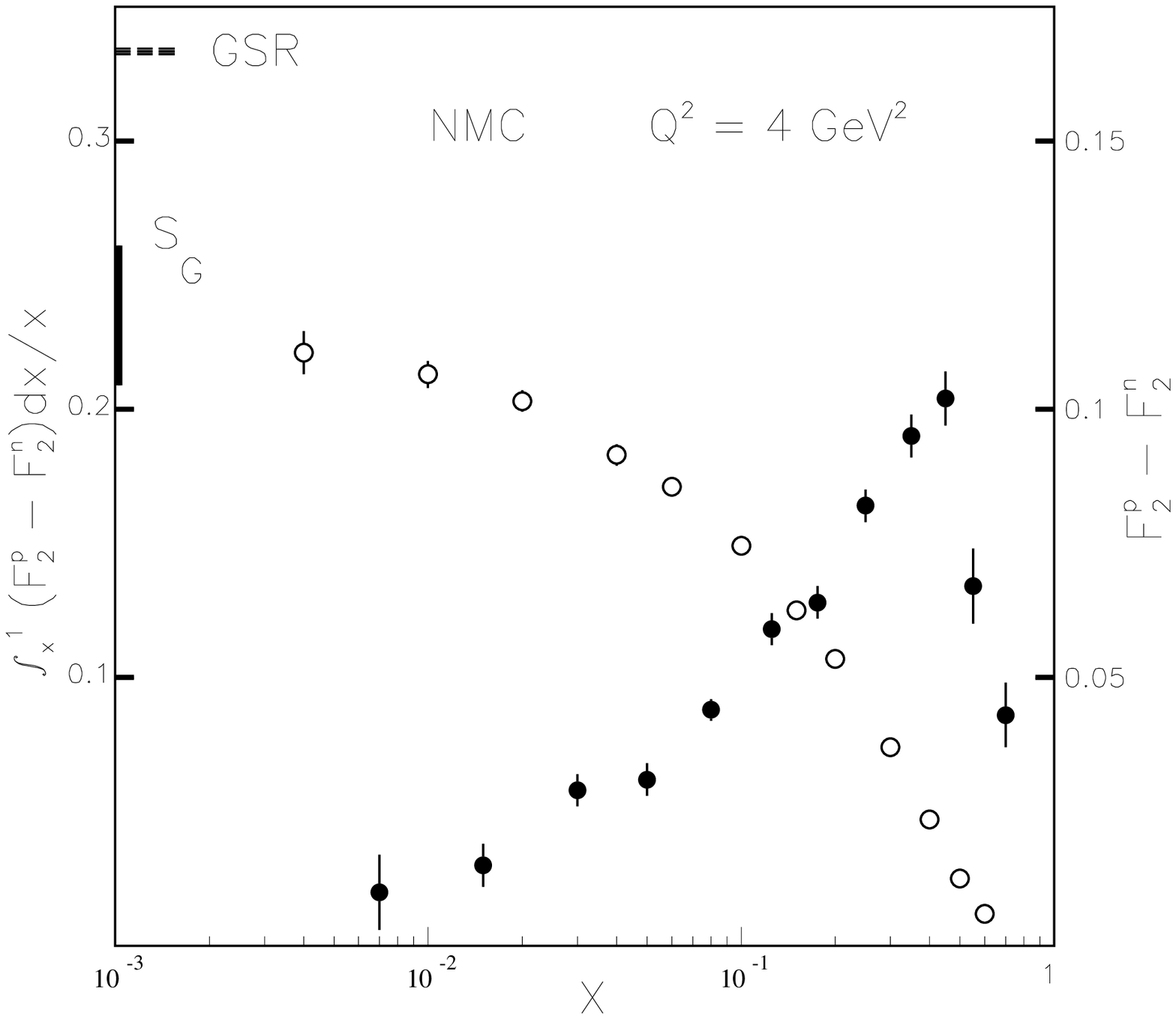}}
  \end{center}       
  \caption{The difference $F^p_2 - F^n_2$ (solid symbols and the scale
		to the right) and $\int_x^1 (F^p_2 - F^n_2)dx/x$ (open symbols
		and the scale to the left) at $Q^2 = 4$~GeV$^2$ as a function of 
		$x$.  The extrapolated integral result, $S_G$ is indicated by the bar.  The 
		traditional Gottfried Sum Rule result, $GSR$ is also shown.}             
  \label{fig:nmc2}
\end{figure}

The NMC measurement was used with a parameterization of the absolute
deuteron structure function, $F^d_2$, to obtain
\begin{equation}
F^p_2 - F^n_2 = 2F^d_2 \frac{1 - F^n_2/F^p_2}{1 + F^n_2/F^p_2}.
\end{equation}
The solid points in Fig. \ref{fig:nmc2} show the derived values 
of $F^{p}_2 - F^{n}_2$ as a function of $x$.  Also shown in this figure is  
$\int^1_x (F^{p}_2 - F^{n}_2)dx/x$ for the same $x$ range of the data 
as well as the extrapolated result ($S_G$) over all values of $x$ ($0\le x \le 1$).
This extrapolated result is
\begin{eqnarray}
S_G \equiv \int^1_0 \left[F^{p}_2 - F^{n}_2\right] \frac{dx}{x} = 0.235 \pm 0.026.
\label{eqn:gsr}
\end{eqnarray}

This result by NMC can be compared to the traditional Gottfried Sum Rule~\cite{GSR}
($GSR$) result of $1/3$. 
To understand this discrepancy between the NMC measurement and the traditional 
result of the Gottfried Sum Rule, it is important to look at the assumptions that 
lead to the traditional result.  Begin with the Gottfried sum as written in
equation \ref{eqn:gsr} and then use equation \ref{eqn:struct} to get
\begin{equation}
S_G = \int^1_0 \sum_i e_i^2 \left[q_i^p(x) + \bar{q}_i^p(x) - q_i^n(x) - \bar{q}_i^n(x)\right]dx.
      \label{eqn:gsr3}
\end{equation}
Next assume charge symmetry\footnote{Charge symmetry implies that $u_p(x)=d_n(x)$, $\bar d_p(x)
= \bar u_n(x)$, etc.} to express the neutron PDF's as proton PDF's and
ignore the heavier quark (e.g. $s,\bar{s},...$) terms to get
\begin{equation}
S_G = \int^1_0 \frac{1}{3} \left[ u + \bar{u} - d - \bar{d} \right] dx, 
\end{equation}
which can be rewritten as
\begin{equation}
S_G = \int^1_0 \frac{1}{3} \left[ u - \bar{u} \right] dx - 
      \int^1_0 \frac{1}{3} \left[ d - \bar{d} \right] dx +
      \int^1_0 \frac{2}{3} \left[ \bar{u} - \bar{d} \right] dx.
      \label{eqn:gsr5}
\end{equation}
The first two integrals are the definition of the valence quarks, which for a 
proton is two up valence quarks and one down valence quark.  Thus 
equation \ref{eqn:gsr5} is reduced to
\begin{equation}
S_G = \frac{1}{3} + \int^1_0 \frac{2}{3} \left[ \bar{u} - \bar{d} \right] dx.
\end{equation}
If it is also assumed that $\bar{d} (x) = \bar{u} (x)$, then one arrives at the 
traditional result of $1/3$.  

To reconcile the NMC measurement and the traditional Gottfried Sum result,
one of the three assumptions made above must be incorrect:  first is the assumption
that the NMC measurement was correctly extrapolated to zero;  
second, that charge symmetry is valid; 
and third, that $\bar{d} (x) = \bar{u} (x)$.

The small $x$ extrapolation was checked by Fermilab E665~\cite{E665}, which made the same 
measurement as NMC except they measured the ratio for \mbox{$10^{-6}\le x \le 0.3$}. 
Over the kinematic range where NMC and E665 overlapped, their measurements 
agree.  However, E665 determined that for $x \le 0.01$ the value of  
$2F^d_2/F^p_2 -1$ was a constant $0.935 \pm 0.008 \pm 0.034$.  
While this could be interpreted as a difference between $F^n_2$ and $F^p_2$,
it is usually thought to be the effect of nuclear shadowing in deuterium~\cite{shadow},
which means that $F^n_2/F^p_2 \ne 2F^d_2/F^p_2 -1$.
Therefore it is not possible to measure $F^n_2/F^p_2$ in a 
model independent way at low $x$, although the same calculations confirm 
$F^n_2/F^p_2 = 2F^d_2/F^p_2 -1$ at higher $x$ values to a good approximation.  
While E665 seems to support the NMC extrapolation, it also highlights 
the difficulty of making and interpreting DIS measurements at small $x$ values.

Charge symmetry is generally assumed to be well respected in strong interactions.
Extensive experimental searches for charge symmetry violation effects 
have shown that charge symmetry holds to the order of the proton-neutron 
mass difference~\cite{chargesymmetry,chargesymmetry2}.  
\footnote{
While it has been recently suggested that charge symmetry violation could 
explain the discrepancy between muon and neutrino DIS~\cite{csv} data,
it can not fully explain the measurement reported here~\cite{csv2}. 
}
Therefore, charge symmetry can not explain
the discrepancy between the NMC measurement and the traditional Gottfried Sum result.   


The only remaining assumption is $\bar{d} (x) = \bar{u} (x)$.  If this assumption 
is solely responsible for the discrepancy between the NMC measurement and the 
traditional Gottfried Sum result, it would imply
\begin{eqnarray}
\int^1_0 [ \bar d(x) - \bar u(x)]dx = 0.148 \pm 0.039.
\end{eqnarray}
The NMC measurement and the above analysis were the
first indications that there were more anti-down quarks in the proton than
anti-up quarks. 

\section{Drell-Yan}

\begin{figure}                                                     
  \begin{center} 
    \mbox{\epsfxsize=3.0in\epsffile{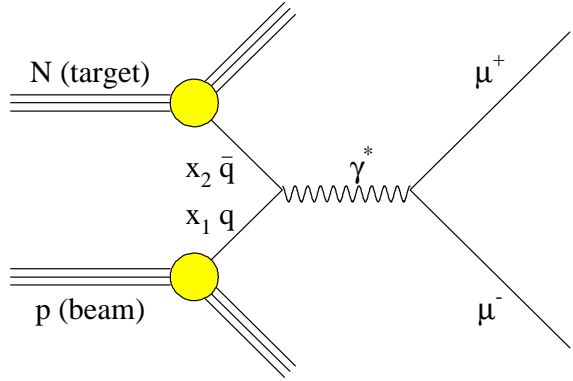}}
    \vspace*{-0.25in}                           
  \end{center}       
  \caption{The leading order Drell-Yan process.}             
  \label{fig:dy}
\end{figure}

Following the publication of the NMC result, it was suggested~\cite{ES} 
that the Drell-Yan process~\cite{DY} could provide 
a more direct probe of the light antiquark asymmetry of
the nucleon.  In its simplest form, shown in Fig. \ref{fig:dy},
the Drell-Yan process is the production 
of a lepton-antilepton pair from a virtual photon produced in a quark-antiquark
annihilation.  Since there are no valence antiquarks in a nucleon-nucleon 
interaction, the Drell-Yan process is a direct probe of the sea antiquarks.

The proton-nucleon cross section for the Drell-Yan process to lowest order in 
the electromagnetic fine structure constant ($\alpha$), as a function of $x$ of 
the initial quark and antiquark, is 
\begin{eqnarray}
{d^2\sigma\over dx_1 dx_2} = \frac{4 \pi \alpha ^2}{9 M^2} 
                             \sum_i e_i^2 \left[f_{i}(x_1)\bar f_{i}(x_2) +
                             \bar f_{i}(x_1) f_{i}(x_2)\right],
\label{eqn:dy}
\end{eqnarray}
where the sum is over all quark flavors and $M$ is the virtual photon or
dilepton mass~\cite{DYeqn}.  Here the subscripts $1$ and $2$ denote 
a parton from the beam and target respectively.
It is important to note that the PDF's in equation \ref{eqn:dy} are the same 
functions used in equation \ref{eqn:struct}.  This means that the DIS process and 
the Drell-Yan process are sensitive to the same antiquark structure of the nucleon.

Some kinematic quantities commonly used to describe Drell-Yan events are the 
$x$ Feynman ($x_F$) and the dilepton mass ($M$), which are defined
\footnote{The approximations in these equations become exact in the infinite 
momentum frame.  The infinite momentum frame is a boosted frame of reference
where momentum transverse to the beam direction is negligible and the quarks 
can be considered to be massless.}
as 
\begin{equation}
x_F = \frac{p^{\gamma}_{||}}{p^{\gamma,max}} \approx \frac{p^{\gamma}_{||}}{\sqrt{s}/2} = x_1 - x_2
\end{equation}
and
\begin{equation}
M^2 = Q^2 \approx x_1 x_2 s
\end{equation}
where $p^{\gamma}_{||}$ is the center of mass longitudinal momentum of the 
virtual photon, $p^{\gamma,max}$ is its maximum possible value, and $s$ is 
the total four momentum squared of the initial nucleons.  The 
momentum transverse to the beam direction ($p_T$) of 
the dileptons is used to complete
\footnote{The complete event description includes production and decay angles
but these will not be discussed in this dissertation.}
the description of the event.

The cross section for producing a Drell-Yan event in a proton-proton 
interaction, $\sigma^{pp}$, or proton-neutron interaction, $\sigma^{pn}$ is
\begin{eqnarray}
\sigma^{pp} \propto \frac{8}{9} u(x)\bar u(x) + \frac{2}{9} d(x)\bar d(x)
\label{eqn:sigpp} \\
\sigma^{pn} \propto \frac{5}{9} u(x)\bar d(x) + \frac{5}{9} d(x)\bar u(x),
\label{eqn:sigpn}
\end{eqnarray}
where the heavier quark terms have been ignored, charge symmetry has been 
invoked so that all the PDF's can be expressed as proton PDF's, 
and $x_F$ is zero\footnote{The reason for this simplification
is so that this discussion can be compared to the NA51 measurement.  This 
simplification is not applicable to E866.}.  
Since there is presently no way of making a free neutron target, a deuterium 
target is 
commonly used as the best approximation to a free neutron target.  
Making the approximation that the proton-deuterium cross section ($\sigma^{pd}$)
can be written as
\begin{eqnarray}
\sigma^{pd} \approx \sigma^{pp} + \sigma^{pn}
\end{eqnarray}
ignores the small 
\footnote{FNAL E772~\cite{E772} searched for nuclear effects in 
Drell-Yan production and found none
except for shadowing at small values of $x$.}
nuclear effects inside the deuterium nucleus.  
Using this approximation the measured cross section ratio can be written as,
\begin{equation}
\label{eqn:xf0}
\left. \frac{\sigma^{pd}}{\sigma^{pp}}\right|_{x_1=x_2} =
\frac{8+5\frac{\bar d}{\bar u} + 5 \frac{d}{u} + 2 \frac{\bar d}{\bar u} \frac{d}{u}}
{8 +  2 \frac{\bar d}{\bar u} \frac{d}{u}}
\end{equation}
where, for simplicity, the notation showing the $x$ dependence has been dropped.  From this 
measurable quantity, one can solve for $\bar d/\bar u$,
\begin{eqnarray}
\frac{\bar d}{\bar u} = 
\frac{8\left(1-\frac{\sigma^{pd}}{\sigma^{pp}}\right) + 5 \frac{d}{u}}
{2\frac{d}{u}\left(\frac{\sigma^{pd}}{\sigma^{pp}}-1\right) -5 }.
\end{eqnarray}
Therefore by measuring the hydrogen and deuterium Drell-Yan cross sections, 
the nucleon antiquark sea can be measured.

The first experiment
\footnote{FNAL E772 \cite{e772du} was the first Drell-Yan experiment 
to look for flavor asymmetry in the nucleon sea, but they measured
cross sections from isoscalar targets and W, which has a large neutron
excess.  This measurement was not as sensitive to $\bar d/\bar u$ as
measurements using deuterium and hydrogen targets.}
to follow up this idea was NA51 at CERN~\cite{NA51}.
It used the primary proton 
beam from the CERN-SPS to supply $450$ GeV/c protons, which interacted in 
one of three identical vessels that contained either liquid hydrogen, liquid
deuterium or a vacuum.  From the almost 
six thousand reconstructed Drell-Yan events with a mass above $4.3$~GeV
it was determined that
\begin{equation}
\frac{\sigma^{pp} - \sigma^{pn}}{\sigma^{pp} + \sigma^{pn}} = -0.09 \pm 0.02 \pm 0.025.
\end{equation}
These events were centered at $x_F = 0$ with both $x_1$ and $x_2$
between $0.08$ and $0.53$.  Since the acceptance of the NA51 spectrometer 
was narrow and statistics were limited, 
the measurement was not reported as a function of $x$, but only 
at the average of $x_2 = 0.18$.  

From the measured cross section
asymmetry, it was determined that
\begin{eqnarray}
\left. \frac{\bar{u}_{p}}{\bar{d}_{p}} \right|_{\langle x \rangle=0.18}
 = 0.51\pm 0.04\pm 0.05.                                         
\end{eqnarray}
This was the first direct determination of the ratio 
of $\bar d$ to $\bar u$, but the limited 
statistics of this measurement restricted it to
a single $x$ value.  

One of the main limitations of the NA51 experiment was the limited 
kinematic coverage that was sharply peaked near $x_F = 0$.  A better
measurement of $\bar d/\bar u$ could have been made if the acceptance was 
largest for $x_F > 0$, since the Drell-Yan cross section ratio is more
sensitive to the antiquark distribution in this kinematic regime.
The increased sensitivity results because in this kinematic
regime the Drell-Yan cross section is dominated by the annihilation 
of a beam quark with a target antiquark.
Using the same assumptions that led to equations \ref{eqn:sigpp} and 
\ref{eqn:sigpn}, except assuming that $x_1 \gg x_2$ instead of $x_1 = x_2$, 
results in the relations
\begin{eqnarray}
\sigma^{pp} \propto \frac{4}{9} u(x_1)\bar u(x_2) + \frac{1}{9} d(x_1)\bar d(x_2)
\label{eqn:sigpp2}
\end{eqnarray}
and
\begin{eqnarray}
\sigma^{pn} \propto \frac{4}{9} u(x_1)\bar d(x_2) + \frac{1}{9} d(x_1)\bar u(x_2).
\label{eqn:sigpn2}
\end{eqnarray} 
From these two equations it is a simple matter to derive 
\begin{equation}
\left. \frac{\sigma^{pd}}                            
           {2\sigma^{pp}}
\right|_{x_1\gg x_2} \approx\frac{1}{2}
\frac{\left( 1 + \frac{1}{4}\frac{d(x_1)}{u(x_1)} \right)}
     {\left( 1 + \frac{1}{4}\frac{d(x_1)}{u(x_1)}
           \frac{\bar{d}(x_2)}{\bar{u}(x_2)} \right)}
      \left( 1 + \frac{\bar{d}(x_2)}{\bar{u}(x_2)} \right).
\label{eqn:xf1}
\end{equation}
This expression can be further simplified by making the additional 
assumption that $d(x) \ll 4u(x)$ which yields 
\begin{equation}                                                            
\left. \frac{\sigma^{pd}}
           {2\sigma^{pp}}
\right|_{x_1\gg x_2} \approx\frac{1}{2}
      \left( 1 + \frac{\bar{d}(x_2)}{\bar{u}(x_2)} \right).
\label{eqn:prl41}                                           
\end{equation}  
This equation illustrates the sensitivity of the Drell-Yan cross 
section ratio to $\bar d/\bar u$ but is valid only for $x_1 \gg x_2$.

\section{Global Fits}

\begin{figure}                                                     
  \begin{center} 
    \mbox{\epsfxsize=4.0in\epsffile{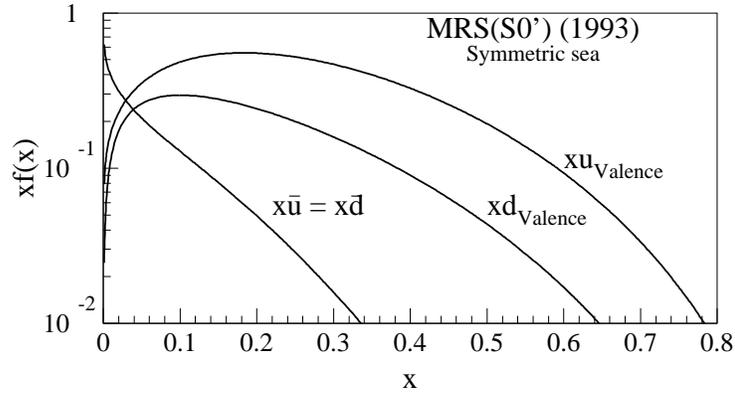}}
    \vspace*{-0.25in}                           
  \end{center}       
  \caption{An old parameterization of the proton by MRS assuming a symmetric 
		sea~\cite{MRSs0}.  These parton distribution functions are
		shown for $Q^2 = 54$~GeV$^2$.}             
  \label{fig:mrs_s0}
\end{figure}

Several groups have applied phenomenological models to 
the results from DIS, Drell-Yan, and other methods
\footnote{These `other methods', which will not be discussed in this dissertation, 
include neutrino induced DIS, W production asymmetry, and Semi-inclusive DIS.} 
of probing the nucleon to generate complete sets of PDF's.  This method of 
fitting all the relevant data is called a global fit.  
Two groups that have generated global
fits parameterizing the nucleon sea are CTEQ~\cite{CTEQ} and MRS~\cite{MRS}.
Prior to the measurements by NMC and NA51, the usual assumption was that 
$\bar {d}(x) = \bar {u}(x)$.  A typical parameterization of the proton 
with this assumption is shown in Fig. \ref{fig:mrs_s0}.

\begin{figure}                                                     
  \begin{center} 
    \mbox{\epsfxsize=4.0in\epsffile{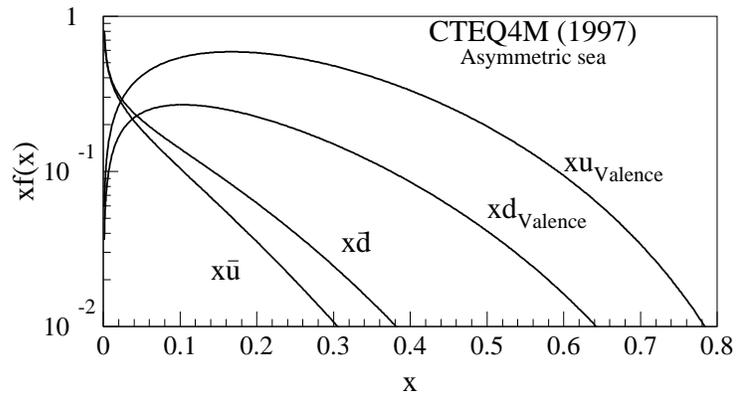}}
    \vspace*{-0.25in}                           
  \end{center}       
  \caption{An old parameterization of the proton published by CTEQ~\cite{CTEQ} 
		prior to FNAL E866/NuSea.  These parton distribution functions are
		shown for $Q^2 = 54$~GeV$^2$.}             
  \label{fig:cteq4m}
\end{figure}

Prior to FNAL E866/NuSea, the main constraints 
on the global fits of $\bar d(x)$ and $\bar u(x)$ were the 
NMC and NA51 results.    While these measurements were important
to show that $\bar d \ne \bar u$, neither imposed rigid
constraints on the global fits of $\bar d(x)$ and $\bar u(x)$. 
The NA51 measurement only constrained the ratio of $\bar d(x)$ to $\bar u(x)$
at $x = 0.18$, while the NMC measurement only constrained the integral
of the difference between $\bar d(x)$ and $\bar u(x)$, which
primarily constrains the PDF's at small values of $x$.
This poorly constrained parameterization,
shown in Fig. \ref{fig:cteq4m}, constituted the world's best guess at the structure
of the nucleon sea prior to the measurement reported here.

The goal of FNAL E866/NuSea was to measure the ratio of the Drell-Yan 
cross section for proton-deuteron interactions to proton-proton interactions
over the kinematic range of $0.04 \le x \le 0.3$ with a systematic uncertainty
of less than 1.5\%.  This measurement would provide stringent
constraints on the ratio of $\bar d(x)$ to $\bar u(x)$ over the same
$x$ region.



\chapter{Experimental Setup}

The goal of this experiment was to measure the ratio of Drell-Yan events from 
proton-proton interactions to proton-deuteron interactions.  To count the number
of Drell-Yan events it was necessary to detect pairs of high energy oppositely
charged muons.  The kinematics of the 
interaction were determined from the four momentum of each muon produced
in the Drell-Yan process.  This was accomplished for 
each muon in a pair of oppositely charged muons (called a dimuon event) 
by reconstructing the path of each muon through a series of tracking stations and 
dipole magnets.

This chapter will describe the experimental apparatus used to make this measurement.
The equipment will be described in the order through which a particle would traverse the
setup, starting with a description of the beam, then the targets, and finally
the spectrometer.  The coordinate system used 
throughout this experiment has the positive z-axis 
in the direction of the incoming beam,  
the y-axis oriented with the positive direction pointing up, and the x-axis
is chosen to complete a right-handed coordinate system.

\section{Beamline and Beam Monitors}			

The beam used in this experiment was an 800 GeV/c proton beam extracted
from the Fermilab Tevatron accelerator.  The beam was extracted slowly over
a twenty second time period, once per minute.  This slow extraction is referred 
to as a spill.  Within each spill, the beam arrived in small bunches, called
"buckets", every 19ns.  The frequency of the buckets was determined by the 
radio frequency (RF) of the accelerator, which was 53 MHz.  

As the beam was transported down the east beamline in the Meson experimental
hall, its size, position, and intensity were monitored with several different detectors.  
The beam position and size were measured using RF cavities and segmented 
wire ion chambers (SWIC's).  The last 
check of the beam, about 70 inches upstream of the targets, was performed by a 
movable SWIC.  
This SWIC had wire spacing of 2mm
in the horizontal direction and 0.5mm in the vertical direction.  
The size of the beam at the SWIC was typically about 6mm wide and 1mm high
(full-width at half-maximum).  

The beam intensity was also monitored by several different detectors.  The primary monitor
was a secondary emission monitor (SEM) located about 100m upstream of the targets. 
In addition to the SEM, the beam intensity was monitored with a quarter-wave 
RF cavity and an ion chamber (IC3).  The SEM calibration was determined to be 
$7.9 \pm 0.6 \times 10^7$ protons per SEM count\cite{sem} using the methods described
in Section \ref{sect:special}.  The nominal maximum beam intensity~\footnote{
The nominal maximum beam intensity for the low mass data (see section \ref{sect:data})
was $5 \times 10^{11}$ protons per spill.} was $2 \times 10^{12}$
protons per spill.

Approximately $85^\circ$ from the beam direction in the x-z plane were a pair 
of four-element scintillator telescopes 
called AMON and WMON.  These scintillator telescopes viewed the target through a hole
in the heavy shielding enclosing the targets.  The purpose of these 
detectors was to monitor the luminosity of the beam hitting the target, 
which was then used to determine the beam duty factor, data acquisition livetime
and the target position.

\section{Targets}

\begin{figure}                                                     
  \begin{center} 
    \mbox{\epsfysize=7.1in\epsffile{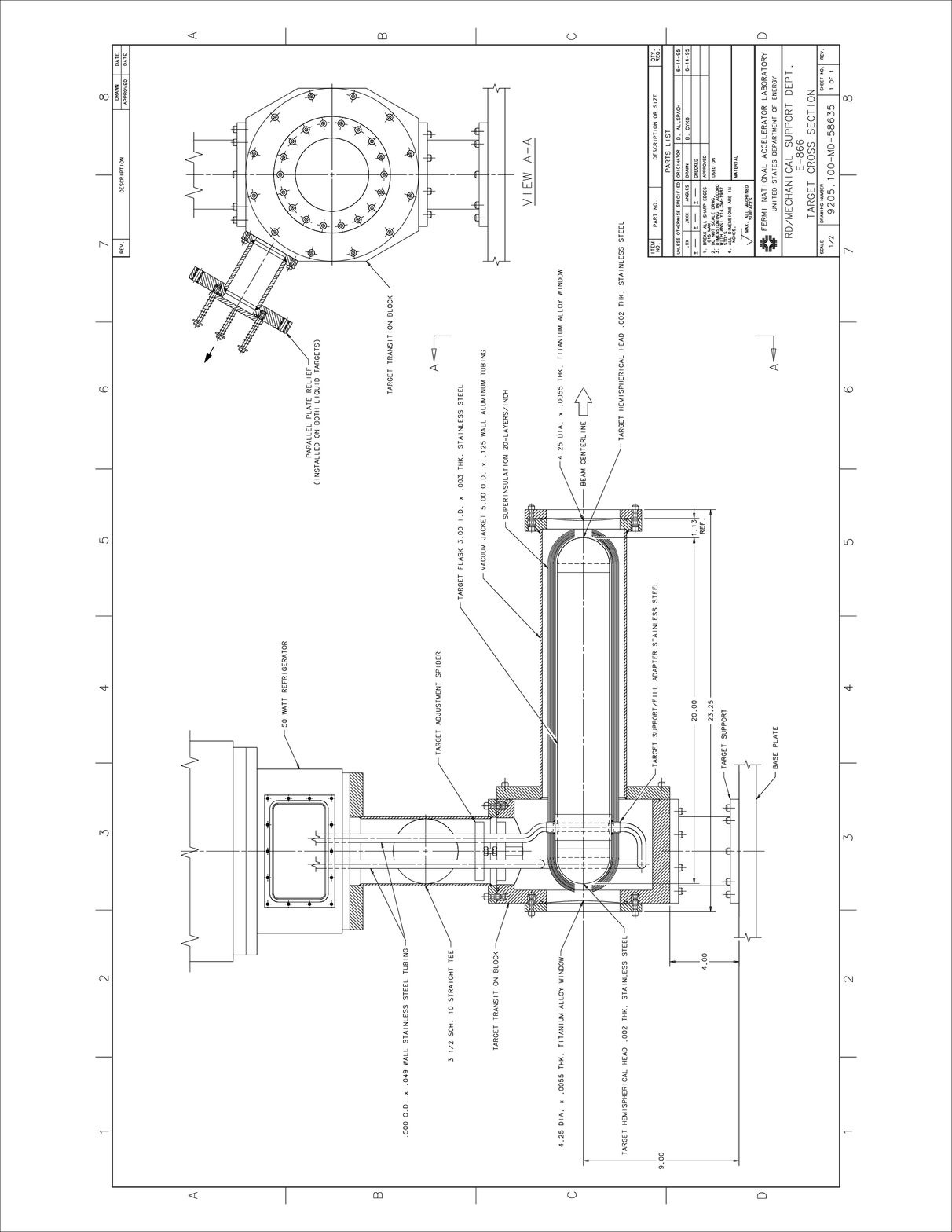}}
    \vspace*{-0.25in}                           
  \end{center}       
  \caption{One of the three identical E866 target flasks.}             
  \label{fig:target}
\end{figure}

The 800 GeV/c extracted proton beam passed through 
one of three target flasks.  Figure \ref{fig:target} shows one of the three
identical twenty inch long flasks which were cylindrical in 
shape with hemispherical ends.  
The flasks were constructed of thin stainless steel with an insulated vacuum jacket
around each flask.  One flask was filled
with liquid deuterium, another was filled with liquid hydrogen, and the third was 
evacuated.  The hydrogen target was 7\% of an interaction length and 6\% of a 
radiation length.  The deuterium target was 15\% of an interaction length 
and 7\% of a radiation length.  The evacuated target was less than 0.2\%
of an interaction length and 1.4\% of a radiation length.  
Both the temperature and the vapor pressure of the 
filled flasks were monitored.  

All three flasks were mounted on a movable table so that the target could be
changed between spills.  The normal target cycle was twelve 
spills long with five spills on the deuterium target, one spill 
on the empty flask, five spills on the hydrogen 
target and another spill on the empty flask.  The electronics that automatically 
cycled the targets checked that the spill had at least $4 \times 10^{10}$ protons
before it counted the spill.  
This prevented the target from cycling while the accelerator
was off.  
This frequent cycling of the targets minimized many of the 
systematic uncertainties between the targets when the cross
section ratio was determined.

\section{E866 Spectrometer}			%
\begin{figure}                                                     
  \begin{center} 
    \mbox{\epsfxsize=5.5in\epsffile{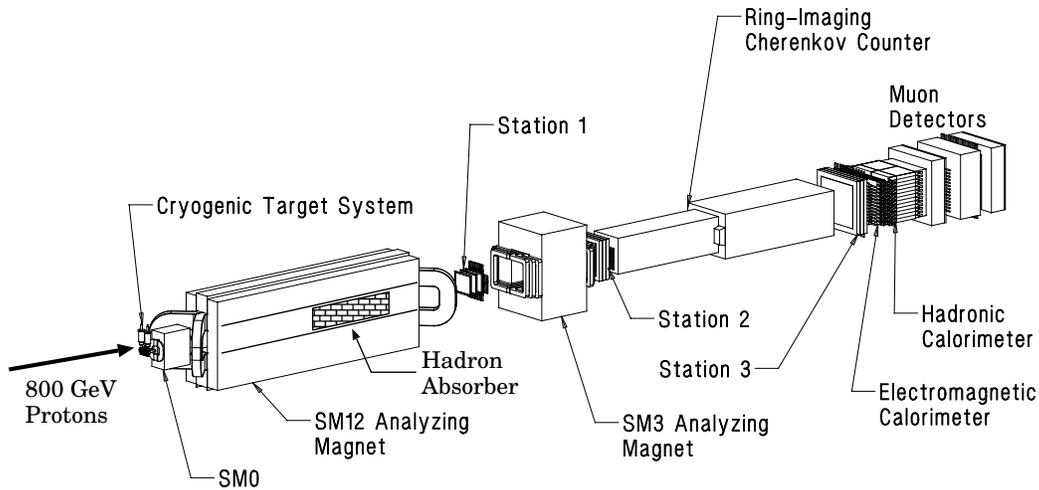}}
    \vspace*{-0.25in}                           
  \end{center}       
  \caption{The FNAL E866/NuSea Spectrometer}             
  \label{fig:spec}
\end{figure}

The primary apparatus used in this experiment was the dimuon 
spectrometer~\cite{moreno} shown in Fig. \ref{fig:spec}.  This was the
fourth experiment to use this spectrometer that was located on the east side of 
the Meson experimental building at Fermilab.  While changes have been made to the 
spectrometer for each experiment, the basic design has remained the same since the 
spectrometer was first used for E605 in the early 1980's.  

Downstream of the targets the particles encountered the first of three magnets.
All three magnets used in this spectrometer were dipole magnets with the magnetic field
aligned in the x direction so that they caused charged particles moving in the 
z direction to bend in the plus or minus y direction.

The first dipole magnet (SM0) was the smallest, measuring 72 inches long.  
It provided a transverse momentum deflection of 0.94 GeV/c when operated at a maximum 
current of 2100 amps.  This magnet was used
to increase the opening angle of dimuon pairs when taking data with 
the magnets configured to have the lowest mass
acceptance.  The aperture of this magnet was filled with a helium bag.  
The purpose of this and all other helium bags was 
to minimize the material that the muons traversed in order
to minimize their multiple scattering.

Following the first magnet was the largest magnet (SM12),  
measuring over 47 feet long.  When operated at a maximum current of 4000 amps,
it provided a transverse momentum deflection of 7 GeV/c.
This magnet focused the dimuons through the 
remainder of the spectrometer.  Sixty-eight inches into  
the aperture of this magnet was a beam dump.  
The protons that did not interact in the target
interacted in the 129 inch long copper beam dump.  The beam dump was
about 22 interaction lengths or 230 radiation lengths.
The dump prevented the beam from passing through
the active elements of the spectrometer.  The beam dump filled the aperture in the 
x direction for most of its length, but was a maximum of ten inches 
thick in the y direction.  This allowed many of the 
dimuons of interest to travel above and below the beam dump.

\begin{figure}                                                     
  \begin{center} 
    \mbox{\epsfxsize=4.0in\epsffile{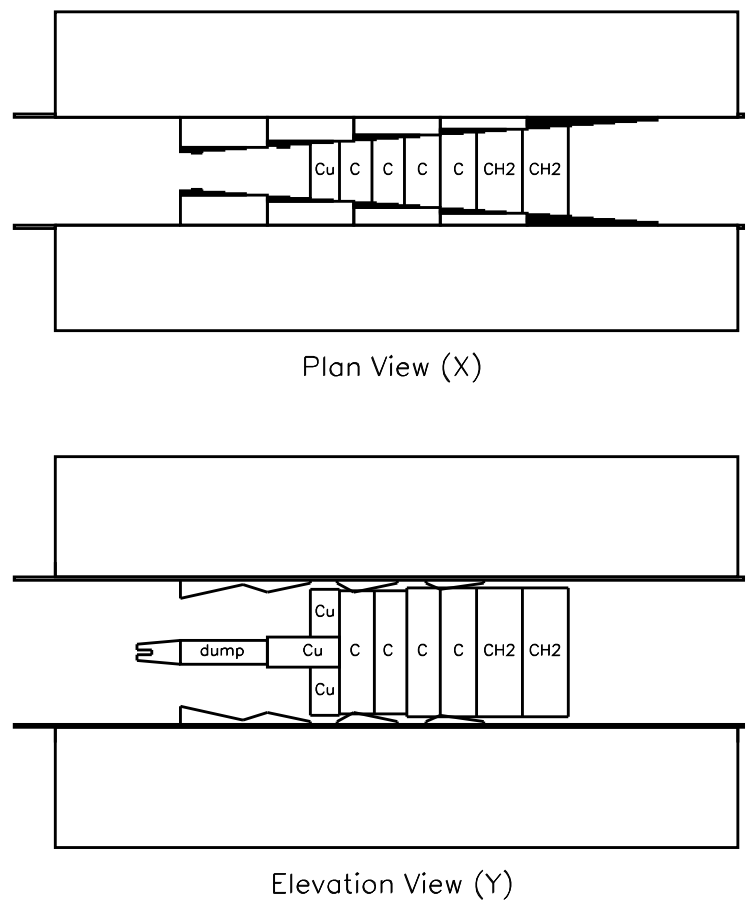}}
    \vspace*{-0.25in}                           
  \end{center}       
  \caption{The beam dump and absorber wall inside the aperture of SM12.  The 
	   beam dump is not shown in the plan view.}             
  \label{fig:dump}
\end{figure}

At the downstream end of the beam dump was an absorber wall 
that completely filled the aperture
of the magnet in both the x and y directions.  This wall, shown in Fig. \ref{fig:dump},
was built in seven layers.  The first layer
was 24~inches of copper, followed by four layers of carbon that were each
27~inches thick.  
The last two layers were each composed of 36~inches of borated polyethylene.  
The effect of this wall, which was over thirteen interaction lengths and
sixty radiation lengths long, was to absorb 
most of the hadrons, electrons, and gammas that were produced in the target or the beam dump.
Essentially, only muons were allowed to traverse the active elements of the spectrometer, allowing 
a higher beam intensity to be used while still keeping the instantaneous number of hits in 
each drift chamber at acceptable levels.
Following the absorber wall and still inside the aperture of SM12 was another helium bag.

Downstream of SM12 was the first of three similar tracking stations (station one).  
Station one consisted of three pairs of drift chambers.  The first pair of planes, called
U1 and U1$'$, had sense wires which were 
oriented at approximately $+14^\circ$ ($\tan \theta = 0.25$) in the x-y plane.  The second 
pair of planes, called Y1 and Y1$'$, were horizontal so that they 
measured the position of the particle in the y direction.  
The third pair of planes, called V and V$'$, were 
oriented at approximately $-14^\circ$ in the x-y plane.
The primed planes were offset in the direction perpendicular to the wires 
by half a drift cell to help resolve the ambiguity of the drift
direction.  These six drift chambers 
were used to track particles accurately and had enough redundancy 
to continue adequate performance
even if some of the channels were inoperative.

While the drift chambers had excellent position 
resolution, they could not distinguish between beam buckets within a spill
and were too slow to use as input to a trigger.  Therefore,
in addition to the drift chambers, planes of scintillators called hodoscopes were used.  
The two hodoscope planes at station one were made of counters which 
were oriented perpendicular to one another.  The plane 
with horizontal paddles
measured the y position of the muons and therefore was called the Y1 hodoscope.
The paddles in this plane were split at x equal to zero leaving  
a $0.47$ inch gap.  
The other plane measured the position of the muons 
in the x direction and was called the X1 hodoscope.  The paddles in 
X1 were split at y equal to zero leaving a $0.38$ inch gap.

Following station one was the final dipole magnet (SM3).  
This magnet had a large aperture, measuring about 
six feet high and five feet wide, which was filled 
with another helium bag.
This magnet was always operated at a current of 4230 Amps, 
which provided a transverse momentum 
deflection of $0.91$ GeV/c.  Unlike the 
first two magnets whose primary purpose was to focus the dimuons into 
the active elements of the spectrometer, this magnet primarily
was a momentum analyzing magnet.  
The bend in the muon tracks as they passed through this magnet was used
to determine the momentum and the sign of the electric charge of the muon.

The second tracking station, which followed SM3, was a slightly larger version of the 
first tracking station.  Like the first station it also had three pairs of drift chambers (called
U2, U2$'$, Y2, Y2$'$, V2, and V2$'$) but, 
unlike the first station, the second station had only one 
hodoscope (Y2) that measured the y position of the muons.  

As shown in Fig. \ref{fig:spec}, the second tracking station was 
followed by a ring imaging Cherenkov counter (RICH).  
In previous experiments the RICH was used for particle identification,
but it was not needed for this experiment. Therefore, it
was filled with helium to reduce multiple scattering of the muons.

Downstream of the RICH was the third tracking station.  
Except for being larger in size, station three
was identical to station one.  It contained three pairs of 
drift chamber planes (U3, U3$'$, Y3, Y3$'$, V3, and V3$'$) 
and two planes of hodoscopes (Y3 and X3).  

Like the RICH, the calorimeters that followed station three 
were not used for this experiment.  
The calorimeter served 
only as additional shielding before the station four detectors.  
Upstream of and between some of the planes in station four were large amounts
of concrete and zinc shielding.  This shielding served 
as a final barrier to all particles except muons 
(and of course neutrinos), so station four was also known as the muon detectors.  

Since station four was so large and because the additional 
shielding upstream of station four caused multiple
scattering of the muons, the precise position resolution of drift chambers was not practical or 
necessary.  Therefore, poorer resolution proportional tubes were 
used in place of drift chambers in station four.  There were three
planes of proportional tubes.  Two planes measured the y position (PTY1 and PTY2) and one plane measured the
x position (PTX).  Station four also had two hodoscope planes (Y4 and X4).

The drift chambers and proportional tubes used the same gas mixture.  The mixture
was 50\% argon and 50\% ethane with a small amount of ethanol added by bubbling 
the mixture through ethanol, maintained at a constant 25$^\circ$~F.  The 
ethanol was added to prevent the buildup of deposits on the wires 
in the drift chambers and to act as a quencher to minimize sparking.

A summary of the physical construction of the drift chambers, hodoscopes, and 
proportional tubes can be found in Tables \ref{tab:dc}, \ref{tab:hodo}, and 
\ref{tab:pt} respectively.  Also included in these tables are the operating high
voltages for each plane.  Average drift chamber efficiencies are listed 
in Table \ref{tab:cheff} for each data set.  Typical hodoscope efficiencies
are listed in Table \ref{tab:hodoeff} for each mass setting.

\begin{table}[tbp]                                                             
\caption{Information on the drift chambers.  Drift cell and aperture sizes are in inches.}
\label{tab:dc} 
\begin{center}
\begin{tabular}{ccccc}
\hline \hline
detector & number 	   & drift cell     & aperture		   & operating  \\
	 & of wires	   & size 	    & (x $\times$ y)	   & voltage	\\
\hline
   Y1    &   160      &     0.25 in.   &     48 in.$\times$40 in.     &      +1700 V        \\ 
   Y1$'$ &   160      &     0.25 in.   &     48 in.$\times$40 in.     &      +1700 V       \\ 
   U1    &   200      &     0.25 in.   &     48 in.$\times$40 in.     &      +1700 V        \\ 
   U1$'$ &   200      &     0.25 in.   &     48 in.$\times$40 in.     &      +1700 V       \\ 
   V1    &   200      &     0.25 in.   &     48 in.$\times$40 in.     &      +1700 V       \\ 
   V1$'$ &   200      &     0.25 in.   &     48 in.$\times$40 in.     &      +1700 V       \\
\hline
   Y2    &   128      &     0.40 in.   &     66 in.$\times$51.2 in.   &      -2000 V       \\
   Y2$'$ &   128      &     0.40 in.   &     66 in.$\times$51.2 in.   &      -2000 V       \\
   U2    &   160      &     0.388 in.  &     66 in.$\times$51.2 in.   &      -1950 V       \\
   U2$'$ &   160      &     0.388 in.  &     66 in.$\times$51.2 in.   &      -1975 V       \\
   V2    &   160      &     0.388 in.  &     66 in.$\times$51.2 in.   &      -2000 V       \\
   V2$'$ &   160      &     0.388 in.  &     66 in.$\times$51.2 in.   &      -2000 V       \\
\hline
   Y3    &   112      &     0.82 in.   &    106 in.$\times$91.8 in.   &      -2200 V       \\
   Y3$'$ &   112      &     0.82 in.   &    106 in.$\times$91.8 in.   &      -2200 V       \\
   U3    &   144      &     0.796 in.  &    106 in.$\times$95.5 in.   &      -2200 V       \\
   U3$'$ &   144      &     0.796 in.  &    106 in.$\times$95.5 in.   &      -2200 V       \\
   V3    &   144      &     0.796 in.  &    106 in.$\times$95.5 in.   &      -2200 V       \\
   V3$'$ &   144      &     0.796 in.  &    106 in.$\times$95.5 in.   &      -2150 V       \\
\hline \hline
\end{tabular}
\end{center}
\end{table}

\begin{table}[tbp]
\caption{Information on hodoscopes.  The counter width, aperture sizes, and the 
center gap size are in inches.}
\label{tab:hodo}
\begin{center}
\begin{tabular}{ccccc}
\hline \hline
detector & number of 	 	& counter  	& aperture	 	& center gap   \\
	 & counters		& width		& (x$\times$y)		& width		\\
\hline
Y1  	&      2$\times$16      &     2.50 in.  &  47.50 in.$\times$40.75 in.  & 0.47 in.  \\
X1  	&      12$\times$2	&     4.00 in.  &  47.53 in.$\times$40.78 in.  & 0.38 in.  \\
Y2 	&      2$\times$16	&     3.00 in.	&  64.63 in.$\times$48.63 in.  & 0.66 in.  \\
X3 	&      12$\times$2   	&     8.68 in.	&  105.2 in.$\times$92.00 in.  & 1.00 in.  \\
Y3 	&      2$\times$13	&     7.50 in.	&  104.0 in.$\times$92.00 in.  & 0.00 in.  \\
Y4	&      2$\times$14	&     8.00 in.	&  116.0 in.$\times$100.0 in.  & 0.00 in.  \\
X4 	&      16$\times$2	&     7.125 in.	&  126.0 in.$\times$114.0 in.  & 0.00 in.  \\
\hline \hline
\end{tabular}
\end{center}
\end{table}

\begin{table}[tbp]
\caption{Information on proportional tubes.  Drift cell and aperture sizes are in inches. }
\label{tab:pt}
\begin{center}
\begin{tabular}{ccccc}
\hline \hline
detector & number 	   & drift cell     & aperture		   & operating  \\
	 & of wires	   & size 	    & (x $\times$ y)	   & voltage	\\
\hline
  PT-Y1  &        120      &     1.00 in.      &  117 in.  $\times$120 in.    &      +2500 V       \\
  PT-X   &        135      &     1.00 in.      &  135.4 in.$\times$121.5 in.  &      +2500 V       \\
  PT-Y2  &        143      &     1.00 in.      &  141.5 in.$\times$143 in.    &      +2500 V       \\
\hline \hline
\end{tabular}
\end{center}
\end{table}

\begin{table}[tbp]                                                             
\caption{Drift chamber efficiency in percent.}
\label{tab:cheff} 
\begin{center}
\begin{tabular}{c|cccccc}
\hline \hline
detector & \multicolumn{6}{|c}{data set} \\
	 & 3    & 4    & 5    & 7    & 8    & 11 \\
\hline
   Y1    & 93.0 & 92.4 & 92.0 & 93.0 & 94.6 & 89.0    \\ 
   Y1$'$ & 94.5 & 93.2 & 97.2 & 94.2 & 94.3 & 93.9    \\ 
   U1    & 97.7 & 97.0 & 97.9 & 98.4 & 98.3 & 98.2    \\ 
   U1$'$ & 96.5 & 94.4 & 95.9 & 96.0 & 95.9 & 95.9    \\ 
   V1    & 97.7 & 96.5 & 96.6 & 98.4 & 98.4 & 95.9    \\ 
   V1$'$ & 98.1 & 97.2 & 98.6 & 99.0 & 99.0 & 98.7    \\
\hline
   Y2    & 95.8 & 95.1 & 95.0 & 95.9 & 96.2 & 95.8    \\
   Y2$'$ & 96.1 & 95.7 & 96.0 & 96.6 & 96.3 & 95.8    \\
   U2    & 96.2 & 95.8 & 95.9 & 96.5 & 96.6 & 96.5    \\
   U2$'$ & 96.0 & 95.6 & 96.1 & 96.3 & 96.6 & 96.9    \\
   V2    & 96.8 & 96.7 & 96.6 & 97.2 & 97.1 & 97.2    \\
   V2$'$ & 96.7 & 96.6 & 96.6 & 97.0 & 96.5 & 96.8    \\
\hline
   Y3    & 94.6 & 94.5 & 93.5 & 96.6 & 94.9 & 94.5    \\
   Y3$'$ & 95.6 & 95.2 & 95.2 & 97.0 & 96.9 & 96.7    \\
   U3    & 95.8 & 95.2 & 95.5 & 97.1 & 97.1 & 96.9    \\
   U3$'$ & 96.4 & 96.0 & 95.9 & 97.8 & 97.7 & 97.5    \\
   V3    & 95.7 & 95.3 & 95.2 & 97.1 & 97.1 & 96.7    \\
   V3$'$ & 95.1 & 94.7 & 94.3 & 96.2 & 95.8 & 95.0    \\
\hline \hline
\end{tabular}
\end{center}
\end{table}

\begin{table}[tbp]                                                             
\caption{Typical Hodoscope efficiencies in percent.  The Y planes are divided 
at $y=0$ and L (R) denotes the left (right) half plane.  The X planes are divided
at $x=0$ and U (D) denotes the upper (lower) half plane.}
\label{tab:hodoeff} 
\begin{center}
\begin{tabular}{c|ccc}
\hline \hline
detector & \multicolumn{3}{|c}{mass setting} \\
	 & low  & int. & high \\
\hline
   Y1L   & 99.9   & 99.1  & 99.6    \\ 
   Y1R   & 99.8   & 99.8  & 99.7    \\ 
   X1U   & 100.0  & 99.8  & 99.7    \\ 
   X1D   & 99.5   & 100.0 & 100.0    \\ 
\hline
   Y2L   & 97.0   & 97.5  & 97.8    \\
   Y2R   & 98.2   & 99.4  & 98.8    \\
\hline
   Y3L   & 100.0  & 99.9  & 99.9    \\
   Y3R   & 100.0  & 99.9  & 100.0    \\
   X3U   & 99.5   & 99.8  & 99.5    \\
   X3D   & 99.5   & 96.8  & 98.7    \\
\hline
   Y4L   & 100.0  & 100.0 & 100.0    \\
   Y4R   & 100.0  & 100.0 & 100.0    \\
   X4U   & 99.9   & 99.1  & 100.0    \\
   X4D   & 100.0  & 99.7  & 100.0    \\
\hline \hline
\end{tabular}
\end{center}
\end{table}

\section{Data Acquisition System}
Although much of the data acquisition system (DAQ) was several years old, it was a 
very fast system.  When the trigger was satisfied, the DAQ would record information
from the spectrometer.  Additionally, the DAQ would also 
record to tape information about each beam spill.  
The following subsections will describe when,
how, and what information was recorded to data tapes.

\subsection{Trigger}
\label{ch:trig}
Events were only written to tape when the trigger was satisfied and only if the DAQ was
not already busy reading out a previous event.  
The trigger was optimized to detect events that consisted 
of two muons originating from the target. 
A new trigger system was implemented for E866 that
used only the signals from the hodoscopes to determine if the event should be written 
to tape \cite{trig,hawker}.  

The signals from the Y1, Y2, and Y4 hodoscopes were compared with the contents of 
a three dimensional look-up table.  This table was generated by Monte Carlo studies 
of dimuon events from the target.  If the hits in the scintillators matched one of
the possible dimuon trajectories through the spectrometer according to the look-up table, the 
trigger was satisfied.  
Once the trigger was satisfied, the readout busy signal was set until the event was
completely recorded in the VME system memory.

In addition to the standard physics triggers that were optimized to detect 
oppositely charged dimuon events from the targets, other triggers were prescaled
to record a limited number of study events.  These study events included single
muon events,  events that relied only on the x hodoscope planes and other 
diagnostic triggers.  One of the most important study triggers was the 
trigger that was satisfied by two like-sign muons from the target area 
that traveled down opposite sides of $x=0$ in the spectrometer.  
The importance of some of these specialty triggers will
be discussed in section \ref{sect:randoms}.

\subsection{Readout System}

Once the trigger was satisfied, it was the responsibility of the readout system to 
record to tape all of the needed information to properly reconstruct the event.  
The readout system was composed of Nevis Transport electronics~\cite{dan}, 
VME modules, and CAMAC modules.  
All detector subsystems fed data onto the Transport
16-bit bus.   Transport read out the coincidence registers (CR) from 
the hodoscopes and proportional tubes, time-to-digital converters (TDC) from 
the drift chambers, and a variety of information about the trigger condition from
the trigger bit latches.  All of the data fed onto the Transport bus were then
transferred to a VME-based memory buffer.  The VME system would then
format the data and write it to 8mm data tapes without increasing 
the detector deadtime by taking advantage of the forty seconds each minute 
when there was no beam in the spectrometer~\cite{ting}.

In addition to the information from the spectrometer recorded for each event, 
the readout system also recorded information necessary to the analysis from each
beam spill.  This information included the beam intensity, position, size and duty
factor, the pressure, temperature, and position of the targets, magnet voltages and 
currents, and various monitors used for calculating the readout deadtime. 
The inclusion of these data with the event data was one of the major
improvements in the DAQ VME system originally built for E789.

%



\chapter{Experimental Procedures}

This chapter focuses on the procedures used in this experiment.
Preparation for this experiment started years before
the first data were taken, and work continued throughout the year of data collecting.
This chapter is a description of this work, in which the author
played a significant role.  The first two sections detail
what was done to recommission the spectrometer before the 
accelerator was turned on.  The third section will describe the work that was done
to tune up the spectrometer after receiving beam, but before data could
be taken.  Section four will explain the standard data taking procedures, and
the last section will describe special procedures that were required periodically. 



\section{Recommisioning the Spectrometer}

Many of the detectors, cables, and readout electronics were unchanged from previous 
experiments that used this spectrometer.  However, this equipment lay idle for 
five years in a dirty environment that was subject to extreme 
temperature and humidity fluctuations.  Therefore, many months 
of work were necessary to restore the spectrometer to operating conditions.  

The cables that carried the signals from the drift chambers to the TDC's 
were 100 wire ribbon cables that were several hundred feet long called Ansley cables.  
These cables were extra long so that the signals from the drift chambers
would be delayed while the trigger information was being processed.
The connectors at the ends of these cables were large paddles with exposed wires.
The rough treatment these cables experienced since they were last used 
damaged many of them.  Each cable was untangled, each wire was tested, 
and many repairs were necessary.  Finally the cables were reattached 
to the detectors and the readout electronics.

The station three and four hodoscopes were all checked for light leaks.  The station 
two hodoscope paddles had become severely misaligned during the past five years and had to be 
remounted.  The signal and high voltage cables were checked, repaired if necessary, and 
reattached.  All of the electronics from the preamplifiers all the way to the tape drives 
were tested and repaired as necessary.

\section{Commissioning the New Parts of the Spectrometer}
The spectrometer was upgraded in many areas for this experiment.  
The station one hodoscopes and drift chambers were replaced with 
larger detectors in the y-direction, to almost double the 
acceptance of the spectrometer.
The drift chambers had to be designed, built, mounted, 
aligned, and tested.  The hodoscopes were expanded by adding
eight new scintillator paddles to Y1 and by replacing all of 
the scintillators in X1 with longer scintillators. 
The scintillating material used in stations one and two 
was 3~mm thick BC-408. 
The new dimensions of the station one detectors are shown in 
Tables \ref{tab:dc} and \ref{tab:hodo}.

The trigger was also completely replaced.  The old trigger processor was removed and several 
new trigger modules were designed, built, installed, and tested~\cite{trig,hawker}.  
The new trigger, described in Section \ref{ch:trig}, performed 
extremely well.

The DAQ underwent a serious overhaul from its previous configuration.  The
previous experiments used a PDP-11 to control parts of the DAQ.\@  
The VME part of the system was enhanced to replace these functions and to 
handle much higher data rates.

Another dipole magnet (SM0) was added to the front of the spectrometer.  
Because of the coupling of the two magnets, the magnetic
fields in SM0 and the upstream end of SM12 had to be measured.  
This was done by measuring the magnetic fields
on a three dimensional grid throughout the aperture of the magnets.

\section{Spectrometer Tune-up with Beam}

Once the accelerator was turned on and beam was delivered to the Meson Center
beam dump, tuning of the spectrometer commenced.  
The proton beam hitting this beam dump, which was about 1400~ft upstream of 
the spectrometer, produced enough muons that some were detected. 
This low flux of muons coming through the spectrometer allowed for the  
alignment of the hodoscopes and drift chambers as well as 
a test of the new trigger and DAQ systems.

When the proton beam finally came down the Meson east beamline, 
the high voltage settings for the 
hodoscopes and drift chambers were rechecked.  
These voltages were optimized to perform at the 
maximum efficiency without causing excessive noise or damaging the detectors.  
The timing of each hodoscope channel 
was also checked and adjusted to ensure that 
all of the hodoscope signals were in
coincidence with each other to within approximately one nanosecond.  
This provided one beam bucket resolution  
and was critical to ensure that the trigger worked properly.

\section{Standard Data Taking Procedures}

Throughout the year of data-taking the spectrometer,
DAQ, beam, targets, magnets, and the gas systems were constantly monitored.  
Twice a day a shift check was performed.  This checked the target pressure and temperature,
the drift chamber and proportional tubes gas flow, gas pressure, and high voltages, beam duty
factor, and magnet settings. 

To better monitor the spectrometer performance and data quality 
during data-taking, a portion of the data was analyzed in real time.
A fraction of the data events that were recorded to tape were 
routed across the network to a Silicon Graphics workstation.  
The on-line analysis of these events provided an excellent 
means of monitoring the spectrometer performance.  

Among the many means of evaluating the spectrometer performance 
provided by the on-line analysis, were hit distributions in the 
hodoscopes, drift chambers, and proportional tubes.
These distributions would quickly show any channels that 
suffered from the two most common modes of failure, which were
not ever showing a hit or showing a hit for every event.
The performance of the readout system was monitored 
by the on-line analysis.  If the format of a data
event was improper, then the analysis generated an error.
The on-line analysis also gave some indication 
about the data quality.  Even though only a fraction of 
the events were analyzed, the reconstructed events were 
sufficient to produce a simple $J/\psi$ or $\Upsilon$ 
peak in the mass spectrum. 

\section{Special Data Taking Procedures}
\label{sect:special}

In addition to the normal daily activities required to take data, some special
measurements were taken periodically.  The SEM counter calibration was checked using
a single beam measurement.  Instead of splitting the beam into several separate channels 
and sending it to different experimental areas at the same time, as was the usual practice,
the beam was delivered only to the Meson east beam line.  This allowed the SEM counter in this
beamline to be calibrated against the beam intensity monitors in the Tevatron.

Another type of special measurement was the hodoscope 
efficiency studies.  Since the normal trigger relied on 
signals from the hodoscopes, it was not very effective 
at monitoring the hodoscope efficiency.  Therefore,
dedicated data taking with a different trigger was necessary.  
These runs were performed for each data set.



\chapter{Analysis}
\label{ch:analysis}

This chapter will explain the process that was used to analyze the data.
First, the data will be described.  Then, the first pass analysis of all of the 
data will be discussed.  Next, the analysis of the data summary tapes (DST),
also referred to as the second pass analysis, will be explained.  The 
final set of cuts implemented on the output of  
the second pass analysis and their justification will be reviewed.  
After the corrections for randoms, rate dependence, and 
target composition have been explained, then the Drell-Yan
cross section ratio will be calculated.  Finally, the ratio of 
$\bar d(x)$ to $\bar u(x)$ will be determined.

\section{Data Sets}
\label{sect:data}
\begin{figure}
  \begin{center}
    \mbox{\epsfxsize=5.0in\epsffile{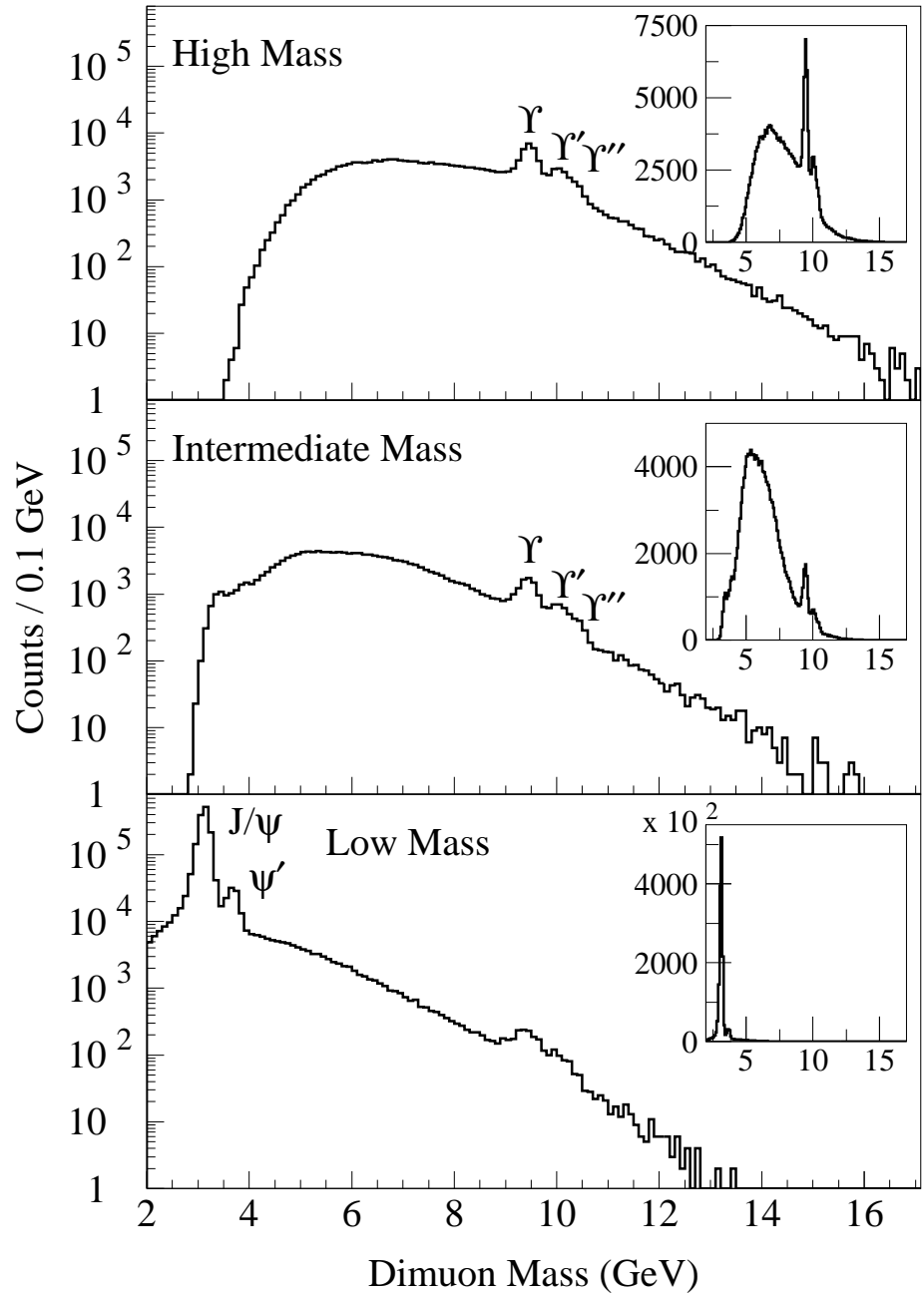}}
    \vspace*{-0.25in}                                
  \end{center}       
  \caption{The dimuon mass distributions for the three different mass settings.
		The inset figures are the same spectra shown on a linear scale.}
  \label{fig:3mass}                    
\end{figure}

The data recorded to tape were categorized according to the dimuon mass 
that had the maximum acceptance.  The spectrometer acceptance was varied
by changing the current and polarities in the first two magnets in the spectrometer.  
These three categories were referred to as high, intermediate, and low mass 
settings.  The motivation for these names can easily be understood
by referring to Fig.~\ref{fig:3mass}, which shows the dimuon mass distributions
for the three mass settings.

The data were further divided into data sets based on the magnet 
polarity, beam duty factor, and target purity.  
Of the eleven data sets recorded, seven 
contained data that were useful to this analysis.
These seven data sets constituted over 90\%
of all the data based on luminosity and are summarized in Table \ref{tab:datasets}.

\begin{table}[tbp]
\caption{Summary of the data sets.  The size of each data set is shown
as the number of Drell-Yan events that passed all data cuts rounded to the nearest thousand.  
All magnet currents are in amperes. 
The deuterium fill refers to the quality of the deuterium in the target
as described in Section \ref{sect:target}.} 
\label{tab:datasets}   
\begin{center}                                                          
\begin{tabular}{cccccc}               
\hline \hline
 mass	 & ~data~   & Drell-Yan & SM0     &  SM12       & deuterium \\
 setting & set    & events    & current & ~current~     &  fill \\ \hline \hline
              &    3     & 78 K &     0       &    2800 A    & first \\
 int.	      &    4     & 50 K &     0       &   -2800 A    & first \\ 
              &    9     & 17 K &     0       &    2800 A    & second \\ \hline
 low	      &    5     & 89 K &   -2100 A   &    2800 A    & first \\ \hline
  	      &    7     & 37 K &     0       &    4000 A    & first \\
 high  	      &    8     & 80 K &     0       &    4000 A    & second \\
              &   11     & 24 K &     0       &   -4000 A    & second \\
\hline \hline
\end{tabular}
\end{center}
\end{table}

The four data sets that were not used in the analysis described here were 
either poor quality data or were not useful to this measurement.  
Data sets one and two were the first data taken and the beam duty factor was
extremely poor.  Data sets six and ten are good quality data, but are not
useful to this analysis.  Data set six was a fourth mass setting that 
focused the spectrometer acceptance below the $J/\psi$ where other production
mechanisms besides Drell-Yan contribute significantly to the continuum.  Data
set ten was a small data set taken in the low mass setting, but used 
the deuterium mixture with the slight hydrogen contamination.  The 
decrease in statistical uncertainty gained by including the events 
from data set ten would be insignificant, while the hydrogen 
contamination would increase the systematic uncertainty.  
Therefore, only the data sets listed in Table \ref{tab:datasets}
were used in this analysis.

\section{First Pass}

The first pass analysis of all the data was done on Fermilab's IBM
parallel computing UNIX farms.  The main analysis code used for both the first and 
second pass analysis was originally written and used for the E605 
experiment.  Since then many changes have been made to the code,
but the basic algorithm has remained unchanged.
Since only about one percent of all the events written to tape 
reconstructed to form a dimuon event from the target, the purpose 
of the first pass analysis was to reduce the raw data tapes
to DST's.

The analysis code started by running initialization and setup routines 
for each data run.  This ensured that the proper spectrometer settings,
such as trigger and magnet maps, were used for the analysis of each run.
Each event was then translated from the output format of the DAQ system to a 
format more suitable for analysis.  Once all of these preliminary tasks
were completed, the code reconstructed the event.  
Reconstructing an event is the task of converting a list of 
drift chamber, hodoscope, and proportional tube hits in the 
spectrometer to tracks of particles responsible for the hits and finally 
combining pairs of tracks to form a dimuon event.      


The reconstruction process began by looking at the hits in the station two
and station three drift chambers.  If at least four 
of the six planes in a station all had 
a hit which could have been produced by a single particle, this position
was considered to be a possible track reconstruction point.  
Some of these points were eliminated by requiring that the location of
the point be consistent with hits in the hodoscopes that satisfied the trigger.
Once all of these
points were determined, an iterative process paired a reconstruction point in
station two with another point from station three and determined if the track segment
generated by connecting the two points could have been produced 
by a single charged particle originating at the target.

Once all of the possible track segments between stations two and three were
determined, they were compared with hits in stations one and four to 
further reduce the number of tracks.  The station two to station three 
track segment was extended upstream to the SM3 bend plane.  Since at this
point in the reconstruction process the sign of the electric charge and
the momentum of the particle are still unknown, a region of possible station
one positions are searched for a possible track reconstruction point.  
If at least
four of the six planes had a hit that could have been produced by a single
charged particle in the region identified by extending the station two to
station three track segment to station one, then station four hits were checked.

The final check that the candidate track was produced by a 
muon used the hits at station four.  At this point in the 
reconstruction process, the track extends from station one
through SM3 to stations two and three.  Since the track went through the 
magnetic field produced by SM3, the charge and momentum of the particle were 
determined.  The track was extended downstream 
to station four.  The absorbing material between station three and station
four ensured that only muons would be detected in the station four detectors. 
To validate a track, at least three of the five detector planes in station four must
have had a hit at the position predicted by the candidate track.  

Once the position, charge, and momentum of the particle were known at station
one, the track was extended upstream to the target.  Between station
one and the target were two dipole magnets, the absorber wall, and the beam
dump.  The track was reconstructed in a series of small variable length steps,
which were typically a few inches long.  At each step the particle's 
trajectory was bent by the magnetic field, and when traveling through 
the absorber wall or the beam dump, energy was added back to the particle 
to correct for the energy that was lost while traversing 
the material.  To correct for the multiple scattering that occurred in
the material, a single bend plane approximation was used.  This correction
slightly modified the particle angle at the effective 
multiple scattering plane to force the track to 
originate at the average interaction point inside the target.

After the individual tracks were fully reconstructed, they were 
paired~\footnote{Less than 0.08\% of all the good events contained more than 2 tracks.} 
to reconstruct the dimuon event.  From the charge and momentum of the 
individual tracks, the kinematics of the event were determined.  

\section{Second Pass}

Once the data had been reduced by almost two orders of magnitude from the raw
data tapes to the data summary tapes, the analysis code was optimized by 
repeated analysis of the DST's.  This second pass analysis was completed 
on Hewlett Packard workstations.  While the basic analysis code was not
changed for the second pass analysis, many small changes were made to 
improve mass resolution and to study systematic effects.

One of the changes that was made to the second pass analysis was the
addition of a small magnetic field in the y-direction inside the aperture
of SM12.  This was added because the data showed a slight focusing 
in the x-direction.    Another improvement to the 
second pass analysis was the placement of the effective 
multiple scattering plane.  Previous analyses had
used a fixed position, but it was determined that the optimal position
depended on how much of the copper beam dump the muon traversed.  If the
muon passed through the entire beam dump, then the effective multiple scattering plane 
should be closer to the front of the beam dump.  If the muon traveled above
or below the beam dump, then the effective multiple scattering 
plane should be closer to the absorber wall.  Therefore, the 
analysis was improved to track the muon to the front of the dump, 
and then the position for the effective multiple scattering plane 
was determined and the muon was retraced using the 
effective multiple scattering plane.
These improvements to the event reconstruction 
improved the mass resolution.

After the second pass analysis was completed, the final results, along with
many intermediate quantities, were stored in large arrays called ntuples~\cite{paw}.  An
ntuple is a two dimensional array where one dimension is equal to the number
of events and the other dimension contains information from each event.  In
addition to the event information stored in these ntuples, information was
recorded about each beam spill.

\section{Ntuple Cuts}
\label{sect:finalcuts}

Once the second pass analysis was completed and ntuples were produced, 
the data were subjected to a final set of necessary conditions known as 
ntuple cuts.  These cuts were carefully studied to remove bad events 
from the data, leaving only good Drell-Yan events.  While the reason 
for these cuts is consistent between data sets, some of the actual values
changed between sets of data. This section describes
these cuts.

Before actual event cuts were made, the beam spill was required to meet 
certain criteria.  The beam duty factor, the readout live time, and the 
beam intensity were all required to exceed some minimum value or
all of the events from that spill, were cut.  These cuts were 
made primarily to remove events that came from spills with poor beam quality
which could introduce additional systematic uncertainties.

Some cuts were made to ensure that the reconstructed individual tracks 
were good.  One cut ensured that the tracks stayed inside of the x aperture 
of SM12 by making cuts on the track angle at the target in the x direction.  
Another cut on the y angle of the track at the target prevented tracks with 
a very small y angle from going on the wrong side of the dump.  The last
cut on tracks ensured that the reconstructed x and y position at the interaction
point was near the center of the target.

The next group of cuts was applied to the reconstructed event.  If the sign
of the electric charge of the two tracks was not different, the event could
not be a Drell-Yan event, so the event was cut.  
The z value of the uniterated reconstructed 
interaction point
\footnote{This point was the z position of the closest approach of the two
tracks in the y-z plane before the multiple scattering correction was made.}
of the pair of tracks was required to be near the center of 
the target.  Also the reconstructed event must satisfy the trigger that fired.
So if the trigger bits that should have been fired by the reconstructed event 
were not present, the event was cut.  
Events with too many total hits were cut to put an upper limit on
the allowable background noise in an event.  

Finally, three cuts were made on the kinematic quantities of the pair.  The
reconstructed momentum of the dimuon pair must not be greater than the momentum of the 
incident proton to conserve momentum.  Any event that appears to violate the 
conservation of momentum was cut.  Another cut limited the maximum transverse 
momentum ($p_T$) of the pair.  The final kinematic cut was on the 
reconstructed dimuon mass.  This cut was used to remove the $J/\psi$ and $\Upsilon$ 
resonance families from the Drell-Yan continuum~\footnote{The typical mass
resolution at the $J/\psi$ was 100 MeV.}.

In addition to all of the above ntuple cuts, which were made on 
all three groups of data and summarized in Table \ref{tab:cuts}, each 
data set had some unique ntuple cuts.  Most of these cuts removed
data not useful for this analysis.  
The low and high mass data required AMON to have at least one
count, and the intermediate mass data required the ratio of the beam 
intensity as measured by IC3 and SEM to be less than eight.  

\begin{table}[tbp]
\caption{Summary of all ntuple cuts.} 
\label{tab:cuts}
\begin{minipage}{5.5in}   
\renewcommand\footnoterule{}
\begin{center}                                                          
\begin{tabular}{lccc}
\\*[.01in]               
\hline \hline
 ntuple cut 				& high 	& int.	 	& low 	\\ \hline
 min beam duty factor		& 25\%	& 50\%		& 25\%	\\ 
 min DAQ livetime			& 0.9	& 0.9		& 0.9	\\ 
 min beam intensity measured by IC3 & 5000	& 10000		& 5000  \\ \hline
max magnitude x angle at target & 0.026 & 0.026 & 0.026 \\ 
min magnitude y angle at target & 0.001 & 0.001 & 0.0023 \\
max radial distance from center of target & 2.45 in. & 2.45 in.	& 2.45 in. \\ \hline
min z of the uniterated interaction point & -70 in. & -60 in. & -55 in. 
	\footnote{This value was raised to -50 inches if the dimuon mass was less than 4.5 GeV.}
									\\
max z of the uniterated interaction point & 90 in. & 75 in. 	& 80 in.	\\
max noise \footnote{This quantity is twelve times the number of hodoscope hits plus the
			total event length, measured by the number of six digit octal words
			in the raw event.}
					& 1300	& 1400 		& 1400	\\ \hline
max dimuon momentum (GeV/c)		& 800	& 800		& 800   \\
max dimuon $p_T$ (GeV/c)		& 7.0	& 6.0		& 6.0   \\
min dimuon mass	 (GeV)			& 4.5	& 4.3		& 4.0	\\ \hline
max mass allowed below $\Upsilon$ (GeV)	& 9.0 	& 8.8		& 8.8	\\ 
min mass allowed above $\Upsilon$ (GeV)	& 10.7	& 10.8		& none
		\footnote{No events above the $\Upsilon$ are included in the low mass data.}
									\\ \hline
\hline
\end{tabular}
\end{center}
\end{minipage}
\renewcommand\footnoterule{%
  \kern-3\p@
  \hrule\@width.4\columnwidth
  \kern2.6\p@}
\end{table}

\section{Randoms Correction}
\label{sect:randoms}

A background that must be corrected for was the random coincidence of two unrelated 
oppositely charged muons produced in the target by different 
protons during the same beam bucket.
Since the production mechanism for these dimuons is not the Drell-Yan
process, but the random coincidence of two single muons, these events
are called randoms.  To determine the number and kinematic distribution
of these events, two of the specialty triggers described in Section
\ref{ch:trig} were used.

The first of these specialty triggers recorded a fraction of the
events that appeared to be a single muon from the target.  
These single muon events were analyzed just like a single track in
a dimuon event.  Then these single tracks were randomly combined to
form a dimuon event.  To check the kinematic distribution of these
combined singles events and to normalize them so that they could be 
used to remove random events from the data required the second 
specialty trigger.

The second specialty trigger was satisfied by two muons from 
the target area that were
of the same sign electric charge and traveled down opposite sides of 
$x = 0$ in the spectrometer.  
These events were analyzed like all of the oppositely charged 
dimuon events once one of the tracks was reflected around $y = 0$.
These events, which are a subset of all the random events,
provided a standard by which the combined singles events 
could be normalized.  

The data were corrected for random dimuons by subtracting 
the normalized combined singles events from the dimuon events.
Since most of the combined singles events reconstructed to a low
effective dimuon mass, the randoms correction was largest 
in the low mass data and smallest in the high mass data.  
The average correction for each mass setting is shown in
Table \ref{tab:randoms}.

\begin{table}[tbp]
\caption{Size of the randoms correction for each mass setting.} 
\label{tab:randoms}
\begin{center}                                                          
\begin{tabular}{ccc}               
\hline \hline
mass setting	& percent of random events & average mass of randoms  \\ \hline
low		& $4.1$\% 	&  4.5 GeV \\ 
intermediate	& $2.9$\% 	&  5.1 GeV \\ 
high		& $0.2$\%  	&  5.4 GeV \\ \hline \hline
\end{tabular}
\end{center}
\end{table}
 
\section{Rate Dependence Correction}

The rate dependence correction is made to remove the effect of any
inefficiency in our ability to detect and reconstruct events that
occurs as a function of beam intensity.  The primary source of this
inefficiency is believed to be lost drift chamber hits due to the single hit TDC's
that were used.  In simple terms,
\begin{equation}
N_m(s) = N_t(s) \times e(s)
\label{eqn:rateproblem}
\end{equation}
where $s$ is the beam intensity for the spill, $N_m(s)$ is the number of events
that are reconstructed at a given beam intensity, and $N_t(s)$ is the number of
true events that actually occurred and would have been accepted at
that beam intensity.  ($N_t(s)/s$ is actually a constant, independent of beam intensity.)
Finally, $e(s)$ is the efficiency for detection and reconstruction as
a function of beam intensity.  

\begin{figure}
  \begin{center}
    \mbox{\epsffile{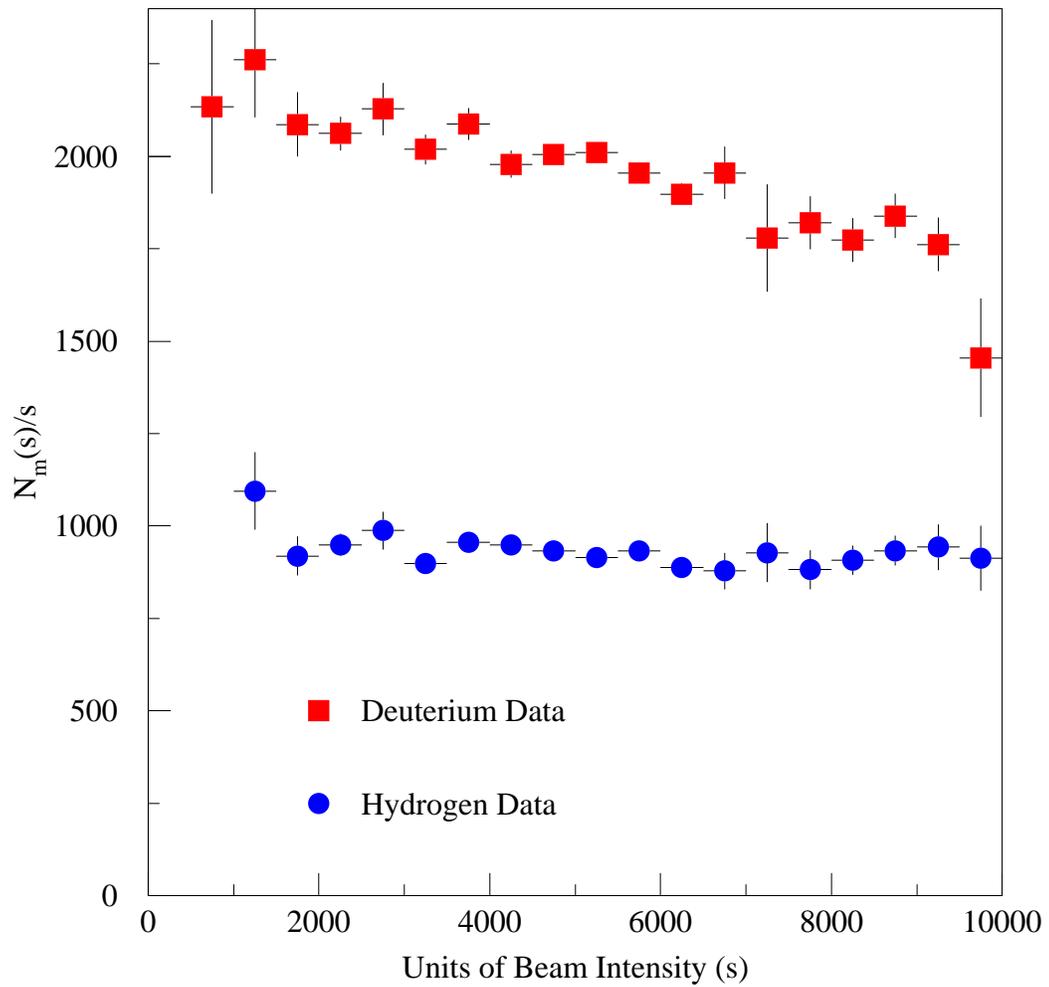}}
    \vspace*{-0.25in}                                
  \end{center}       
  \caption{The rate dependence problem shown here for the low mass data.  
           The yield of Drell-Yan events per unit of beam intensity
	   is shown versus the beam intensity for both the hydrogen and deuterium events.}
  \label{fig:rate_problem}                    
\end{figure}

The drop in reconstruction efficiency 
as the beam intensity increases can clearly be seen in 
the low mass data shown in Fig.~\ref{fig:rate_problem}.
This figure shows that the yield of Drell-Yan events per unit beam intensity decreases
as the beam intensity increases.  But of even more importance to the measurement 
of the cross section ratio is the fact that the deuterium events suffer from a larger 
inefficiency than do the hydrogen events. 

In order to correct the data, the reconstruction efficiency as a 
function of the beam intensity ($e(s)$) must be determined. 
To determine $e(s)$, fits were made to $N_m(s)/s$ as a function of beam intensity.  
So that no unfounded assumptions would be made, the functional form of 
the rate dependence was studied at length.  

\subsection{Functional Form of the Rate Dependence}

The data suggest that the reconstruction efficiency drops 
in a linear manner and this basic assumption was justified by extensive 
Monte Carlo simulations.  
The reason that Monte Carlo simulations were necessary is because
of the lack of sufficient statistics in the data for this study 
as well as the need to verify
that the functional form was not changing at the limits of our data.

\begin{figure}
  \begin{center}
    \mbox{\epsffile{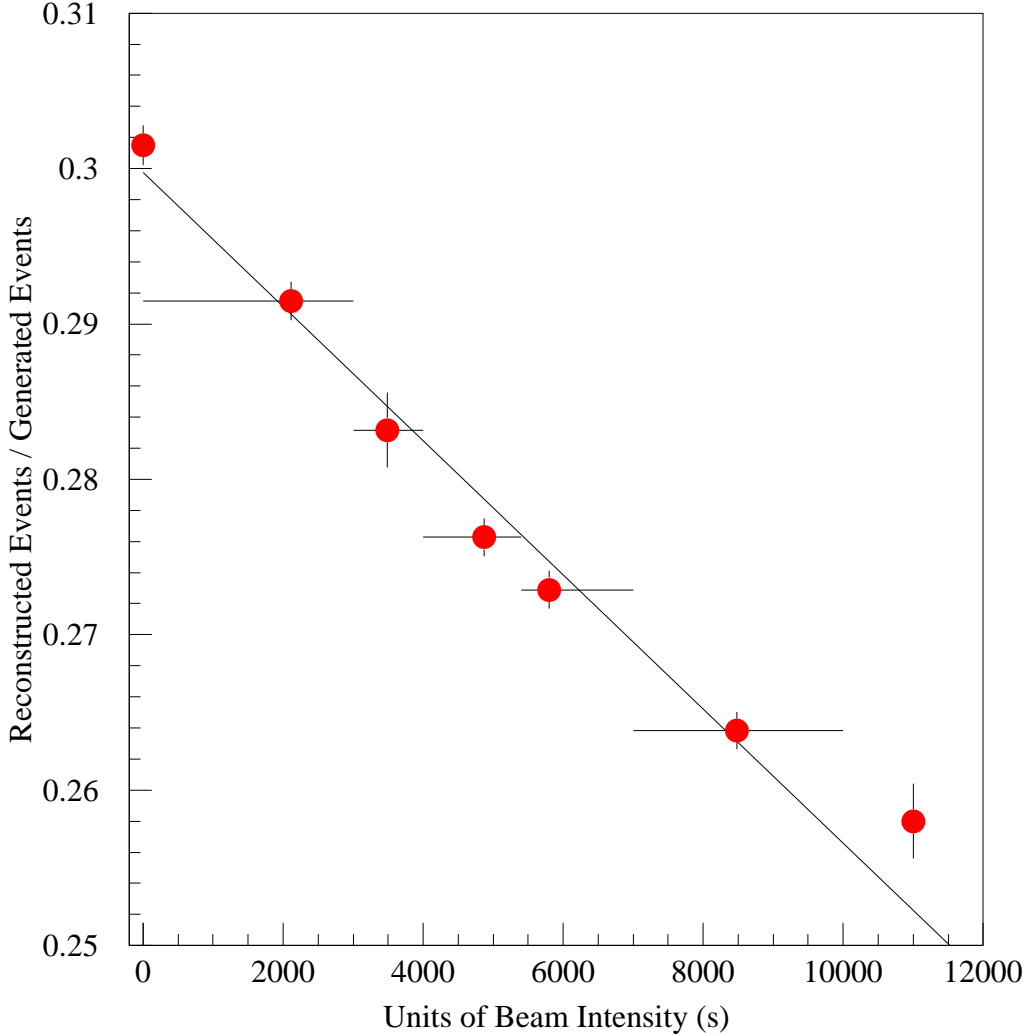}}
    \vspace*{-0.25in}                                
  \end{center}       
  \caption{Monte Carlo study of the rate dependence.  The horizontal error
	   bars indicate where the data were divided into bins of beam intensity.
           These points are plotted at the average beam intensity within the bin.
	   The highest and lowest points in beam intensity do not have horizontal 
	   error bars because there were no data at these beam intensities.}
  \label{fig:mcrate}                    
\end{figure}

The data were broken into different bins based on beam intensity so that
Monte Carlo events could be generated for different beam intensities.  For
each bin, the noise in the drift chambers was carefully studied.  Specifically,
the number, position, correlation between planes within a station, and 
the correlation between stations of the noise hits in the drift chambers
were determined.  This allowed Monte Carlo events to be generated with 
the proper background of chamber noise hits for each different bin in beam
intensity.  In addition to generating Monte Carlo events over the 
entire range of beam intensity covered by the data, the chamber noise 
was extrapolated to both higher and lower beam intensities.  This 
ensured that the functional form was valid over the entire range of 
the data.
These Monte Carlo events were then analyzed using the same analysis methods
and cuts as the data.  The results showed that the reconstruction 
efficiency did decrease linearly with increasing beam intensity as shown
in Fig.~\ref{fig:mcrate}.  
This Monte Carlo study was performed  
for both $J/\psi$ and Drell-Yan events with consistent results.

Another concern was that the rate dependence might also be a function
of the kinematics of the dimuon event.  This potential dependence was 
searched for in both the data and Monte Carlo events by breaking the
events into two or three bins based on the kinematic variable being studied.
The rate dependence was then calculated for each bin.  These studies 
were performed for Monte Carlo Drell-Yan events broken into $x_2$ and
mass bins, Monte Carlo $J/\psi$ events broken into $p_T$ and $x_F$ bins,
data Drell-Yan events broken into $x_2$ bins, data $J/\psi$ events 
broken into $x_F$ bins, and data events broken into mass bins.  
Despite this extensive search for some kinematical dependence to the 
rate correction, no dependence was found within the limits 
of the statistical uncertainty.  
Therefore, the rate dependence correction was determined to be 
a linear function dependent only on the beam intensity.

If each data set were treated independently, there 
would be 28 parameters (a slope and an intercept for each 
target and for each of seven data sets).  
However, these parameters
are not independent and therefore the number of parameters
can be reduced to 16.

The reconstruction efficiency function was determined for each mass 
setting: high, intermediate, and low.
Within each of these mass settings, the intercept of the 
deuterium data~($D_i$) and the hydrogen data for each data 
set~($i$) was related by a single 
factor common to all data sets~($E$).
Additionally the relative slopes of the deuterium data from 
all data sets within a mass setting are the same~($R_d$)\@.  Likewise the hydrogen
data share a common relative slope~($R_h$)\@.  This means that for each data set $i$ within
a mass setting, the deuterium data can be fit to
\begin{equation}
\frac{N_m^d(s)}{s} = D_i \times \left(1 + R_d s \right)
\label{eqn:rated}
\end{equation}
and the hydrogen to
\begin{equation}
\frac{N_m^h(s)}{s} = D_i E \times \left(1 + R_h s \right).
\label{eqn:rateh}
\end{equation}  
From these equations and equation \ref{eqn:rateproblem}, it is
easy to identify the efficiency functions as 
\begin{equation}
e_d(s) = 1 + R_d s
\label{eqn:ed1}
\end{equation}
and
\begin{equation}
e_h(s) = 1 + R_h s.
\label{eqn:eh1}
\end{equation}

\begin{figure}
  \begin{center}
    \mbox{\epsffile{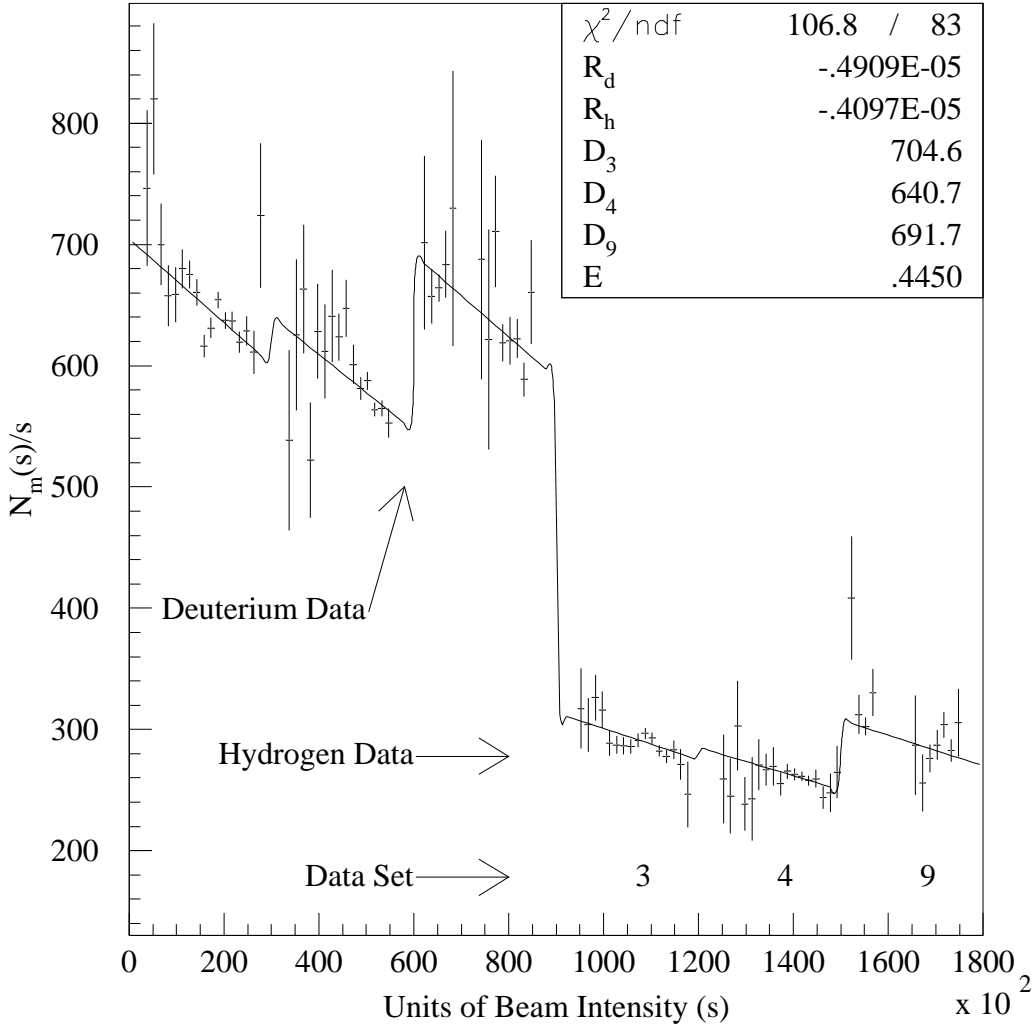}}
    \vspace*{-0.25in}                                
  \end{center}       
  \caption{Number of deuterium (left) and hydrogen
		(right) events per unit of beam intensity as a function of the beam intensity.  
		Both the deuterium and hydrogen data for all three of the 
		intermediate mass data sets are shown here.  The hydrogen
		data are offset by 90,000 units of beam intensity.  Within
		the deuterium and hydrogen data, data sets three, four and
		nine are offset by 0, 30,000 and 60,000 units of beam intensity
		respectively.}
  \label{fig:rateuncor}                    
\end{figure}

Figure \ref{fig:rateuncor} shows the intermediate mass data 
fit with equations \ref{eqn:rated} and \ref{eqn:rateh}.
The hydrogen data are offset by 90,000 units of beam intensity.
Within the hydrogen and deuterium data, data sets three, four, and nine 
are offset by 0, 30,000 and 60,000 units of beam intensity respectively.
This figure shows how the twelve parameters for the intermediate mass
data have been reduced to only six.  However there remains one
more constraint that should be imposed on this fit. 
The slopes $R_d$ and $R_h$ are correlated.
The explanation of this correlation will be discussed in 
detail in the following subsection.

\subsection{Relation Between the Rate Dependence in the Two Targets}

The unwanted tracks through the spectrometer, which caused us to lose the 
tracks we want, came from the target or the dump.  Obviously when we change the 
target, the dump does not change.  The only change in the events from the 
dump when we change the target is due to the different fraction of the beam 
that makes it through the target.  When the target is hydrogen, 93\% of the incident beam 
reaches the dump, while only 85\% makes it to the dump when the target is deuterium.  
Let
\begin{equation}
N_t(s) = A \times s,
\end{equation}
\begin{equation}
e_d(s) = 1 - (0.85B_{\rm dump} + B_d) s,
\label{eqn:ed2}
\end{equation}
and
\begin{equation}
e_h(s) = 1 - (0.93B_{\rm dump} + B_h) s.
\label{eqn:eh2}
\end{equation}
Here $A$ is the yield per beam intensity with no rate dependence and 
$B_{\rm dump}$, $B_d$, and $B_h$ are the beam intensity dependent backgrounds that cause us to lose 
events and originate in the dump, deuterium, and hydrogen respectively. 

Then using equation \ref{eqn:rateproblem} yields
\begin{equation}
\frac{N_m^d(s)}{s} = A_d \left[ 1 - (0.85B_{\rm dump} + B_d) s\right]
\end{equation}
and
\begin{equation}
\frac{N_m^h(s)}{s} = A_h \left[1 - (0.93B_{\rm dump} + B_h) s\right],
\end{equation}
so the relative rate dependence is
\begin{equation}
\frac{N_m^d(s)/s}{2 N_m^h(s)/s} = \frac{A_d \left[1 - (0.85B_{\rm dump} + B_d) s\right]}
{2 A_h \left[1 - (0.93B_{\rm dump} + B_h) s\right]}.
\end{equation}
In the limit that the inefficiency is small (so that $1/(1-x) \approx 1 + x$) this reduces to
\begin{equation}
\frac{N_m^d(s)/s}{2 N_m^h(s)/s} = 
\frac{A_d}{2 A_h} \left[1 + (0.08B_{\rm dump} + B_h - B_d) s\right].
\end{equation}

Since the acceptance is approximately the same for both 
hydrogen and deuterium, $B_d = 2.13 \times B_h$,  
where the factor of 2.13 is the ratio of interaction lengths 
for the deuterium and hydrogen targets.  If the 
acceptance for dump events was the same as for target events, 
then $B_d = 2.13 \times B_h  = .15 \times B_{\rm dump}$ and 
the rate dependence vanishes as it should.
However, since the acceptance is not the same for target 
and dump events, there is a rate dependence
that must be corrected.  

It should be noted that the important quantity is not the absolute 
rate dependence inefficiency, but rather the difference between the 
inefficiencies for the two different targets.  A simple way of 
approximating the size of the correction to the cross section ratio ($C$) is to compare
it to the absolute correction for deuterium events which is, 
\begin{equation}
\frac{C\times s}{R_d\times s}=\frac{C}{R_d} = \frac{0.08B_{\rm dump} + -0.53 B_d}{-0.85B_{\rm dump} - B_d}.
\end{equation}
This equation shows that the rate dependence correction
to the cross section ratio is considerably smaller than the 
absolute correction in the deuterium events.

The ratio of extra drift chamber hits in an average deuterium event
compared to the same number in an average hydrogen event ($F$) can 
be determined from the data for each mass setting.
This ratio should be the same as the ratio of the average number
of background hits in a deuterium event to the average number of 
background hits in a hydrogen event. So 
\begin{equation}
F = \frac{0.85B_{\rm dump} + B_d}{0.93B_{\rm dump} + B_h}.
\end{equation}

Comparing equations \ref{eqn:ed1} and \ref{eqn:eh1} with equations 
\ref{eqn:ed2} and \ref{eqn:eh2}, clearly indicates that
\begin{equation}
\frac{0.85B_{\rm dump} + B_d}{0.93B_{\rm dump} + B_h} = \frac{R_d}{R_h}.
\end{equation}
This means that
\begin{equation}
F = \frac{R_d}{R_h}.
\end{equation}
So the number of independent parameters introduced to fit
$N_m(s)/s$, has been reduced again.  The only difficulty that eliminating
$R_h$ from the fit introduces is the determination of $F$ from the data.
This is the subject of the next section.

\subsection{Determination of $F$}

The previous section explained why the determination of $F$ was important.  
This section will expound on how $F$ was determined.  Start by  
defining $F$ to be
\begin{equation}
F \equiv \frac{\rm {extra~drift~chamber~hits~in~an~average~deuterium~event}}
	 {\rm {extra~drift~chamber~hits~in~an~average~hydrogen~event}}.
\end{equation}
Here `extra' means that the drift chamber hits 
produced by the reconstructed dimuon event 
will not be included.  The numerator (denominator) is merely the 
average multiplicity of a drift chamber minus the number of hits from 
the reconstructed dimuon in a deuterium (hydrogen) event.
The number of hits from the reconstructed dimuon is two times the chamber efficiency.

Since there are eighteen drift chambers, $F$ can be determined eighteen times 
for each of the three mass settings.  The six calculated values of $F$
for each station of drift chambers for each mass setting are very similar.
However, there are differences between stations and mass settings 
as shown in Table \ref{tab:f}.  The $F$ values shown in Table \ref{tab:f}
have been averaged over each station and the uncertainty is the standard
deviation of these six measurements.

\begin{table}[tbp]
\caption{Calculated $F$ values for each station and each mass setting.} 
\label{tab:f}
\begin{center}                                                          
\begin{tabular}{c|ccc}               
\hline \hline
		& \multicolumn{3}{c}{station}   \\
mass setting	& one 			& two			& three			\\ \hline
low		& 1.83	$\pm$ 0.023	& 1.65	$\pm$ 0.034	& 1.83	$\pm$ 0.013	\\ 
intermediate	& 1.11	$\pm$ 0.014	& 1.07	$\pm$ 0.020	& 1.26	$\pm$ 0.011	\\ 
high		& 1.49	$\pm$ 0.015	& 1.42	$\pm$ 0.017	& 1.56  $\pm$ 0.022	\\ \hline \hline
\end{tabular}
\end{center}
\end{table}

While the variation of $F$ between mass settings is to be expected, the 
reason for the variation of $F$ between stations is less clear.  Monte Carlo
generated events were used to determine which station's extra hits 
affected the rate dependence the most.  By turning off the noise in one
station at a time, it was determined that station three contributed the most
to the rate dependence while station one contributed the least.  
The relative magnitudes of the effect on the rate dependence from station
one to station two to station three is $36$:$73$:$135$.  Based on this 
relationship, a weighted average over the stations produced the final $F$ 
values shown in Table \ref{tab:f2}.  The uncertainty on $F$ is a conservative
estimate based primarily on the magnitude of the variation of $F$ between 
stations.

\begin{table}[tbp]
\caption{Calculated $F$ values for each mass setting obtained by taking a weighted average
	 over all three stations.} 
\label{tab:f2}
\begin{center}                                                          
\begin{tabular}{cc}               
\hline \hline
mass setting	& $F$   \\ \hline
low		& $1.78	\pm 0.10$ \\ 
intermediate	& $1.18	\pm 0.10$ \\ 
high		& $1.51 \pm 0.10$ \\ \hline \hline
\end{tabular}
\end{center}
\end{table}

\begin{figure}
  \begin{center}
    \mbox{\epsffile{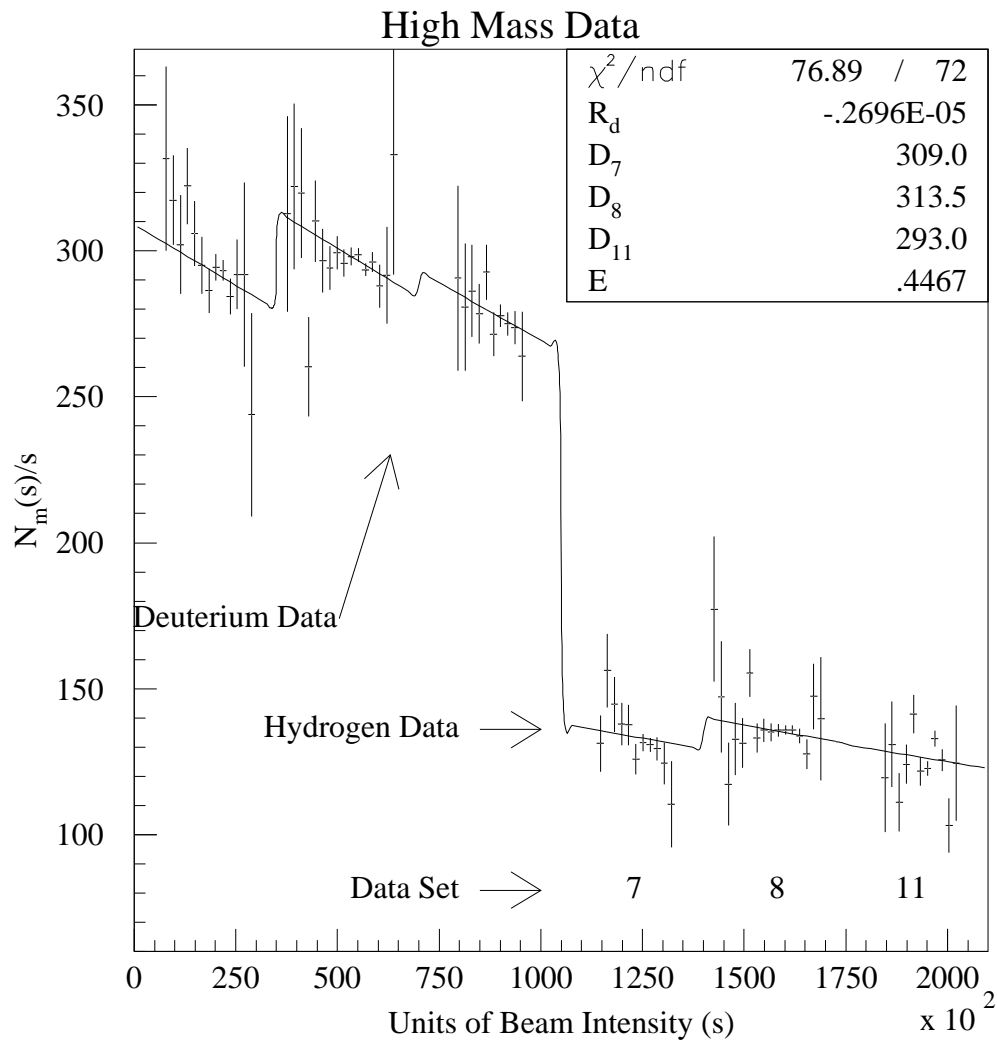}}
    \vspace*{-0.25in}                                
  \end{center}       
  \caption{Number of deuterium (left) and hydrogen
		(right) events per unit of beam intensity as a function of the beam intensity.  
		Both the deuterium and hydrogen data for all three of the 
		high mass data sets are shown here.  The hydrogen
		data are offset by 105,000 units of beam intensity.  Within
		the deuterium and hydrogen data, data sets three, four and
		nine are offset by 0, 35,000 and 70,000 units of beam intensity
		respectively.}
  \label{fig:ratehigh}                    
\end{figure}

\begin{figure}
  \begin{center}
    \mbox{\epsffile{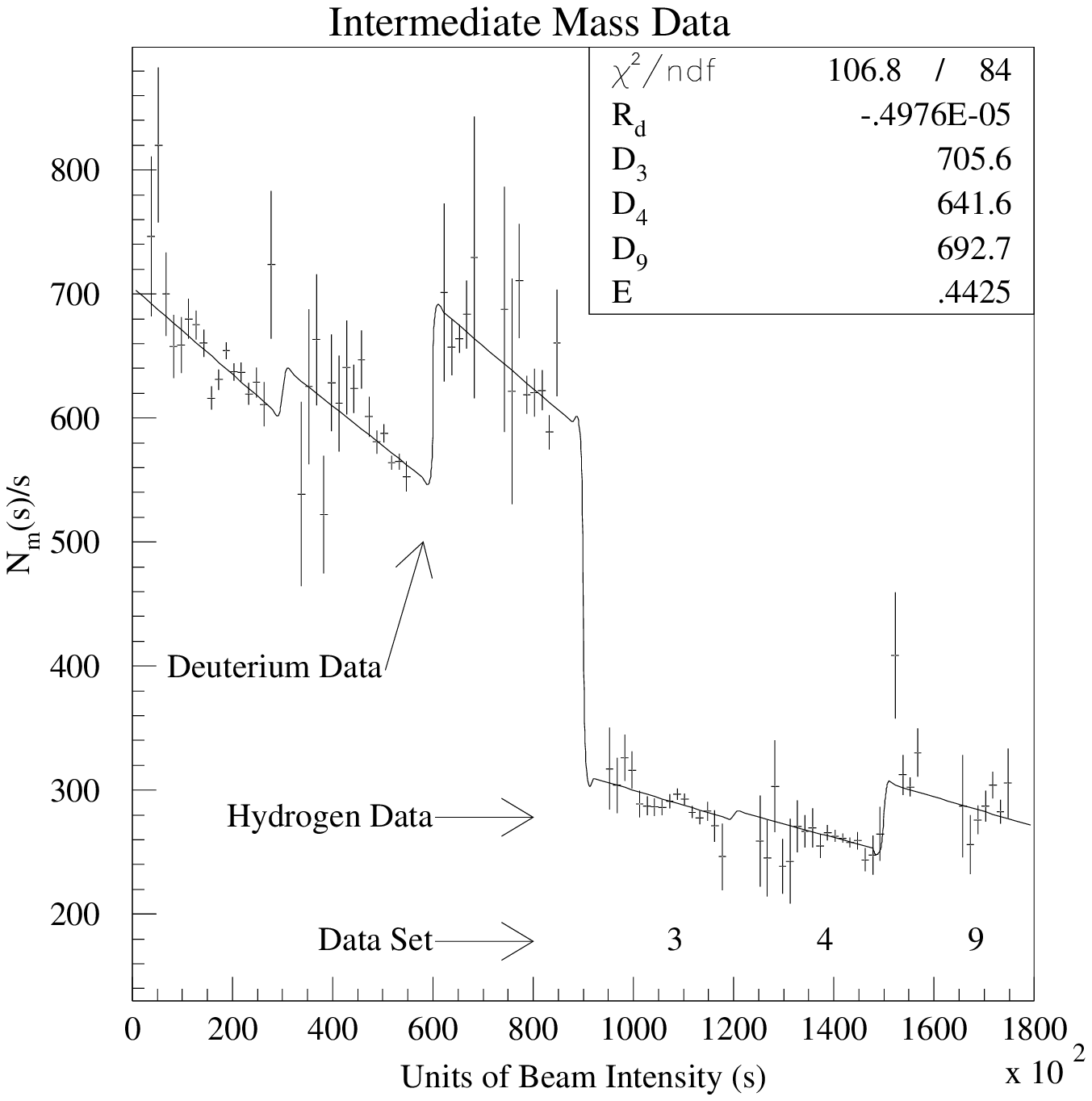}}
    \vspace*{-0.25in}                                
  \end{center}       
  \caption{Number of deuterium (left) and hydrogen
		(right) events per unit of beam intensity as a function of the beam intensity.  
		Both the deuterium and hydrogen data for all three of the 
		intermediate mass data sets are shown here.  The hydrogen
		data are offset by 90,000 units of beam intensity.  Within
		the deuterium and hydrogen data, data sets three, four and
		nine are offset by 0, 30,000 and 60,000 units of beam intensity
		respectively.}
  \label{fig:rateint}                    
\end{figure}

\begin{figure}
  \begin{center}
    \mbox{\epsffile{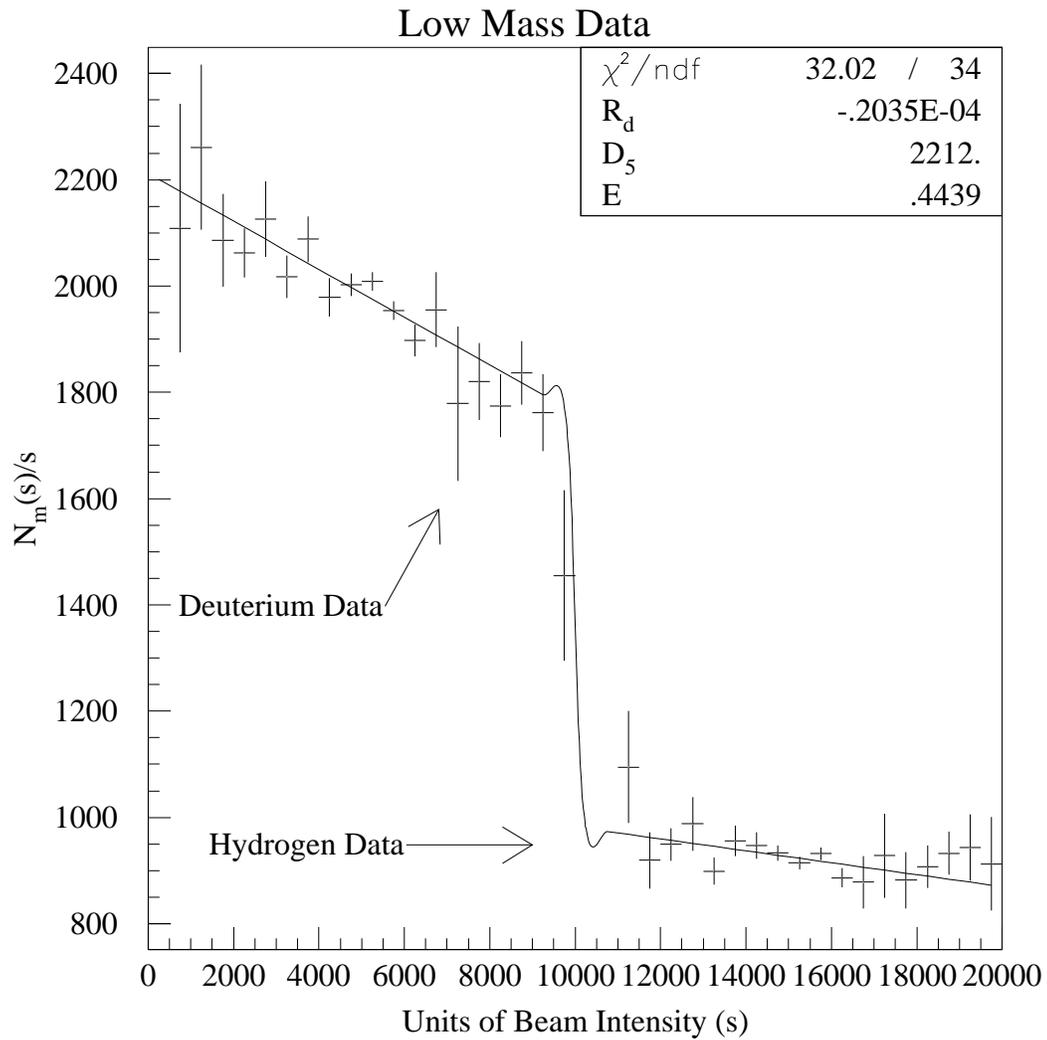}}
    \vspace*{-0.25in}                                
  \end{center}       
  \caption{Number of deuterium (left) and hydrogen
		(right) events per unit of beam intensity as a function of the beam intensity.  
		Both the deuterium and hydrogen data from the low mass 
		data set are shown here.  The hydrogen
		data are offset by 10,000 units of beam intensity.}
  \label{fig:ratelow}                    
\end{figure}

Once $F$ was determined for each of the mass settings, the 
deuterium data were fit to
\begin{equation}
\frac{N_m^d(s)}{s} = D_i \times \left(1 + R_d s \right)
\label{eqn:rated2}
\end{equation}
and the hydrogen data were fit to
\begin{equation}
\frac{N_m^h(s)}{s} = D_i E \times \left(1 + \frac{R_d}{F} s \right).
\label{eqn:rateh2}
\end{equation}  
These fits to the high, intermediate, and low mass data 
are shown in Fig.~\ref{fig:ratehigh}, \ref{fig:rateint}, and 
\ref{fig:ratelow} respectively.  From these fits the data can be 
corrected for the rate dependence by weighting each deuterium 
event by 
\begin{equation}
\frac{1}{e_d(s)} = \frac{1}{1 + R_ds}
\end{equation}
and each hydrogen event by
\begin{equation}
\frac{1}{e_h(s)} = \frac{1}{1 + \frac{R_d}{F}s}.
\end{equation}
The final correction to $\sigma^{pd}/2\sigma^{pp}$ due to the rate dependence
is summarized in Table \ref{tab:finalrate}.

\begin{table}[tbp]
\caption{Correction to $\sigma^{pd}/2\sigma^{pp}$ due to the rate dependence.} 
\label{tab:finalrate}
\begin{center}                                                          
\begin{tabular}{cc}               
\hline \hline
mass setting	& percent correction to $\sigma^{pd}/2\sigma^{pp}$   \\ \hline
low		& $5.45\%	\pm 0.82\%$ \\ 
intermediate	& $1.06\%	\pm 0.89\%$ \\ 
high		& $1.76\%       \pm 0.69\%$  \\ \hline \hline
\end{tabular}
\end{center}
\end{table}

\section{Target Composition, Density, and Attenuation}
\label{sect:target}

Since the data included in this analysis were taken over
a period of five months, there were several small but 
important changes to the target composition and density.
These changes and their effect on the calculation of the 
cross section ratio will be discussed in this section.

The deuterium target was filled twice during the acquisition
of these data.  Both of the deuterium mixtures that were used to fill
the target were analyzed for contaminations.  The analysis of 
the first fill indicated that the deuterium was 99.99\% pure.  
The second fill was of slightly lesser quality and was 
therefore analyzed twice.  The first sample of the second fill was taken 
directly from the storage flask, while the second sample was 
taken half way through the process of emptying the target flask.
The analyses of both samples of the second fill are shown in
Table \ref{tab:ld2}.

\begin{table}[tbp]
\caption{Results of the gas analyses of the second fill of the deuterium
	 target.  The results shown are in percent volume.} 
\label{tab:ld2}
\begin{center}                                                          
\begin{tabular}{ccc}               
\hline \hline
material	& target sample		& storage sample   	\\ \hline
D$_2$		& $93.8 \%\pm 0.7\%$	& $92.7\% \pm 0.8\%$ 	\\ 
HD		& $5.80	\%\pm 0.58\%$	& $6.89\% \pm 0.69\%$ 	\\ 
H$_2$		& $0.053\% \pm 0.011\%$  	& $0.147\% \pm 0.015\%$	\\ 
N$_2$		& $0.327\% \pm 0.033\%$	& $0.245\% \pm 0.024\%$	\\
Ar		& $0.003\% \pm 0.002\%$	&  ---			\\
CO$_2$		& $0.006\% \pm 0.003\%$	& $0.0039\% \pm 0.0008\%$ 	\\
\hline \hline
\end{tabular}
\end{center}
\end{table}
 
All samples were taken when the deuterium was 
in a gaseous state so the analyses are not a perfect indication
of what was present in the target where the deuterium was in a 
liquid state.  As the deuterium was cooled, the heavier
components (Ar, CO$_2$, and N$_2$) should have frozen out of the 
liquid while the lighter components (H$_2$ and HD) should have
been somewhat distilled out of the liquid into the gas
above the liquid in the cryogenic system.  Another concern when
interpreting the information in Table \ref{tab:ld2} is which of
the samples better represents what was in the target.  Considering
how and when the samples were taken, the sample taken while
emptying the flask should be more accurate.  Based on these 
considerations, Table \ref{tab:ld22} shows the best estimate 
of the second fill deuterium composition.

\begin{table}[tbp]
\caption{Best estimate of the composition of the second fill 
	 deuterium.  The results shown are in percent volume.} 
\label{tab:ld22}
\begin{center}                                                          
\begin{tabular}{cc}               
\hline \hline
material	& percent volume   \\ \hline
D$_2$		& $94.05\% \pm 0.6\%$ \\ 
HD		& $5.90\%	\pm 0.6\%$  \\ 
H$_2$		& $0.05\% \pm 0.01\%$  \\ \hline 
deuterium	& $97.0\% \pm 0.6\%$   \\
hydrogen	& $3.0\%  \pm 0.6\%$   \\
\hline \hline
\end{tabular}
\end{center}
\end{table}
 
The density of the target material was determined from
the vapor pressure of the gas above the liquid in both cryogenic systems.
These pressures were constantly monitored and recorded in a database.
They were also manually checked at least twice a day and recorded 
during the standard shift check.  Also recorded during the shift check
was the temperature of each flask.  From these shift checks, 
the average pressure was determined 
for each target and for each data set.  These average pressures
are shown in Table \ref{tab:pressure}.

\begin{table}[tbp]
\caption{Average pressure in psi of each liquid target for each data set.} 
\label{tab:pressure}
\begin{center}                                                          
\begin{tabular}{ccc}               
\hline \hline
data set	& hydrogen		& deuterium   	\\ \hline
3		& $ 15.12$ psi	& $15.06$ psi 	\\ 
4		& $ 15.05$ psi	& $14.92$ psi 	\\ 
5		& $ 14.97$ psi  & $14.92$ psi	\\ 
7		& $ 15.04$ psi	& $14.96$ psi	\\
8		& $ 15.11$ psi	& $15.17$ psi	\\
11		& $ 15.15$ psi	& $15.21$ psi 	\\
\hline \hline
\end{tabular}
\end{center}
\end{table}

Cryogenic data Tables~\cite{ld2} for hydrogen and deuterium were
used to convert vapor pressure to mass density.
For liquid H$_2$, the relationship between vapor pressure ($P$ in psi)
and density ($\rho_h$ in g/cm$^3$) is
\begin{equation}
\frac{1}{\rho_{h}} = 62.473 \left( 0.2115 + 1.171\times10^{-3}P - 1.109
\times 10^{-5}P^2 \right).
\end{equation}
The formula used to determine the density of deuterium ($\rho_d$) is 
\begin{equation}
\rho_{d} = 4.028\times10^{-3} \left[ 43.291 - 3.4176 \frac{P}{14.6959} +
0.5783 \left( \frac{P}{14.6959} \right)^{2} \right].
\end{equation}
From the pressures listed in Table \ref{tab:pressure} and the above equations,
the densities shown in Table \ref{tab:density} were calculated.

\begin{table}[tbp]
\caption{Density in g/cm$^3$ for each liquid target and for each data set.} 
\label{tab:density}
\begin{center}                                                          
\begin{tabular}{ccc}               
\hline \hline
data set	& hydrogen		& deuterium   	\\ \hline
3		& $ 0.07062$ g/cm$^3$	& $0.16272$ g/cm$^3$	\\ 
4		& $ 0.07064$ g/cm$^3$	& $0.16280$ g/cm$^3$	\\ 
5		& $ 0.07066$ g/cm$^3$	& $0.16280$ g/cm$^3$	\\ 
7		& $ 0.07064$ g/cm$^3$	& $0.16278$ g/cm$^3$	\\
8		& $ 0.07062$ g/cm$^3$	& $0.16265$ g/cm$^3$	\\
11		& $ 0.07061$ g/cm$^3$	& $0.16259$ g/cm$^3$	\\
\hline \hline
\end{tabular}
\end{center}
\end{table}

As the beam interacted with the target material, the beam was attenuated.  
Because the hydrogen and deuterium targets had different densities, they
also suffered from different amounts of attenuation.  Because the deuterium
target was more dense, the protons in the beam were more likely to interact
as they traveled through the target.  Therefore, the beam intensity decreased
more rapidly as it passed through the deuterium target than when it 
passed through the hydrogen target.  This means that there was a smaller fraction of the 
protons in the beam that reached the downstream end of the deuterium target
than there was that reached the downstream end of the hydrogen target.
Calculations based on the proton-proton and proton-deuteron cross sections
\cite{attenuation} determined that the ratio of effective luminosity in the
hydrogen target ($A_h$) compared to the effective luminosity in the deuterium
target ($A_d$) is
\begin{equation}
\frac{A_h}{A_d} = 1.042 \pm 0.002.
\end{equation}

The acceptance for the events from the hydrogen and deuterium 
targets was not identical.  Although the target flask construction 
was identical, the attenuation of the beam through the targets 
meant that the average interaction point for the two targets was 
slightly different.  The average interaction point in the deuterium
target was almost $0.2$~inches upstream of the average interaction 
point in the hydrogen target.  

Monte Carlo studies were performed to study the effect 
of beam attenuation through the target on the acceptance. 
These studies showed a slight $x_2$ dependence to the 
acceptance correction, which was most important at the 
edges of the acceptance.  The maximum size of this correction 
was about one percent at the highest $x_2$ data points in
the low and intermediate mass data.  The typical correction
was an order of magnitude smaller.

\section{Calculation of $\sigma^{pd}/2\sigma^{pp}$}

This experiment counted the number of dimuon events from the 
hydrogen target ($N_h$), the deuterium target ($N_d$) and the
empty target ($N_e$).  To compare the yields from these different 
targets, the beam intensity for each spill was recorded and the 
integrated beam intensity ($I_{\rm target}$) for each target 
was determined.  Using the many small corrections already 
described in this chapter, the number of raw hydrogen dimuon events is
\begin{equation}
N_h = I_h A_h t_h \rho_h \frac{H}{g} \frac{d\sigma^{pp}}{d\Omega} 
      \Delta\Omega_h e_h + N_h^{BG},
\end{equation}
and the number of raw deuterium events is
\begin{equation}                                                      
N_d = I_d A_d t_d \rho_d \frac{D}{g} \frac{d\sigma^{pd}}{d\Omega}     
      \Delta\Omega_d e_d + N_d^{BG}.                                    
\end{equation}                                                         
In the previous two equations $t_{\rm target}$ is the target length, 
$H/g$ ($D/g$) is the number of hydrogen (deuterium) atoms per gram,
$\Delta\Omega_{\rm target}$ is the spectrometer acceptance for a 
given target, and $N_{\rm target}^{BG}$ is the number of background 
events from a given target.
Using these equations, it is easy to get
\begin{equation}
\label{eqn:main}
\frac{\sigma^{pd}}{2\sigma^{pp}} =
\frac{1}{2}
\frac{N_d - N_d^{BG}}{N_h - N_h^{BG}}~
\frac{I_h}{I_d}
\frac{A_h}{A_d}
\frac{t_h}{t_d}
\frac{\rho_h}{\rho_d}
\frac{H/g}{D/g}
\frac{\Delta\Omega_h}{\Delta\Omega_d}
\frac{e_h}{e_d}.
\end{equation}

The number of background events is the sum of two separate production
mechanisms.  There were background Drell-Yan events produced when the 
beam interacted with the target flask windows or other non-target materials.
The number of these events was determined by normalizing the yields off
of the empty target.  To properly normalize the number of 
empty target events from downstream of the center of the target, attenuation 
of the beam through the target must be included.
The second source of background events are the
randoms ($N_{\rm target}^{randoms}$) 
that were described in Section \ref{sect:randoms}.
Combining these two sources of background events gives
\begin{equation}
N_h^{BG} = \left(N_e^{upstream} + 0.93*N_e^{downstream}\right)
	   \frac{I_h}{I_e} + N_h^{randoms}
\end{equation}
for the hydrogen target background and 
\begin{equation}
N_d^{BG} = \left(N_e^{upstream} + 0.85*N_e^{downstream}\right)
	   \frac{I_d}{I_e} + N_d^{randoms}
\end{equation}
for the deuterium target background.  
In the previous two equations the superscript on $N_e$ designates 
whether the empty target event originated from upstream or downstream
of the center of the target.  

The output of the second pass analysis was subjected to the 
final set of cuts described in Section \ref{sect:finalcuts}.
The events that passed this final set of cuts were used with
the corrections discussed in this chapter and the above equations\footnote{
Equation \ref{eqn:main} was modified for data sets eight and eleven 
to include corrections needed to account for the small hydrogen 
content in the deuterium target.  Details can be found in reference \cite{hawker}.}
to determine $\sigma^{pd}/2\sigma^{pp}$ as a function of $x$ of 
the target parton.  These results are shown in
Tables \ref{tab:highratio}, \ref{tab:intratio}, and \ref{tab:lowratio}.
The results shown for the high mass data are slightly different 
then the results first published in reference \cite{prl} due to
improvements made to the rate dependence and acceptance calculations.  
The average value of $x_2$, $x_F$, $p_T$, and dimuon mass are shown
for each $x_2$ bin for the high, intermediate, and low mass settings
in Tables \ref{tab:hidhigh}, \ref{tab:hidint}, and \ref{tab:hidlow} 
respectively.
\begin{table}[tbp]
\caption{The cross section ratio calculated for each data set of the high
	mass setting and the final high mass result for each $x_2$ bin.
	The uncertainty shown here is the statistical uncertainty.} 
\label{tab:highratio}
\begin{center}
\begin{tabular}{c|cccc}
\hline \hline
\multicolumn{1}{c}{$x_2$ range} &
\multicolumn{4}{|c}{$\sigma^{pd}/2\sigma^{pp}$}  \\
  min-max   &      data set 7   &  data set 8       & data set 11       &  final result  \\
\hline
0.020-0.045 & 1.044 $\pm$ 0.033 & 1.053 $\pm$ 0.022 & 1.045 $\pm$ 0.042 & 1.049 $\pm$ 0.017 \\
0.045-0.070 & 1.061 $\pm$ 0.024 & 1.104 $\pm$ 0.018 & 1.086 $\pm$ 0.032 & 1.088 $\pm$ 0.013 \\
0.070-0.095 & 1.123 $\pm$ 0.030 & 1.114 $\pm$ 0.020 & 1.149 $\pm$ 0.039 & 1.122 $\pm$ 0.016 \\
0.095-0.120 & 1.114 $\pm$ 0.039 & 1.167 $\pm$ 0.028 & 1.113 $\pm$ 0.048 & 1.142 $\pm$ 0.021 \\
0.120-0.145 & 1.182 $\pm$ 0.053 & 1.219 $\pm$ 0.040 & 1.203 $\pm$ 0.071 & 1.205 $\pm$ 0.029 \\
0.145-0.170 & 1.143 $\pm$ 0.069 & 1.132 $\pm$ 0.047 & 1.128 $\pm$ 0.081 & 1.134 $\pm$ 0.035 \\
0.170-0.195 & 1.117 $\pm$ 0.083 & 1.142 $\pm$ 0.060 & 0.981 $\pm$ 0.091 & 1.100 $\pm$ 0.043 \\
0.195-0.220 & 1.090 $\pm$ 0.101 & 1.207 $\pm$ 0.083 & 0.954 $\pm$ 0.111 & 1.108 $\pm$ 0.056 \\
0.220-0.245 & 0.980 $\pm$ 0.126 & 1.070 $\pm$ 0.087 & 1.241 $\pm$ 0.196 & 1.065 $\pm$ 0.067 \\
0.245-0.295 & 0.826 $\pm$ 0.102 & 1.110 $\pm$ 0.104 & 1.022 $\pm$ 0.173 & 0.974 $\pm$ 0.067 \\
0.295-0.345 & 0.843 $\pm$ 0.277 & 0.873 $\pm$ 0.179 & 1.251 $\pm$ 0.448 & 0.903 $\pm$ 0.143 \\
\hline \hline
\end{tabular}
\end{center}
\end{table}

\begin{table}[tbp]
\caption{The cross section ratio calculated for each data set of the intermediate
	mass setting and the final intermediate mass result for each $x_2$ bin.
	The uncertainty shown here is the statistical uncertainty.} 
\label{tab:intratio}
\begin{center}
\begin{tabular}{c|ccc}
\hline \hline
\multicolumn{1}{c}{$x_2$ range} &
\multicolumn{3}{|c}{$\sigma^{pd}/2\sigma^{pp}$}  \\
  min-max   &      data set 3  &  data set 4   &  final result  \\
\hline
0.015-0.040 & 1.038 $\pm$ 0.031 & 1.047 $\pm$ 0.039 & 1.042 $\pm$ 0.024 \\
0.040-0.065 & 1.080 $\pm$ 0.019 & 1.066 $\pm$ 0.023 & 1.074 $\pm$ 0.015 \\
0.065-0.090 & 1.090 $\pm$ 0.020 & 1.068 $\pm$ 0.024 & 1.081 $\pm$ 0.015 \\
0.090-0.115 & 1.095 $\pm$ 0.025 & 1.068 $\pm$ 0.029 & 1.084 $\pm$ 0.019 \\
0.115-0.140 & 1.114 $\pm$ 0.034 & 1.161 $\pm$ 0.042 & 1.132 $\pm$ 0.026 \\
0.140-0.190 & 1.118 $\pm$ 0.038 & 1.114 $\pm$ 0.045 & 1.116 $\pm$ 0.029 \\
0.190-0.240 & 1.019 $\pm$ 0.072 & 1.109 $\pm$ 0.104 & 1.048 $\pm$ 0.059 \\
0.240-0.290 & 1.267 $\pm$ 0.405 & 0.755 $\pm$ 0.196 & 0.852 $\pm$ 0.176 \\
\hline \hline
\end{tabular}
\end{center}
\end{table}


\begin{table}[tbp]                   
\caption{The cross section ratio calculated from the low
	mass setting data for each $x_2$ bin.
	The uncertainty shown here is the statistical uncertainty.
	The average values for kinematic variables is also shown.} 
\label{tab:hidlow}                                                           
\label{tab:lowratio}
\begin{center}                           
\begin{tabular}{c|ccccc}
\hline \hline           
$x_2$ range &         &         & $\langle p_T\rangle$ & $\langle M_{\mu^+\mu^-}\rangle$ & \\
  min-max   & $\langle x_2\rangle$ & $\langle x_F\rangle$ & (GeV/c) &    (GeV/c$^2$)  &
$\sigma^{pd}/2\sigma^{pp}$ \\ \hline
0.015-0.040 &  0.032  &  0.415  &   1.02  &      4.5  & 1.053 $\pm$ 0.018 \\
0.040-0.065 &  0.052  &  0.280  &   1.07  &      5.0  & 1.107 $\pm$ 0.015 \\
0.065-0.090 &  0.076  &  0.221  &   1.07  &      5.7  & 1.099 $\pm$ 0.020 \\
0.090-0.115 &  0.100  &  0.174  &   1.05  &      6.3  & 1.148 $\pm$ 0.034 \\
0.115-0.165 &  0.129  &  0.129  &   1.06  &      7.0  & 1.108 $\pm$ 0.055 \\
\hline \hline
\end{tabular}
\end{center}
\end{table}

\begin{table}[tbp]                   
\caption{The average values for kinematic variables in the high mass data 
	for each $x_2$ bin.  The uncertainty shown here for the cross section ratio 
	is the statistical uncertainty.} 
\label{tab:hidhigh}                                                           
\begin{center}                           
\begin{tabular}{c|ccccc}
\hline \hline           
$x_2$ range &         &         & $\langle p_T\rangle$ & $\langle M_{\mu^+\mu^-}\rangle$ &  \\
  min-max   & $\langle x_2\rangle$ & $\langle x_F\rangle$ & (GeV/c) &    (GeV/c$^2$)  
&  $\sigma^{pd}/2\sigma^{pp}$\\ \hline
0.020-0.045 &  0.036  &  0.537  &   0.92  &      5.5  & 1.049 $\pm$ 0.017 \\
0.045-0.070 &  0.057  &  0.441  &   1.03  &      6.5  & 1.088 $\pm$ 0.013 \\
0.070-0.095 &  0.082  &  0.369  &   1.13  &      7.4  & 1.122 $\pm$ 0.016 \\
0.095-0.120 &  0.106  &  0.294  &   1.18  &      7.9  & 1.142 $\pm$ 0.021 \\
0.120-0.145 &  0.131  &  0.244  &   1.21  &      8.5  & 1.205 $\pm$ 0.029 \\
0.145-0.170 &  0.156  &  0.220  &   1.21  &      9.3  & 1.134 $\pm$ 0.035 \\
0.170-0.195 &  0.182  &  0.192  &   1.20  &      9.9  & 1.100 $\pm$ 0.043 \\
0.195-0.220 &  0.207  &  0.166  &   1.19  &     10.6  & 1.108 $\pm$ 0.056 \\
0.220-0.245 &  0.231  &  0.134  &   1.18  &     11.1  & 1.065 $\pm$ 0.067 \\
0.245-0.295 &  0.264  &  0.097  &   1.18  &     11.8  & 0.974 $\pm$ 0.067 \\
0.295-0.345 &  0.312  &  0.052  &   1.14  &     12.8  & 0.903 $\pm$ 0.142 \\
\hline \hline
\end{tabular}
\end{center}
\end{table}

\begin{table}[tbp]                   
\caption{The average values for kinematic variables in the intermediate mass data 
	for each $x_2$ bin.  The uncertainty shown here for the cross section ratio 
	is the statistical uncertainty.} 
\label{tab:hidint}                                                           
\begin{center}                           
\begin{tabular}{c|ccccc}
\hline \hline           
$x_2$ range &         &         & $\langle p_T\rangle$ & $\langle M_{\mu^+\mu^-}\rangle$ & \\
  min-max   & $\langle x_2\rangle$ & $\langle x_F\rangle$ & (GeV/c) &    (GeV/c$^2$) &
$\sigma^{pd}/2\sigma^{pp}$ \\ \hline
0.015-0.040 &  0.034  &  0.427  &   1.21  &      4.8  & 1.042 $\pm$ 0.024 \\
0.040-0.065 &  0.053  &  0.331  &   1.16  &      5.5  & 1.074 $\pm$ 0.015 \\
0.065-0.090 &  0.077  &  0.277  &   1.14  &      6.3  & 1.081 $\pm$ 0.015 \\
0.090-0.115 &  0.101  &  0.221  &   1.15  &      6.9  & 1.084 $\pm$ 0.019 \\
0.115-0.140 &  0.126  &  0.163  &   1.15  &      7.4  & 1.132 $\pm$ 0.026 \\
0.140-0.190 &  0.159  &  0.111  &   1.09  &      7.9  & 1.116 $\pm$ 0.029 \\
0.190-0.240 &  0.207  &  0.071  &   1.06  &      9.1  & 1.048 $\pm$ 0.059 \\
0.240-0.290 &  0.254  &  0.145  &   1.17  &     12.0  & 0.852 $\pm$ 0.176 \\
\hline \hline
\end{tabular}
\end{center}
\end{table}

\begin{figure}
  \begin{center}
    \mbox{\epsffile{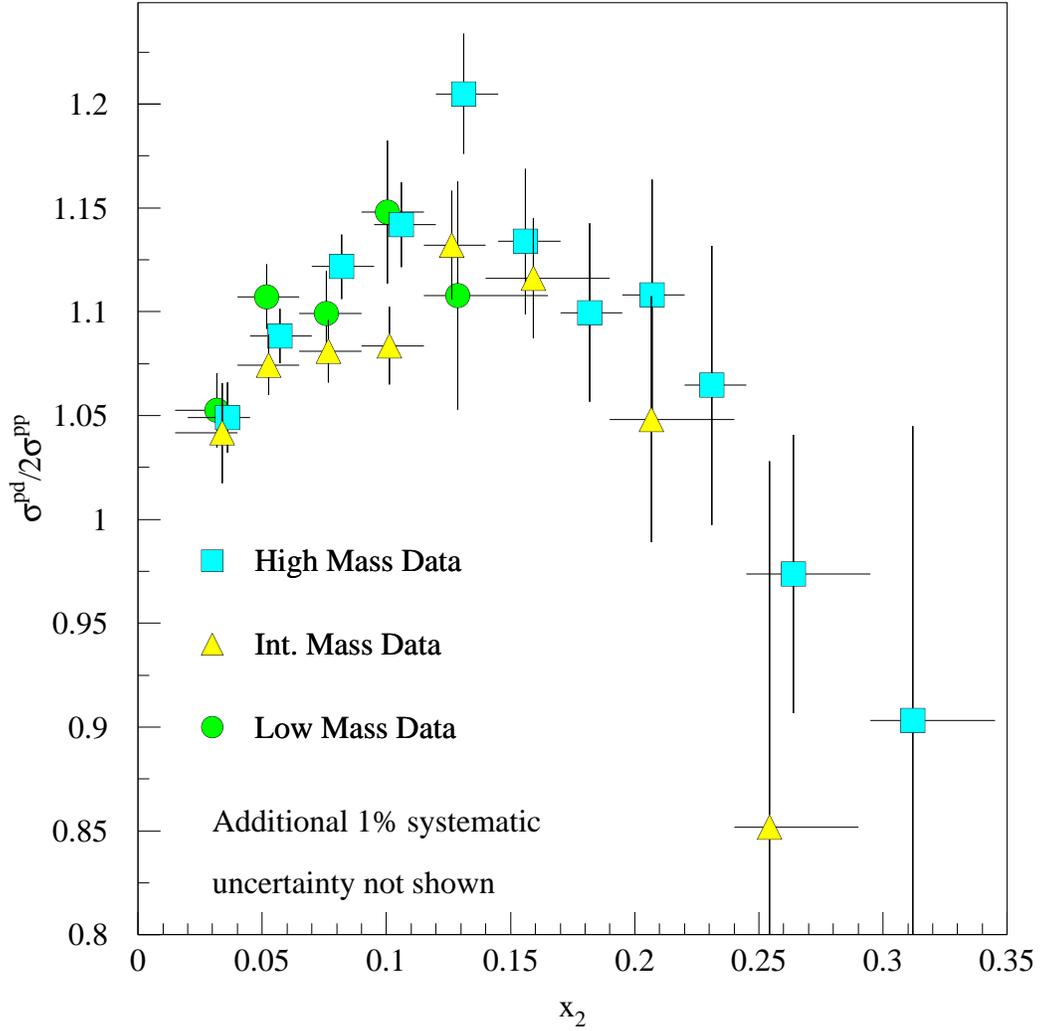}}
    \vspace*{-0.25in}                                
  \end{center}       
  \caption{The Drell-Yan cross section ratio versus $x$ of the target parton.
	   The results from all three mass settings are shown.  The error bars
	   represent the statistical uncertainty.  An additional one percent 
	   systematic uncertainty is common to all points within a mass setting.}
  \label{fig:ratio}                    
\end{figure}

\begin{figure}
  \begin{center}
    \mbox{\epsffile{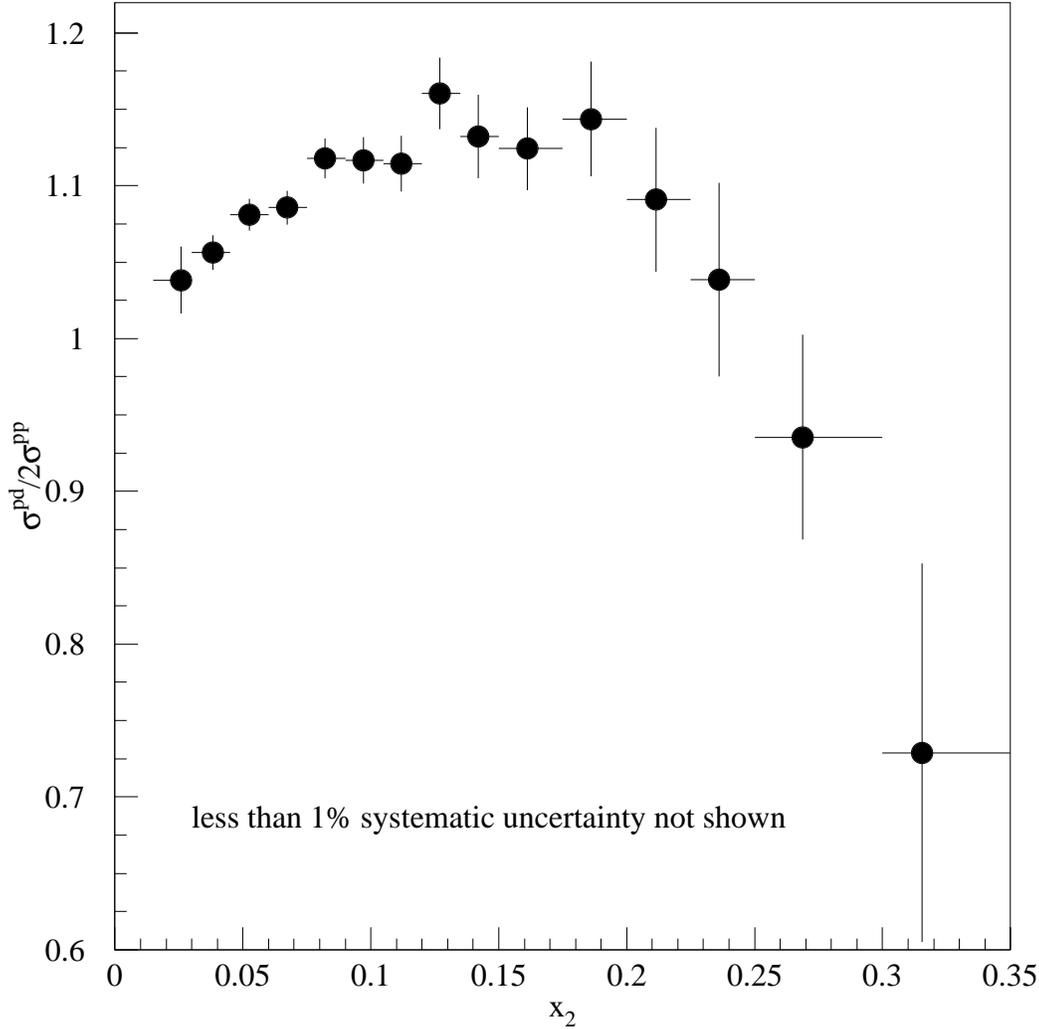}}
    \vspace*{-0.25in}                                
  \end{center}       
  \caption{The Drell-Yan cross section ratio versus $x$ of the target parton.
	   The combined result from all data sets is shown.  The error bars
	   represent the statistical uncertainty.  There is a less than one percent 
	   systematic uncertainty common to all points.}
  \label{fig:ratiocom}        
\end{figure}

The intermediate and high mass data were divided into different 
data sets.  The cross section ratio ($r_i \pm \delta_i$) determined from 
each data set ($i$) within a mass setting, were combined to 
give an average cross section ratio for the mass setting ($R \pm \Delta$)
using 
\begin{equation}
R \pm \Delta = \frac{\displaystyle \sum_i {r_i}/{\delta_i^2}}{\displaystyle \sum_i {1}/{\delta_i^2}}
	       \pm \left[\frac{1}{\displaystyle \sum_i {1}/{\delta_i^2}}\right]^{1/2}.
\label{eqn:combine}
\end{equation}
The results
obtained from this equation are shown in the `final result' column of Table 
\ref{tab:highratio} and Table \ref{tab:intratio}.  The final 
cross section ratio as determined by each mass setting is shown
in Fig.~\ref{fig:ratio}.  Equation \ref{eqn:combine} can also
be applied to the final results of each mass setting to produce
a single cross section ratio for all data sets.  This 
result is shown in Fig.~\ref{fig:ratiocom}.  
The average value of $x_2$, $x_1$, $p_T$, and dimuon mass are shown
along with the cross section ratio for each $x_2$ bin for the combined result
in Table \ref{tab:ratiocom}.

\begin{table}[tbp]                   
\caption{The cross section ratio calculated from all data sets
	for each $x_2$ bin.
	The uncertainty shown here is the statistical uncertainty.
	The average values for kinematic variables is also shown.} 
\label{tab:ratiocom}                                                           
\begin{center}                           
\begin{tabular}{c|ccccc}
\hline \hline           
$x_2$ range &         &         & $\langle p_T\rangle$ & $\langle M_{\mu^+\mu^-}\rangle$ & \\
  min-max   & $\langle x_2\rangle$ & $\langle x_1\rangle$ & (GeV/c) &    (GeV/c$^2$)  &
$\sigma^{pd}/2\sigma^{pp}$ \\ \hline
0.015-0.030 &  0.026  &  0.559  &   1.00  &  4.6    & 1.038 $\pm$ 0.022 \\
0.030-0.045 &  0.038  &  0.454  &   1.05  &  5.1    & 1.056 $\pm$ 0.011 \\
0.045-0.060 &  0.052  &  0.408  &   1.08  &  5.6    & 1.081 $\pm$ 0.010 \\
0.060-0.075 &  0.067  &  0.393  &   1.10  &  6.2    & 1.086 $\pm$ 0.011 \\
0.075-0.090 &  0.082  &  0.378  &   1.12  &  6.8    & 1.118 $\pm$ 0.013 \\
0.090-0.105 &  0.097  &  0.358  &   1.14  &  7.2    & 1.116 $\pm$ 0.015 \\
0.105-0.120 &  0.112  &  0.339  &   1.16  &  7.5    & 1.115 $\pm$ 0.018 \\
0.120-0.135 &  0.127  &  0.326  &   1.17  &  7.8    & 1.161 $\pm$ 0.023 \\
0.135-0.150 &  0.142  &  0.324  &   1.16  &  8.2    & 1.132 $\pm$ 0.027 \\
0.150-0.175 &  0.161  &  0.325  &   1.16  &  8.7    & 1.124 $\pm$ 0.027 \\
0.175-0.200 &  0.186  &  0.333  &   1.15  &  9.5    & 1.144 $\pm$ 0.038 \\
0.200-0.225 &  0.211  &  0.345  &   1.15  &  10.3   & 1.091 $\pm$ 0.047 \\
0.225-0.250 &  0.236  &  0.356  &   1.18  &  11.1   & 1.039 $\pm$ 0.063 \\
0.250-0.300 &  0.269  &  0.366  &   1.18  &  12.0   & 0.935 $\pm$ 0.067 \\
0.300-0.350 &  0.315  &  0.361  &   1.08  &  12.9   & 0.729 $\pm$ 0.124 \\
\hline \hline
\end{tabular}
\end{center}
\end{table}

\begin{figure}
  \begin{center}
    \mbox{\epsffile{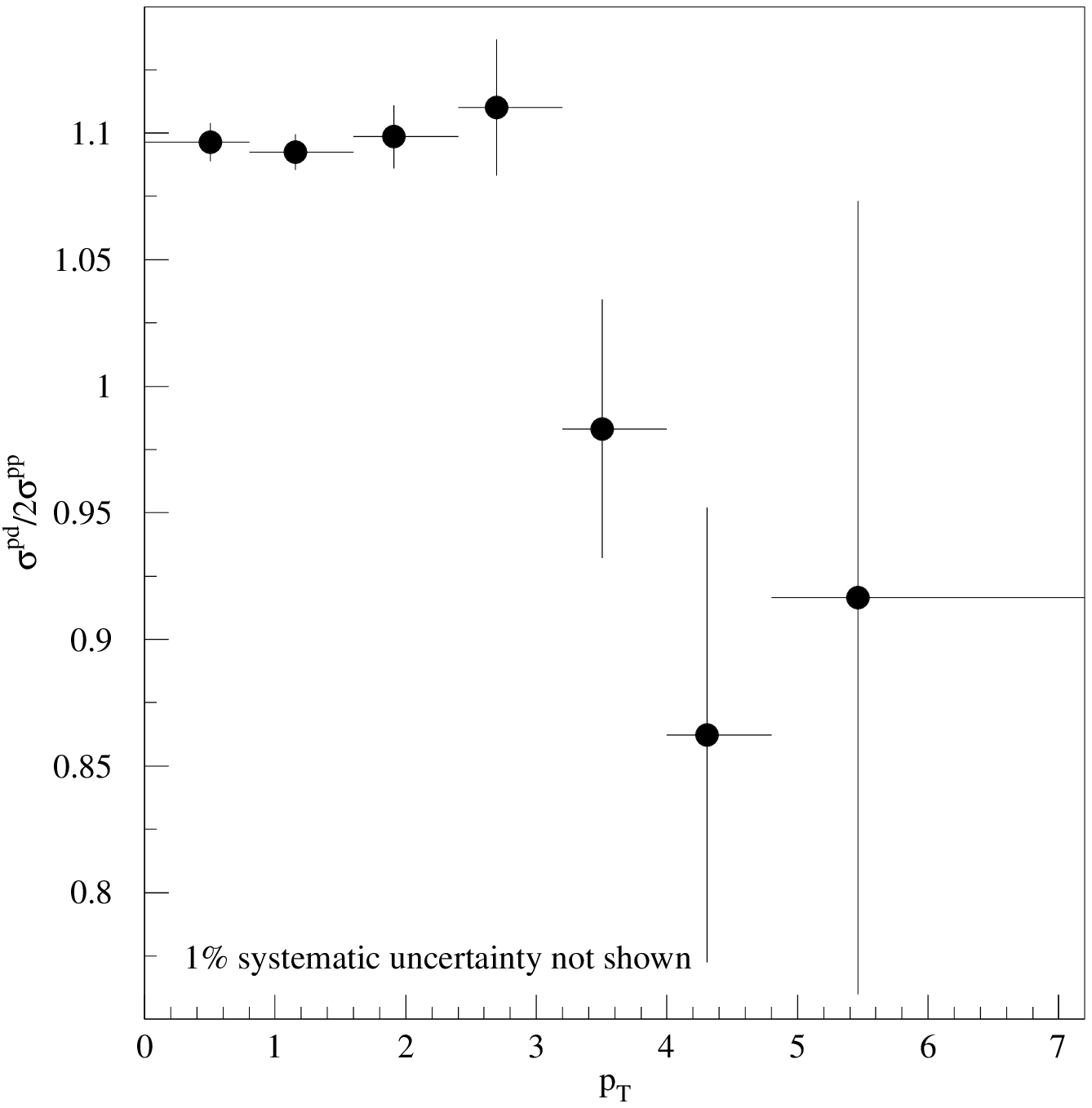}}
    \vspace*{-0.25in}                                
  \end{center}       
  \caption{The Drell-Yan cross section ratio versus $p_T$.
	   The combined result from all data sets is shown.  The error bars
	   represent the statistical uncertainty.  There is a one percent 
	   systematic uncertainty common to all points.}
  \label{fig:ratiopt}        
\end{figure}

The cross section ratio can also be determined as a function of other 
kinematic quantities.  Figure \ref{fig:ratiopt} shows the cross section
ratio as a function of the transverse momentum of the dileptons.

\section{Systematic Uncertainty in $\sigma^{pd}/2\sigma^{pp}$}

Many of the possible sources of systematic errors in calculating
the cross section can be ignored when calculating the cross section 
ratio.  So the only sources of systematic uncertainty that must be 
considered are sources that affect the two targets differently.   
Because the targets were changed every few minutes, effects such
as changes in detector efficiency or beam quality will affect 
both groups of data equally.

The sources of systematic uncertainty that can not be neglected include
the rate dependence, length of the target flask, target composition, beam attenuation,
and acceptance differences.  Table~\ref{tab:systematics} shows the 
sources of systematic uncertainty in the cross section ratio 
for each mass setting.  Clearly the rate dependence is the dominant 
systematic uncertainty except in the data taken with a slight hydrogen
contamination in the deuterium target.  By adding all of the 
sources of systematic uncertainty in quadrature, the 
total systematic uncertainty in the measured cross section
ratio is determined to be less than one percent.

\begin{table}[tbp]
\caption{Systematic uncertainties in measurement of $\sigma^{pd}/2\sigma^{pp}$.}
\label{tab:systematics}
\begin{center}
\begin{tabular}{cccc}
\hline \hline
\multicolumn{1}{c}{source of} &
\multicolumn{3}{c}{uncertainty in mass setting}  \\
 uncertainty  	& high		& intermediate	& low  \\
\hline
rate dependence	& 0.69 \%	& 0.89 \%	& 0.82 \% \\
target length	& 0.2 \%	& 0.2 \% 	& 0.2 \% \\
beam intensity 	& 0.1 \%	& 0.1 \%	& 0.1 \% \\
attenuation/acceptance & 0.2 \% & 0.2 \% 	& 0.2 \% \\
deuterium composition & 0.61 \%	&  ---		& ---     \\
\hline \hline
\end{tabular}
\end{center}
\end{table}

The systematic uncertainty from the target length is 
due to a known slight difference between the length of the 
target flasks.  What is unknown, is which flask is longer.
The beam intensity systematic uncertainty is based on 
the difference between the ratio of the integrated beam intensity
on the two targets as measured by the many different monitors.  Since the
low and intermediate mass data used to calculate the
cross section ratio was all from the first fill of 
the deuterium target, they do not suffer from the 
systematic uncertainty due to the deuterium composition.


\section{Extraction of $\bar d(x)/\bar u(x)$}

From the discussion in Chapter \ref{ch:theory} it is clear 
that $\sigma^{pd}/2\sigma^{pp}$ is closely related to $\bar d/\bar u$.
However, the simple approximations that lead to equations
\ref{eqn:xf0} and \ref{eqn:xf1} are based on mutually exclusive kinematic 
conditions.  Since the data shown in Fig.~\ref{fig:ratio} includes 
events that fall into both of these kinematic regions, as well as in 
between where neither approximation is valid, neither equation 
can be used to extract $\bar d/\bar u$.  Therefore, an iterative
process was used to extract $\bar d/\bar u$ from the cross section ratio
that did not make any assumptions about the kinematics of the data.

\begin{figure}
  \begin{center}
    \mbox{\epsffile{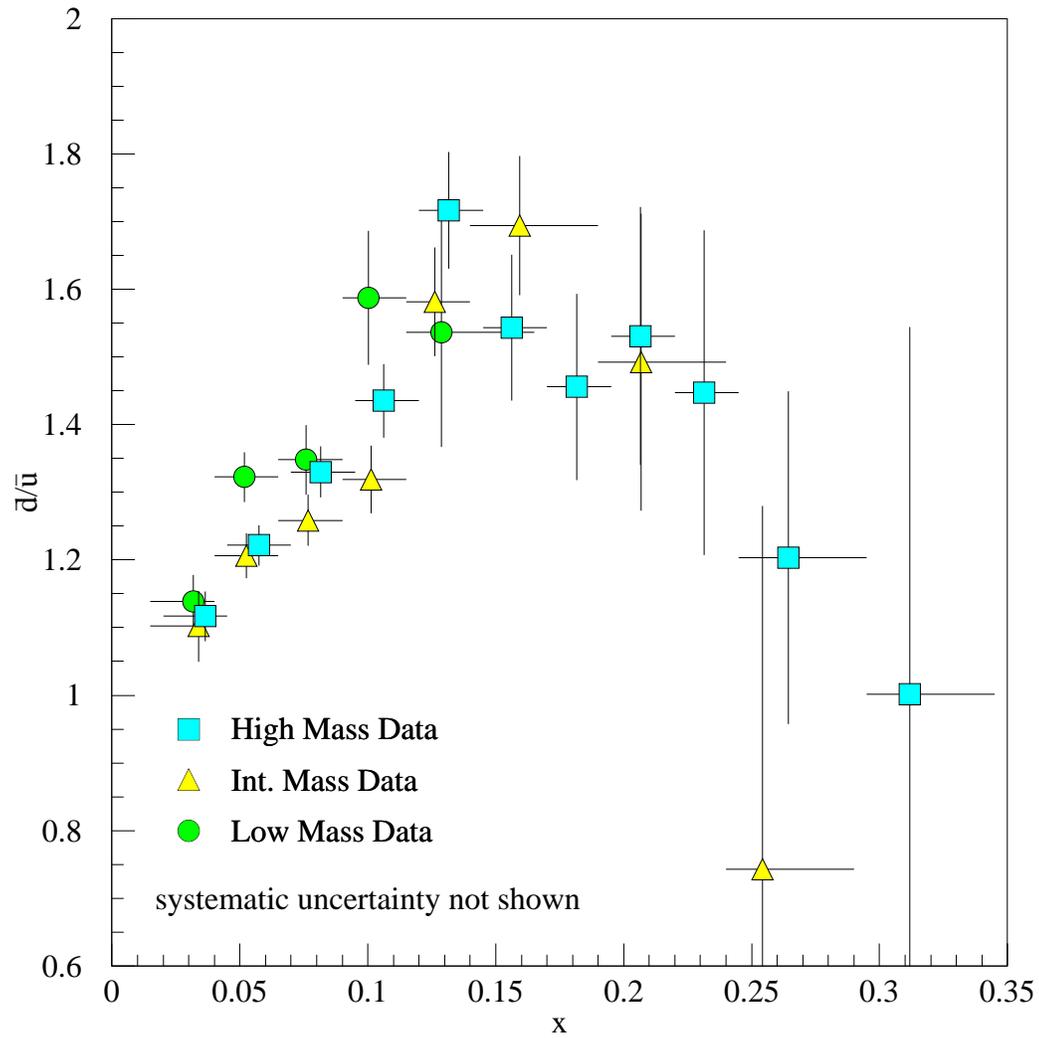}}
    \vspace*{-0.25in}                                
  \end{center}       
  \caption{$\bar d/\bar u$ versus $x$.  The results 
	    from all three mass settings are shown.
		The error bars represent the statistical uncertainty.
		The systematic uncertainty, which varies from about 2\%
		at low $x$ to a maximum of 3.5\% at high $x$, is not shown.}
  \label{fig:du3}                    
\end{figure}

This iterative process calculated $\sigma^{pd}/2\sigma^{pp}$, 
compared this calculated quantity with the measured quantity, 
adjusted $\bar d/\bar u$ that goes into the calculation of
$\sigma^{pd}/2\sigma^{pp}$, and repeated.  This process continued 
until the calculated $\sigma^{pd}/2\sigma^{pp}$ agreed with the measured
ratio.  The results of this method, for each mass setting treated independently,
are shown in Fig.~\ref{fig:du3}.  

\begin{figure}
  \begin{center}
    \mbox{\epsffile{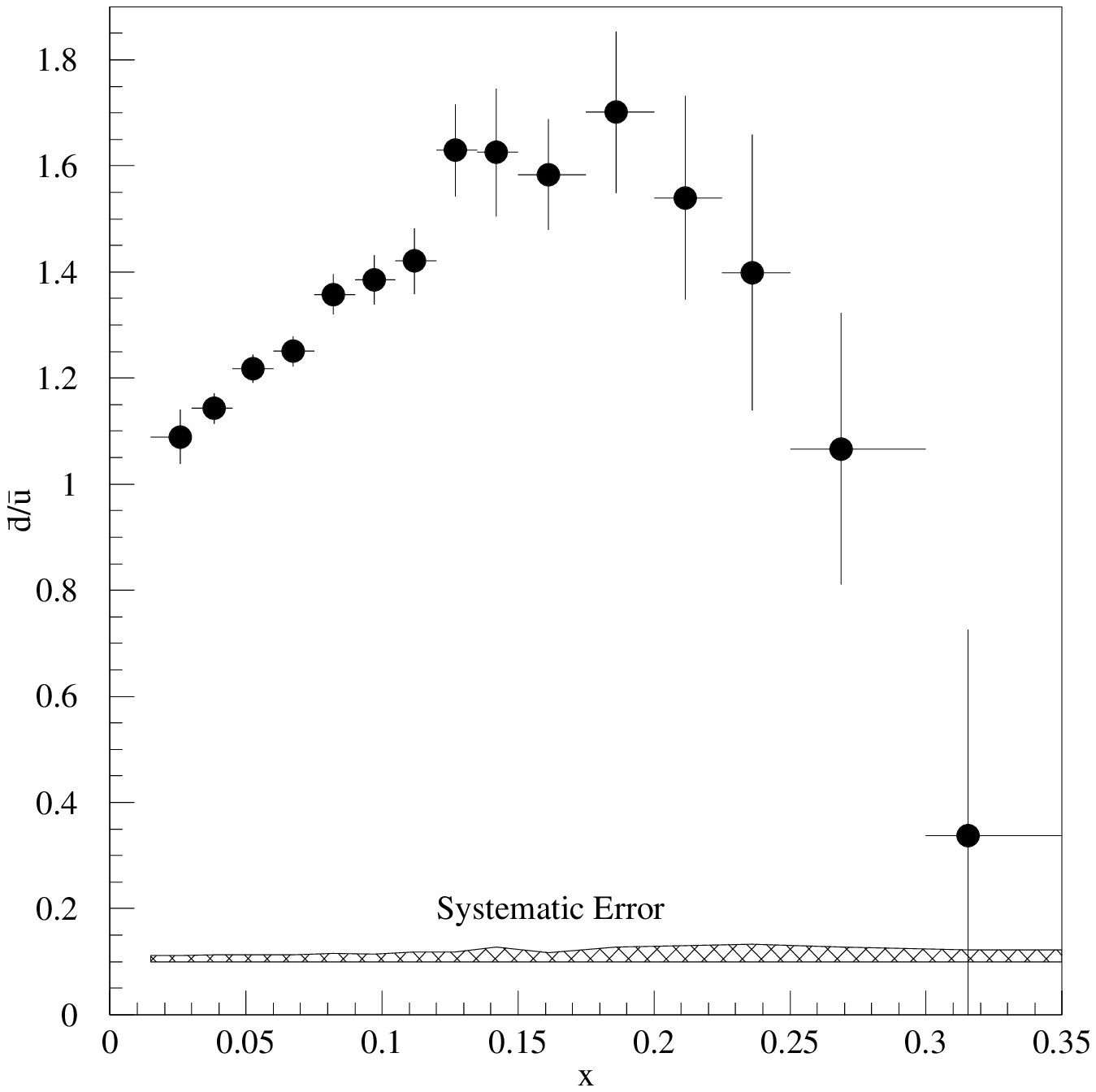}}
    \vspace*{-0.25in}                                
  \end{center}       
  \caption{$\bar d/\bar u$ versus $x$.  The combined result 
	    from all three mass settings is shown.}
  \label{fig:du}                    
\end{figure}

In the combined $\bar d/\bar u$ ratio results shown in
Fig.~\ref{fig:du} and Table~\ref{tab:du}, convergence of the iterative
process was determined by selecting the $\bar d/\bar u$ which
minimized
\begin{equation}
\chi^2 = \sum_{i} \left[
    \frac{\left.r_i\right|_{(\mathrm{meas})} -
          \left.r_i\right|_{(\mathrm{calc})}}
         {\delta_{(\mathrm{meas})}} \right]^2
\end{equation}
for each data point, where $r_i$ and $\delta_i$ are as defined in
equation~\ref{eqn:combine}.

These extracted quantities have been scaled to a common
$Q^2$ valued of 41 GeV$^2$.  The details important to this iterative
extraction process will be discussed in the remainder of this section.


\begin{table}[tbp]                   
\caption{The value of $\bar d(x)/\bar u(x)$ and $\bar d(x) - \bar u(x)$ 
	as determined from all data sets
	for each $x_2$ bin.
	The statistical uncertainty is listed first followed by the systematic uncertainty.}
\label{tab:du}                             
\begin{center}                           
\begin{tabular}{c|cc}
\hline \hline           
$x_2$ range &                   &   \\
  min-max   & $\bar d/\bar u$   & $\bar d(x) - \bar u(x)$   \\ \hline
0.015-0.030 & 1.089 $\pm$.051 $\pm$0.011 & 0.882 $\pm$0.486 $\pm$0.104  \\
0.030-0.045 & 1.143 $\pm$.028 $\pm$0.012 & 0.780 $\pm$0.145 $\pm$0.063  \\
0.045-0.060 & 1.217 $\pm$.026 $\pm$0.012 & 0.715 $\pm$0.078 $\pm$0.037  \\ 
0.060-0.075 & 1.250 $\pm$.028 $\pm$0.013 & 0.543 $\pm$0.055 $\pm$0.025  \\
0.075-0.090 & 1.358 $\pm$.038 $\pm$0.015 & 0.523 $\pm$0.047 $\pm$0.019  \\
0.090-0.105 & 1.385 $\pm$.047 $\pm$0.014 & 0.409 $\pm$0.042 $\pm$0.012  \\
0.105-0.120 & 1.421 $\pm$.062 $\pm$0.018 & 0.332 $\pm$0.040 $\pm$0.012  \\
0.120-0.135 & 1.629 $\pm$.087 $\pm$0.018 & 0.350 $\pm$0.037 $\pm$0.008  \\
0.135-0.150 & 1.625 $\pm$.120 $\pm$0.028 & 0.271 $\pm$0.040 $\pm$0.009  \\
0.150-0.175 & 1.584 $\pm$.105 $\pm$0.017 & 0.189 $\pm$0.026 $\pm$0.004  \\
0.175-0.200 & 1.701 $\pm$.152 $\pm$0.027 & 0.148 $\pm$0.024 $\pm$0.004  \\ 
0.200-0.225 & 1.540 $\pm$.193 $\pm$0.030 & 0.083 $\pm$0.023 $\pm$0.004  \\ 
0.225-0.250 & 1.399 $\pm$.260 $\pm$0.033 & 0.045 $\pm$0.025 $\pm$0.003  \\
0.250-0.300 & 1.067 $\pm$.256 $\pm$0.028 & 0.005 $\pm$0.020 $\pm$0.002  \\
0.300-0.350 &  .337 $\pm$.389 $\pm$0.021 &-0.042 $\pm$0.037 $\pm$0.002  \\
\hline \hline
\end{tabular}
\end{center}
\end{table}

From Eq. \ref{eqn:dy} it is obvious that to calculate $\sigma^{pd}/2\sigma^{pp}$
the PDF for each quark and antiquark present in the nucleon must be known.
While calculating $\sigma^{pd}/2\sigma^{pp}$ for the iterative process,
it was assumed that an existing PDF parameterization accurately described 
the quark distributions and the quantity $\bar d(x) + \bar u(x)$ since
these quantities have been constrained by previous measurements.  
The parameterizations used were CTEQ4M~\cite{CTEQ}, MRS(R2)~\cite{MRS}
and MRST~\cite{mrst}.
While the ratio of $\bar d(x)$ to $\bar u(x)$ was very poorly constrained in 
these parameterizations and therefore was adjusted with each iteration, 
the sum of $\bar d(x)$ and $\bar u(x)$ was reasonably well known and
it was held constant.  

So that the calculated $\sigma^{pd}/2\sigma^{pp}$ could be compared to the
measured $\sigma^{pd}/2\sigma^{pp}$, the acceptance of the spectrometer
had to be included in the calculated quantity.  To do this the cross section
ratio was calculated for the $x_1$, $x_2$, and Q$^2$ values of every 
event that passed the analysis cuts.  These calculated cross section
ratios were then averaged over a given $x_2$ bin.  

As $\sigma^{pd}/2\sigma^{pp}$ was calculated for each iteration,
it was assumed that $\bar d/\bar u$ for the beam proton was the
same as $\bar d/\bar u$ for the target proton 
over the $x_2$ range of the data.  For many events however, $x_1$ was 
greater than the maximum $x_2$ in the data, so
something had to be assumed for $\bar d(x_1)/\bar u(x_1)$ above \mbox{$x_1 = 0.35$}.
The affects of several different assumptions were investigated.  The 
extracted $\bar d/\bar u$ was not noticeably affected by any of 
these assumptions except at the highest $x$ values, which changed by less than
five percent.  The final assumption
that was used was to assume that $\bar d(x_1)/\bar u(x_1) = 1.0$ 
in the proton for \mbox{$x_1 > 0.35$}.  

\begin{figure}
  \begin{center}
    \mbox{\epsffile{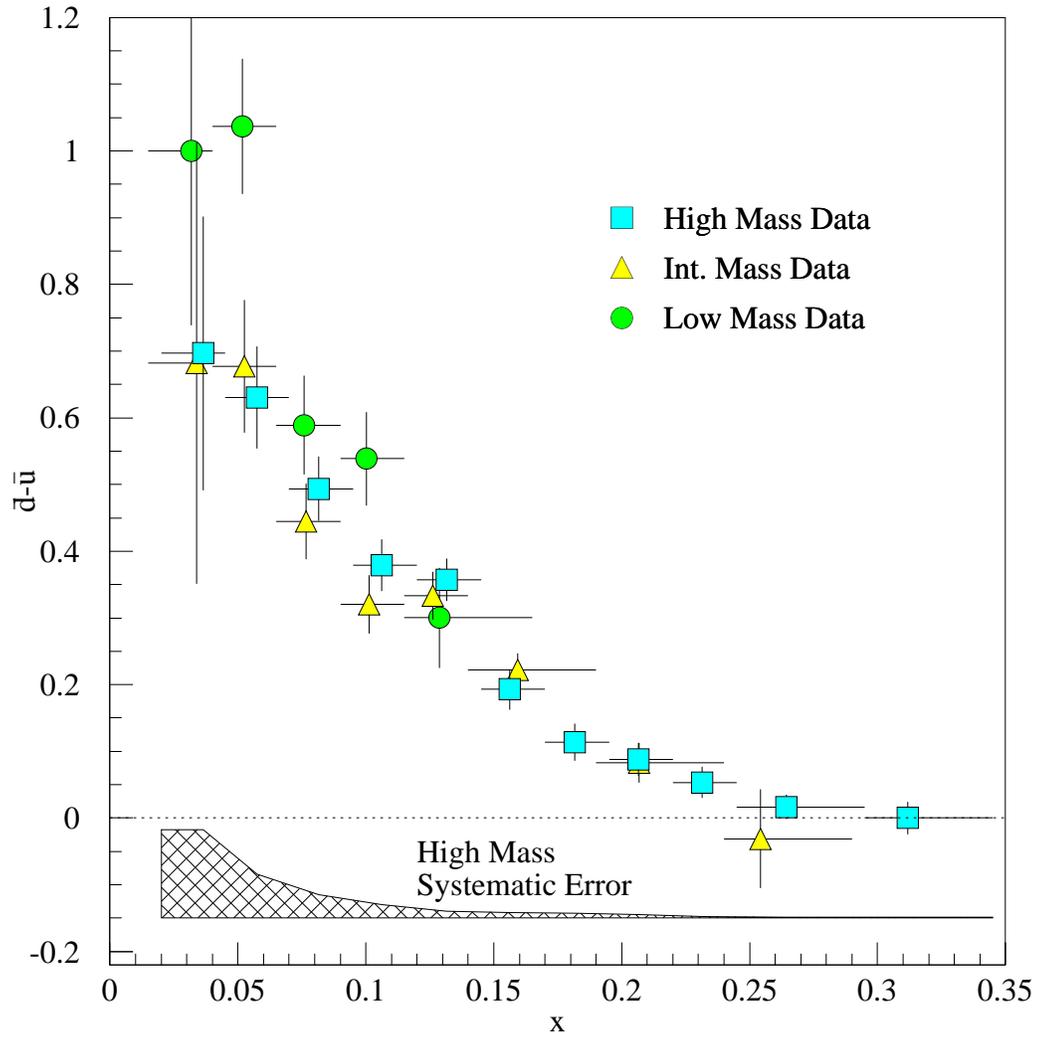}}
    \vspace*{-0.25in}                                
  \end{center}       
  \caption{$\bar d - \bar u$ versus $x$.  The results 
	    from all three mass settings are shown.  
		The systematic uncertainty is shown for the
		high mass data.  The systematic uncertainty for 
		the other data sets are very similar.}
  \label{fig:dmu3}                    
\end{figure}

\begin{figure}
  \begin{center}
    \mbox{\epsffile{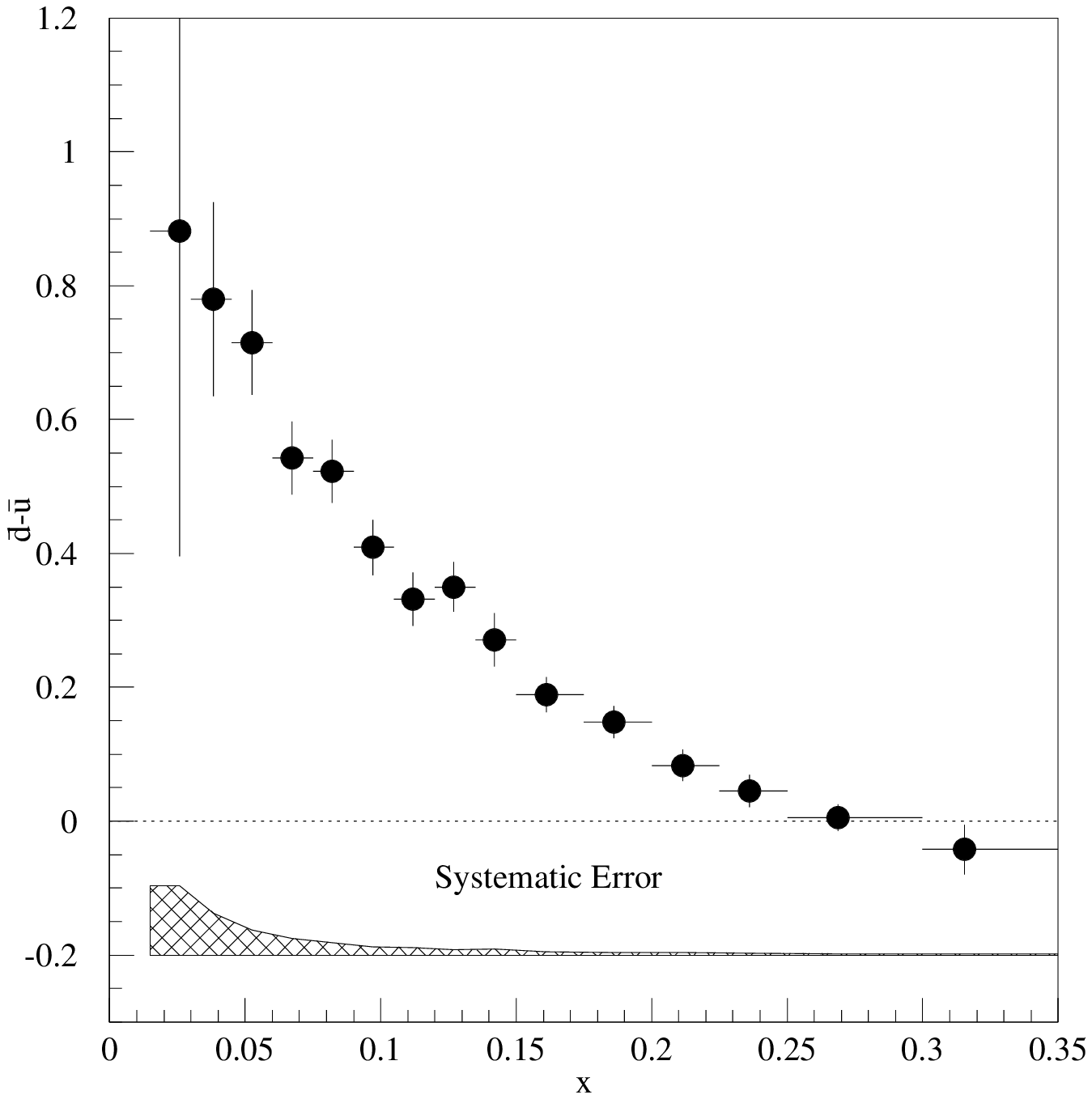}}
    \vspace*{-0.25in}                                
  \end{center}       
  \caption{$\bar d - \bar u$ versus $x$.  The combined result 
	    from all three mass settings is shown.}
  \label{fig:dmucom}                    
\end{figure}

Once the quantity $\bar d(x)/\bar u(x)$ is determined, the
quantity $\bar d(x)-\bar u(x)$ can be calculated using
\begin{equation}
\bar d(x)-\bar u(x) = \frac{\left(\frac{\bar d}{\bar u} - 1\right) }
{\left( \frac{\bar d}{\bar u} + 1\right) } (\bar d + \bar u).
\end{equation}
Here again the quantity $\bar d + \bar u$ is assumed to be 
correctly described by one of the parameterizations.
Figure~\ref{fig:dmu3} shows $\bar d(x)-\bar u(x)$ as a function of $x$ for 
each of the mass settings while the combined result is shown
in Fig.~\ref{fig:dmucom} and Table~\ref{tab:du}.

\begin{figure}
  \begin{center}
    \mbox{\epsffile{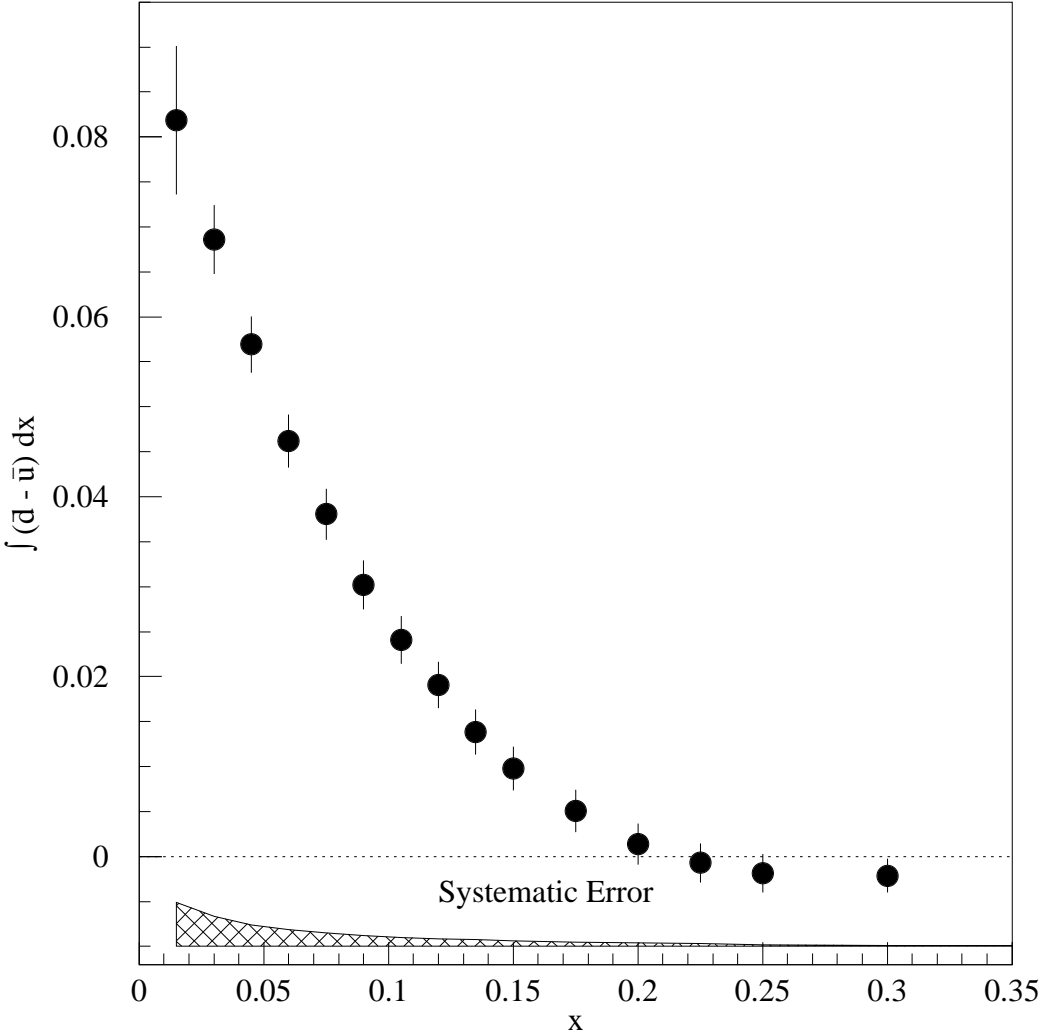}}
    \vspace*{-0.25in}                                
  \end{center}       
  \caption{$\int^{0.35}_{x} ( \bar d(x) - \bar u(x)) dx'$ in the proton versus $x$.}
  \label{fig:idmucom}                    
\end{figure}

Once $\bar d - \bar u$ is determined, then the integral of $\bar d - \bar u$
can be calculated between $x^{min}$ and 0.35.
Figure~\ref{fig:idmucom} shows $\int^{0.35}_{x^{min}} ( \bar d(x) - \bar u(x) ) dx$.
Over the measured region, the value of this integral has been
determined to be
\begin{equation}
\int^{0.35}_{0.015} ( \bar d(x) - \bar u(x) ) dx = 0.0818 \pm 0.0082 \pm 0.0049.
\end{equation}

\chapter{Results and Conclusions}

While previous experiments have indicated that $\bar d >\bar u$, 
FNAL E866/NuSea was the first measurement of the $x$ dependence of
the flavor asymmetry in the nucleon sea.  This measurement has had 
an impact in several different areas.  The global parameterizations 
of the nucleon sea will obviously change to fit these new data.  
Surprisingly, this measurement when used in conjunction with the NMC
measurement, puts new and tighter constraints on the valence PDF's.
This measurement has also provided a means of testing the 
predictions of several nonperturbative models.  Finally, the 
unexpected sharp downturn in $\bar d(x)/\bar u(x)$ apparently back to unity
at the large $x$ limits of this measurement, has motivated a proposal to 
perform a similar experiment focused at higher $x$ values.

\section{E866 Results}

\begin{figure}
  \begin{center}
    \mbox{\epsffile{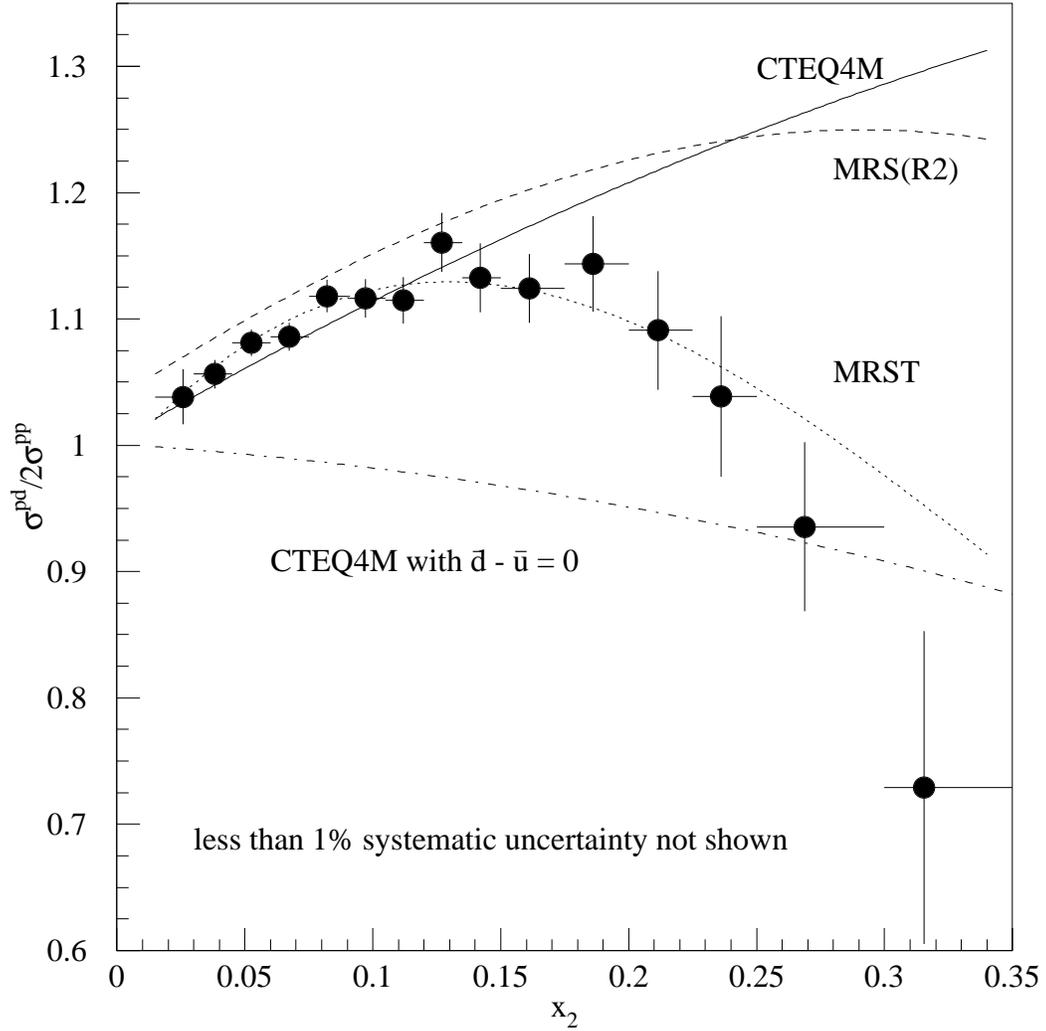}}                           
    \vspace*{-0.25in}                                                      
  \end{center}                                                               
  \caption{The Drell-Yan cross section ratio versus $x$ of the target parton.
           The results from all three mass settings have been combined.  The curves 
	   are the calculated cross section ratio using CTEQ4M~\cite{CTEQ},
	   MRS(R2)~\cite{MRS}, and MRST~\cite{mrst}.  The bottom curve is calculated
	   using CTEQ4M where $\bar d -\bar u$ has been set to zero.}
  \label{fig:ratiowpdf}
\end{figure}

The primary measurement of this experiment was the determination
of $\sigma^{pd}/2\sigma^{pp}$ over a wide kinematic range.
The combined result from all three mass settings is 
shown in Fig.~\ref{fig:ratiowpdf} along with the curves from the 
calculated cross section ratio using various parameterizations.  
The CTEQ4M~\cite{CTEQ} and MRS(R2)~\cite{MRS} parameterizations
do not include this measurement as a constraint and 
do not reproduce the data.  The MRST~\cite{mrst} parameterization
does include the first published results~\cite{prl} from this 
experiment which included the measurement of $\sigma^{pd}/2\sigma^{pp}$
as determined from the high mass data.
 
\begin{figure}
  \begin{center}
    \mbox{\epsffile{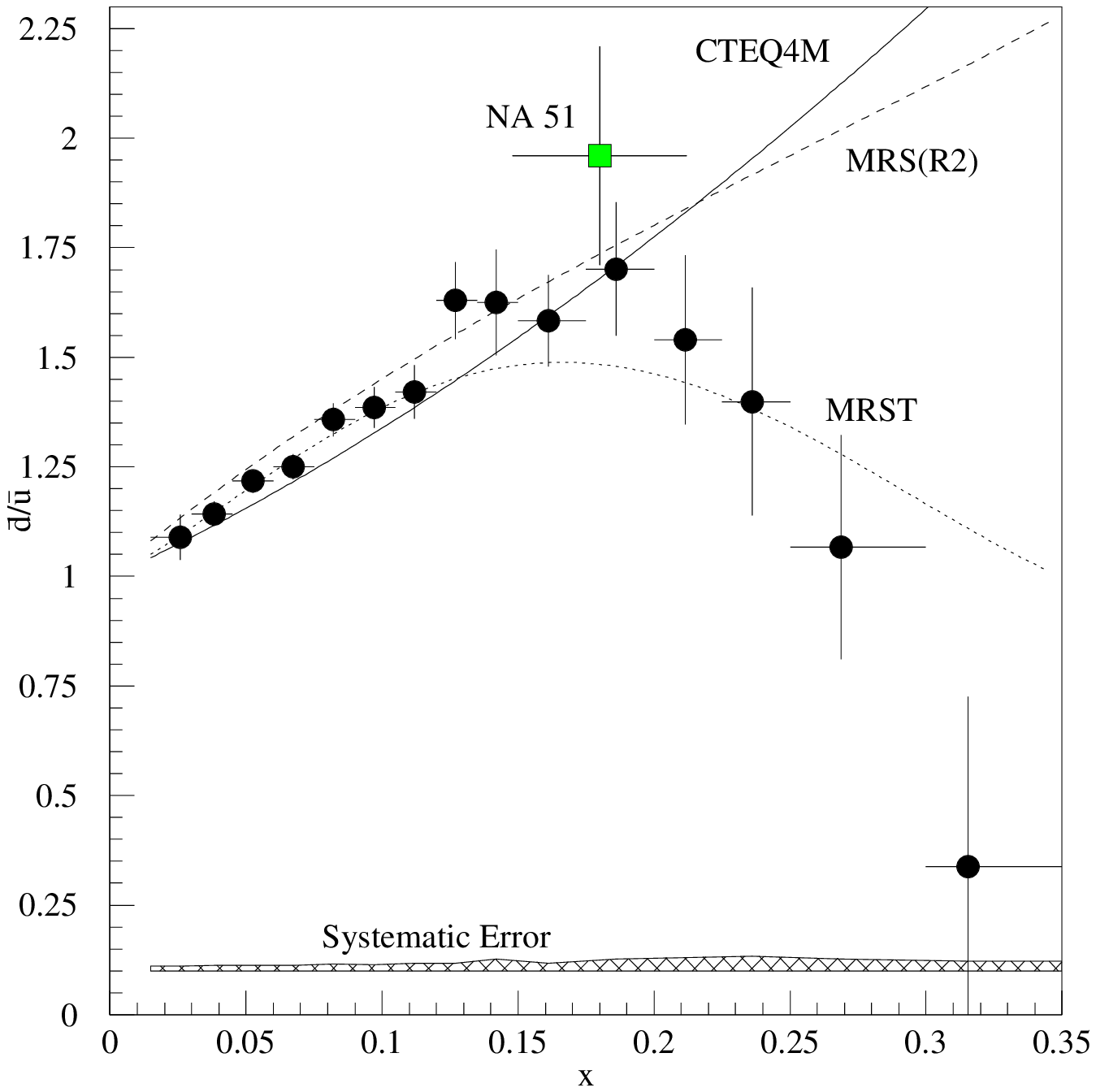}}                           
    \vspace*{-0.25in}                                                      
  \end{center}                                                               
  \caption{$\bar d(x)/\bar u(x)$ versus $x$.  The combined result
            from all three mass settings is shown along with three
	    parameterizations.  The NA51 data point is also shown.}
  \label{fig:duwpdf}
\end{figure}

\begin{figure}
  \begin{center}
    \mbox{\epsffile{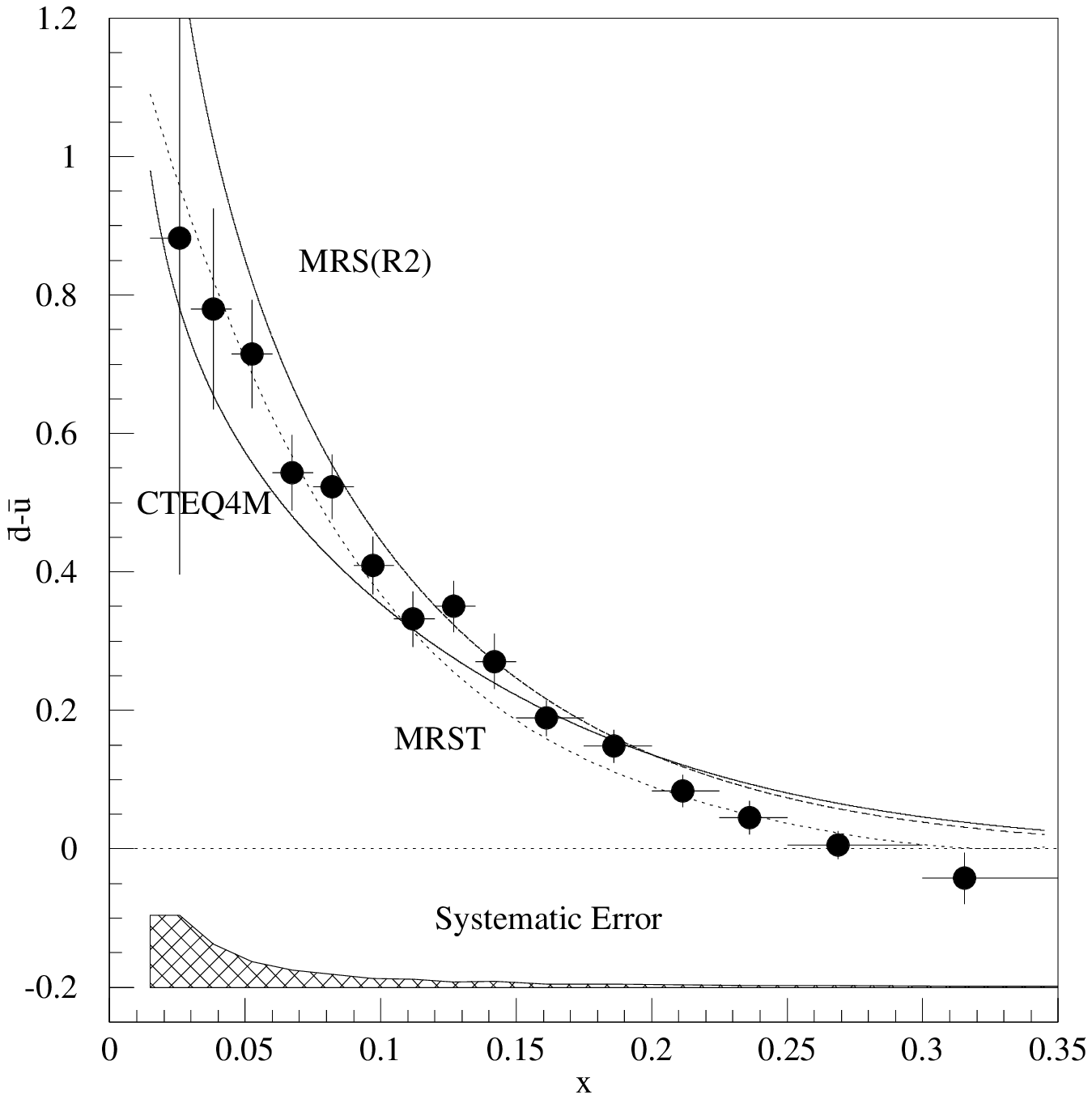}}                           
    \vspace*{-0.25in}                                                      
  \end{center}                                                               
  \caption{$\bar d(x) - \bar u(x)$ versus $x$.  The combined result
            from all three mass settings is shown along with three
	    parameterizations.}
  \label{fig:dmu}
\end{figure}

\begin{figure}
  \begin{center}
    \mbox{\epsffile{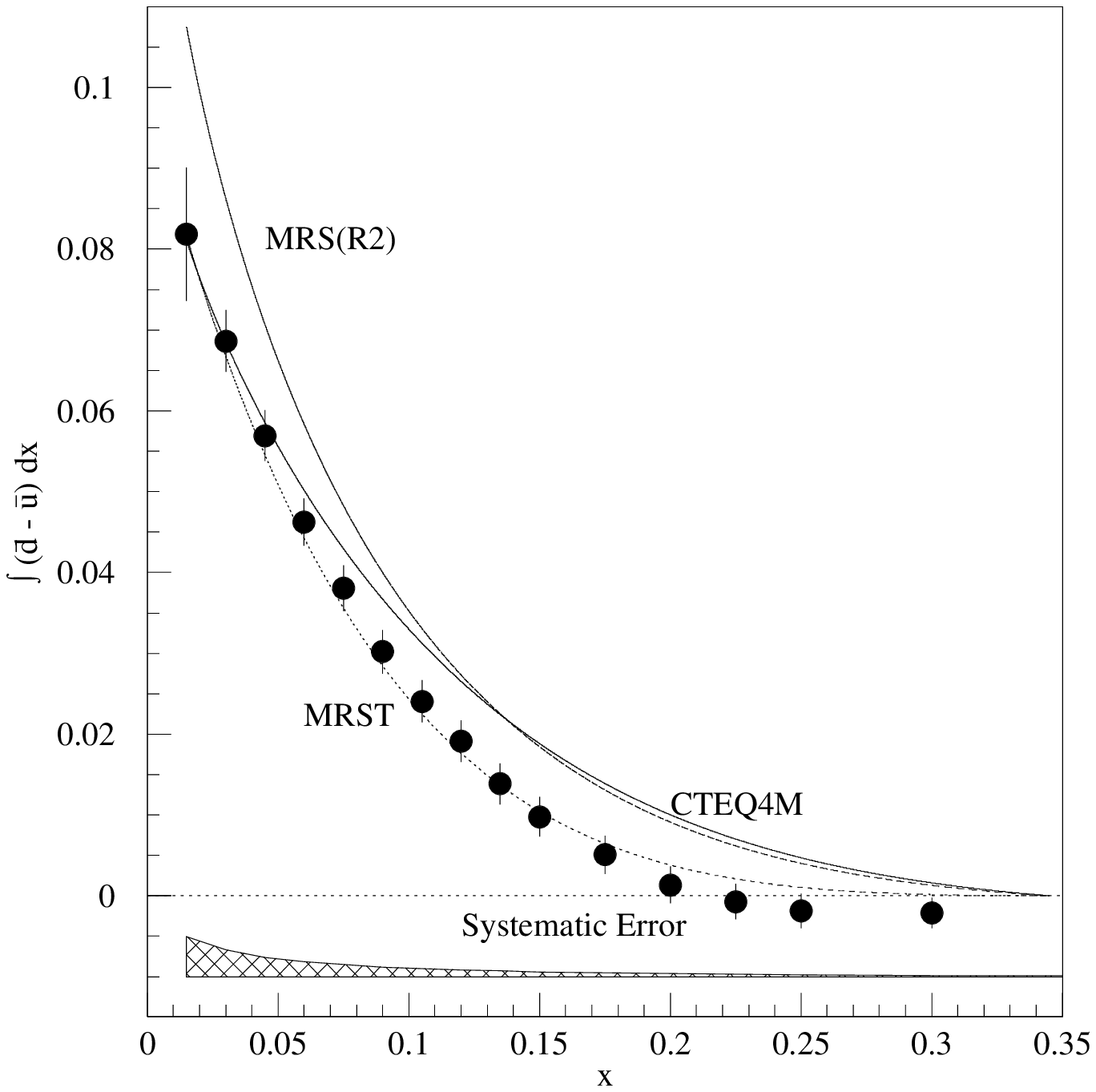}}                           
    \vspace*{-0.25in}                                                      
  \end{center}                                                               
  \caption{$\int^{0.35}_x (\bar d(x) - \bar u(x)) dx'$ versus $x$.
	   The curves are from three different parameterizations.}
  \label{fig:idmu}
\end{figure}

The main physics results of this experiment are \mbox{$\bar d(x)/\bar u(x)$},
\mbox{$\bar d(x) - \bar u(x)$}, and \mbox{$\int (\bar d(x) - \bar u(x)) dx$}
of the proton.
These results are shown in 
Figs.~\ref{fig:duwpdf}, \ref{fig:dmu}, and \ref{fig:idmu}
 along with the curves from the various 
parameterizations.  Again, parameterizations that do not 
include results from this measurement do not fit these results.

To illustrate how this measurement has tightened constraints 
on the valence quark PDF's, it is useful to decompose how the
parameterizations fit the NMC~\cite{NMC} measurement discussed in 
Section \ref{sect:dis}.  The quantity that NMC measured, 
$F_2^p - F_2^n$, can be expressed as the sum of the valence quark
contribution and the sea quark contribution.  
\begin{equation}
F_2^p - F_2^n = \underbrace{\frac{x}{3} (u - d)}_{\rm {\textstyle valence}} 
		+ \underbrace{\frac{2x}{3} (\bar u - \bar d)}_{\rm {\textstyle sea}}
\end{equation}
Since $u>d$ the valence contribution is positive while the sea 
contribution is negative because $\bar d > \bar u$.  

\begin{figure}
  \begin{center}
    \mbox{\epsffile{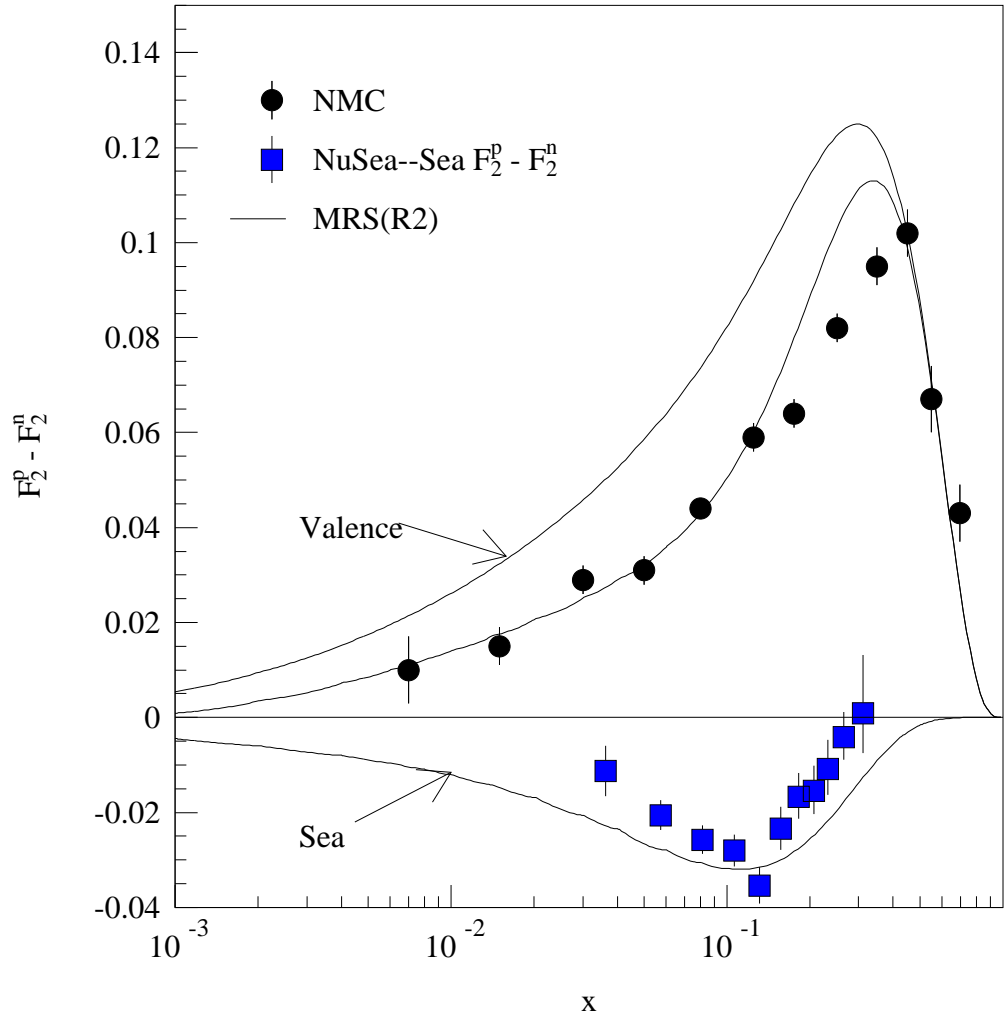}}                           
    \vspace*{-0.25in}                                                      
  \end{center}                                                               
  \caption{$F_2^p - F_2^n$ as measured by NMC at $Q=2$~GeV compared
	    with predictions based on the MRS(R2) parameterization.
	    Also shown are the E866 high mass results, evolved to $Q=2$~GeV,
	    for the sea contribution to $F_2^p - F_2^n$.  The top
	    (bottom) curve is the valence (sea) contribution and
	    the middle curve is the sum of the two.}
  \label{fig:nmce866}
\end{figure}

\begin{figure}
  \begin{center}
    \mbox{\epsffile{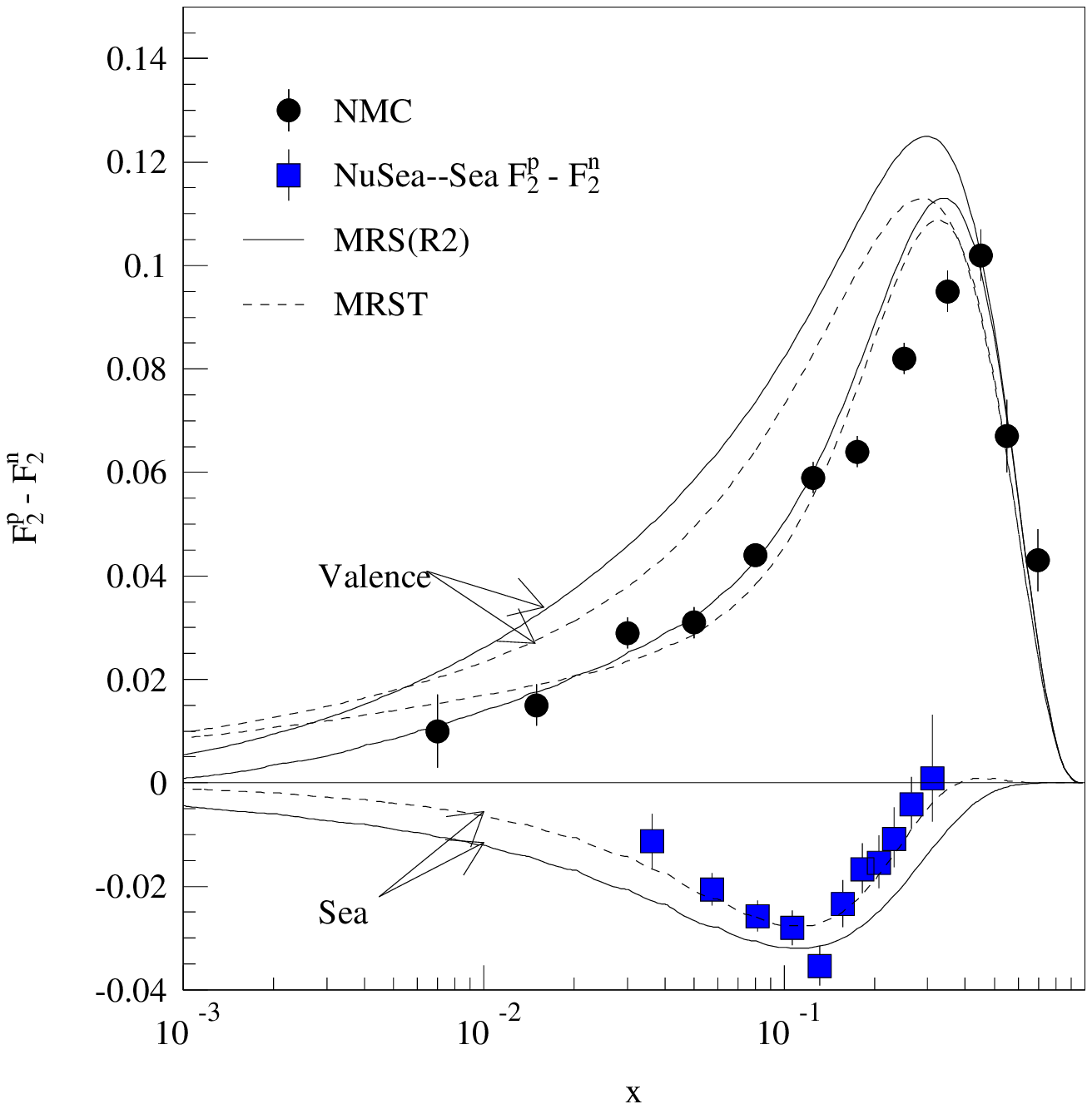}}                           
    \vspace*{-0.25in}                                                      
  \end{center}
  \caption{$F_2^p - F_2^n$ as measured by NMC at $Q=2$~GeV compared
	    with predictions based on the MRS(R2) (solid) and MRST (dashed) parameterizations.
	    Also shown are the E866 high mass results, evolved to $Q=2$~GeV,
	    for the sea contribution to $F_2^p - F_2^n$.  For each prediction, the top
	    (bottom) curve is the valence (sea) contribution and
	    the middle curve is the sum of the two.}
  \label{fig:nmce8662}
\end{figure}

Figure \ref{fig:nmce866} shows the NMC measurement and the fit to the 
data according to the MRS(R2) parameterization.  The NMC 
measurement provides a rigid constraint on the sum of the 
valence and the sea contributions, but does not constrain either
contribution separately.  Also shown in Fig.~\ref{fig:nmce866} 
is the MRS(R2) parameterization decomposed into the valence
and sea contributions.  The sea contribution can be compared 
with the new constraint provided by the high mass data~\footnote{
Only the high mass data is shown in Fig.~\ref{fig:nmce866} and Fig.~\ref{fig:nmce8662}.
This allows for a fair comparison with the MRST parameterization which only 
had the high mass data as a constraint.} from E866.  
This comparison shows that the MRS(R2) parameterization of the sea contribution to 
$F_2^p - F_2^n$ over predicts the magnitude of the contribution.

The next generation global fit done by the MRS group, called MRST, 
does include~\footnote{The MRST parameterization
includes other new measurements which also affect 
the valence parameterization.} the first published 
results from E866~\cite{prl}.  As expected 
this new parameterization reproduces the E866 measured sea
contribution to $F_2^p - F_2^n$.  However, this was not the only 
change in the new global fit due to the E866 measurement.  
To maintain the fit to the NMC measurement while reducing the
magnitude of the sea contribution required a corresponding 
decrease in the magnitude of the valence contribution over
the same range in $x$.  Finally, to maintain the proper number
of valence up quarks,
\begin{equation}
\int_0^1 \left ( u(x) - \bar u(x) \right) dx = 2,
\end{equation} 
and down quarks,
\begin{equation}
\int_0^1 \left ( d(x) - \bar d(x) \right) dx = 1,
\end{equation} 
the magnitude of the valence contribution must increase
over some other $x$ range.  All three of these changes 
can be seen in Fig.~\ref{fig:nmce8662}, which is identical 
to Fig.~\ref{fig:nmce866} except with the addition of the 
MRST parameterization.

\section{Comparison to Other Experiments}

\begin{figure}
  \begin{center}
    \mbox{\epsffile{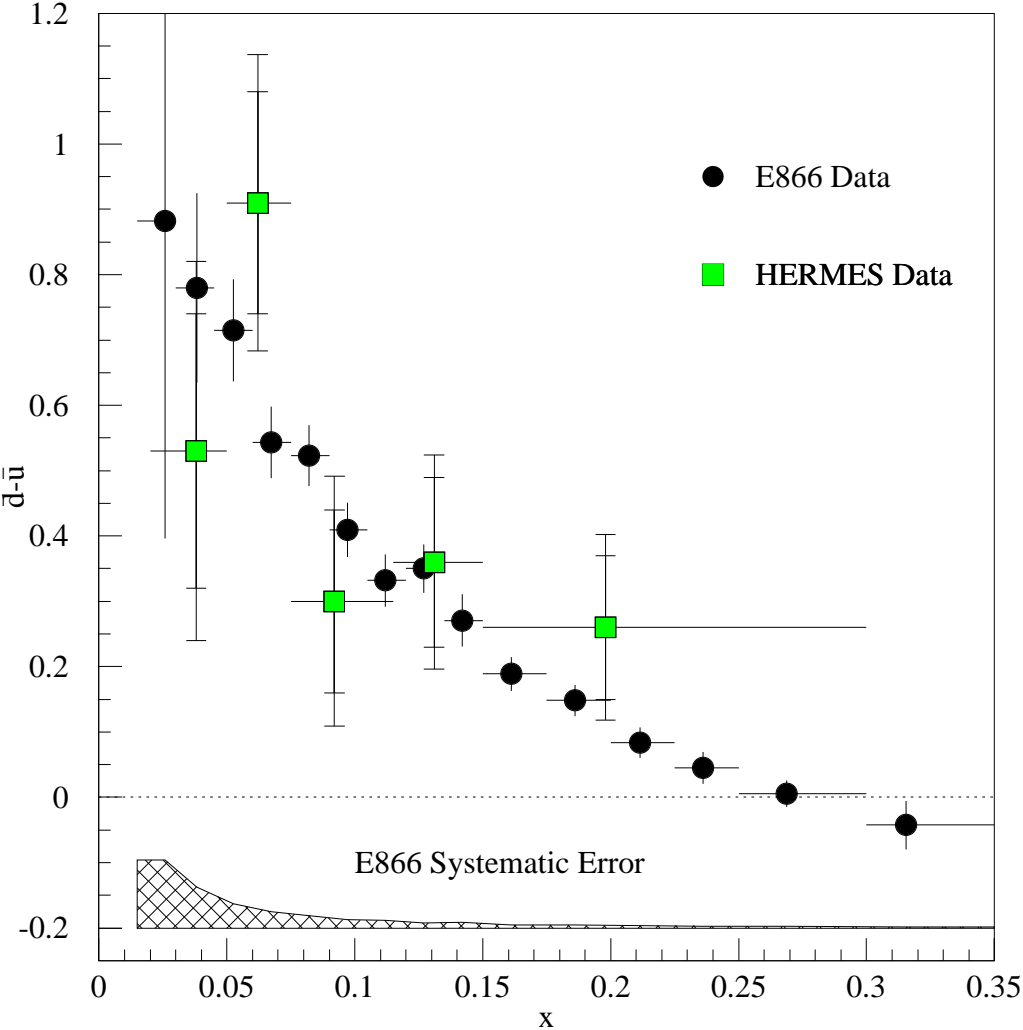}}                           
    \vspace*{-0.25in}                                                      
  \end{center}
  \caption{$\bar d - \bar u$ as a function of $x$.  The E866 (HERMES) results 
		are shown as round (square) data points.  The error bars on the 
		E866 data points represent the statistical uncertainty.  The inner error bars 
		on the HERMES data points represent the statistical uncertainty	
		while the outer error bars represent the statistical and 	
		systematic uncertainty added in quadrature.}
  \label{fig:hermes}
\end{figure}

The results of this experiment are much more extensive
and precise than any other measurement of $\bar d(x)/\bar u(x)$.
Other measurements of $\bar d(x)/\bar u(x)$ include the early 
measurement by NA51 which was discussed in Chapter \ref{ch:theory} and
the recent result from the HERMES collaboration~\cite{hermes} at
DESY.  Both of these measurements are in general agreement with 
the E866 results as seen in Fig.~\ref{fig:duwpdf} and Fig.~\ref{fig:hermes}. 
Even though the average Q$^2$ of these measurements are different,
comparisons can be made between them because the Q$^2$ dependence 
is small.

While the NA51 determination of $\bar d(x)/\bar u(x)$ was very
similar to the method used by E866, the HERMES result was 
based on a measurement of semi-inclusive deep-inelastic scattering.
While their measurement does not have either the coverage
or the precision of E866, it does provide a truly independent 
confirmation of the results.  Many of the systematic effects 
that are common to the NA51 and E866 Drell-Yan experiments do
not affect the HERMES measurement.  


These measurements of $\bar d(x)/\bar u(x)$ can be compared to 
the NMC DIS results by integrating $\bar d(x) - \bar u(x)$.  Table \ref{tab:idmupar}
summarizes the value of this integral over different 
$x$ ranges as parameterized by various global fits and as measured by E866.  
To extrapolate this integral from the measured region to the unmeasured 
region, MRST was used to estimate the contribution for $0 \le x \le 0.015$ and
it was assumed that the contribution for $x \ge 0.35$ was negligible.  
To estimate the uncertainty from this extrapolation, it was assumed that
the MRST contribution contained a 20\% uncertainty.
Table~\ref{tab:idmuexp}
summarizes three different experimental determinations of this integral
over all $x$ values.

\begin{table}[tbp]
\caption{$\int (\bar d(x) - \bar u(x)) dx $ evaluated over different $x$
ranges based on three different parameterizations and as measured by E866.}
\label{tab:idmupar}
\begin{center}
\begin{tabular}{ccccc}
\hline \hline
$x$ range      & CTEQ(4M) & MRS(R2) & MRST    & E866 \\ \hline
$0<x<1$        & 0.10941  & 0.16553 & 0.11482 &    \\
$0.35<x<1$     & 0.00186  & 0.00129 &-0.00031 &    \\
$0.015<x<0.35$ & 0.08163  & 0.10845 & 0.08219 & 0.082 $\pm$ 0.010    \\
$0<x<0.015$    & 0.02592  & 0.05578 & 0.03293 &     \\
\hline \hline
\end{tabular}
\end{center}
\end{table}

\begin{table}[tbp]
\caption{$\int (\bar d(x) - \bar u(x)) dx $ as determined by three experiments.  The 
range of the measurement is shown along with the value of the integral over all $x$.}
\label{tab:idmuexp}
\begin{center}
\begin{tabular}{cccccc}
\hline \hline
Experiment   & $x$ range       & $\int_0^1 (\bar d(x) - \bar u(x)) dx $ \\ \hline
E866         & $0.015<x<0.35$  & $0.115 \pm 0.012$   \\
NMC          & $0.004<x<0.80$  & $0.147\pm 0.039$    \\
HERMES       & $0.020<x<0.30$  & $0.16 \pm 0.03$     \\
\hline \hline
\end{tabular}
\end{center}
\end{table}

\section{Possible Origins of the Nucleon Sea}

The possible production mechanisms that can account for the
sea of quark-antiquark pairs in the nucleon can be categorized 
as either perturbative or nonperturbative.  The next two subsections 
will describe each of these categories.  The primary focus of this section
is to compare each possible production mechanism with 
the observed flavor asymmetry.

\subsection{Perturbative Origins}

The perturbative production mechanism is the 
production of a quark-antiquark pair from gluon splitting. 
This is the simplest method of producing the nucleon sea, 
but it can not produce a substantial asymmetry.

In 1977 Field and Feynman first suggested~\cite{fieldandfeynman} 
a possible means by which gluon splitting may produce an asymmetric sea.
They suggested that Pauli blocking in the proton would suppress 
a gluon splitting into an up-antiup quark pair compared to a
gluon splitting into a down-antidown quark pair. Since there are
two valence up quarks and only one valence down quark in the proton, 
Pauli blocking could produce a flavor asymmetry.

Shortly after it was suggested that Pauli blocking could produce 
a flavor asymmetry in the nucleon, quantitative 
calculations~\cite{rossandsachrajda}
showed that this effect was less than one percent.
More recently, additional calculations~\cite{steffensandthomas}
have shown that while Pauli blocking does produce a 
small effect, it also produces an asymmetry opposite of
what has been observed.  

While gluon splitting is   a source of quark-antiquark
pairs in the nucleon, it can not be the source of the 
large observed flavor asymmetry.  Additional production 
mechanisms must be considered.  As these nonperturbative
methods are discussed in the next subsection, it is important
to remember that they must be considered in addition to 
the perturbative method, which produces a mostly symmetric sea.

\subsection{Nonperturbative Origins}


Meson cloud models~\cite{kumano,prd} describe the production of an asymmetric
nucleon sea by expressing the physical proton as the combination
of a proton with a symmetric sea and a series of virtual 
meson-baryon states.  For example, the physical proton ($|p\rangle$) can be 
expressed as,
\begin{eqnarray}
\label{eqn:mb}
|p\rangle& = & \sqrt{1-|\alpha|^2-|\beta|^2}~|p_0\rangle \nonumber \\ 
   &   &  + ~\alpha\left[\sqrt{2/3}~|n,\pi^+\rangle - ~\sqrt{1/3}~|p,\pi^0\rangle\right] \\
   &   &  + ~\beta\left[\sqrt{1/2} ~|\Delta^{++},\pi^-\rangle
 - ~\sqrt{1/3}~|\Delta^+,\pi^0\rangle + ~\sqrt{1/6} ~|\Delta^0,\pi^+\rangle \right] \nonumber
\end{eqnarray}
where $|p_0\rangle$ is a proton with a symmetric sea and $|\alpha|^2$ ($|\beta|^2$) 
is the probability that a proton is in a virtual 
$|N,\pi\rangle$ ($|\Delta,\pi\rangle$) state.

From Eq. \ref{eqn:mb} it is easy to show that
\begin{equation}
\int_0^1[\bar d(x) - \bar u(x)] dx = (2a - b)/3,
\label{eqn:ab}
\end{equation}
where $a = |\alpha|^2$ and $b = |\beta|^2$.  Calculations of 
the relative probability of these two configurations find 
$a\approx 2 b$~\cite{garvey}.
Using the value for the integral in Eq. \ref{eqn:ab} extracted from 
E866 and assuming that $a=2b$ yields $a = 2b = 0.230 \pm 0.024$.  
From this simple meson model, quark counting would indicate that
\begin{equation}
\frac{\bar d}{\bar u} = \frac{\frac{5}{6}a + \frac{1}{3} b}{\frac{1}{6}a + \frac{2}{3} b} = 2.
\end{equation}
This simple determination of $\bar d/\bar u$ does not include the 
$x$ dependence of the ratio or any 
contribution from the perturbatively produced symmetric sea.

\begin{figure}
  \begin{center}
    \mbox{\epsffile{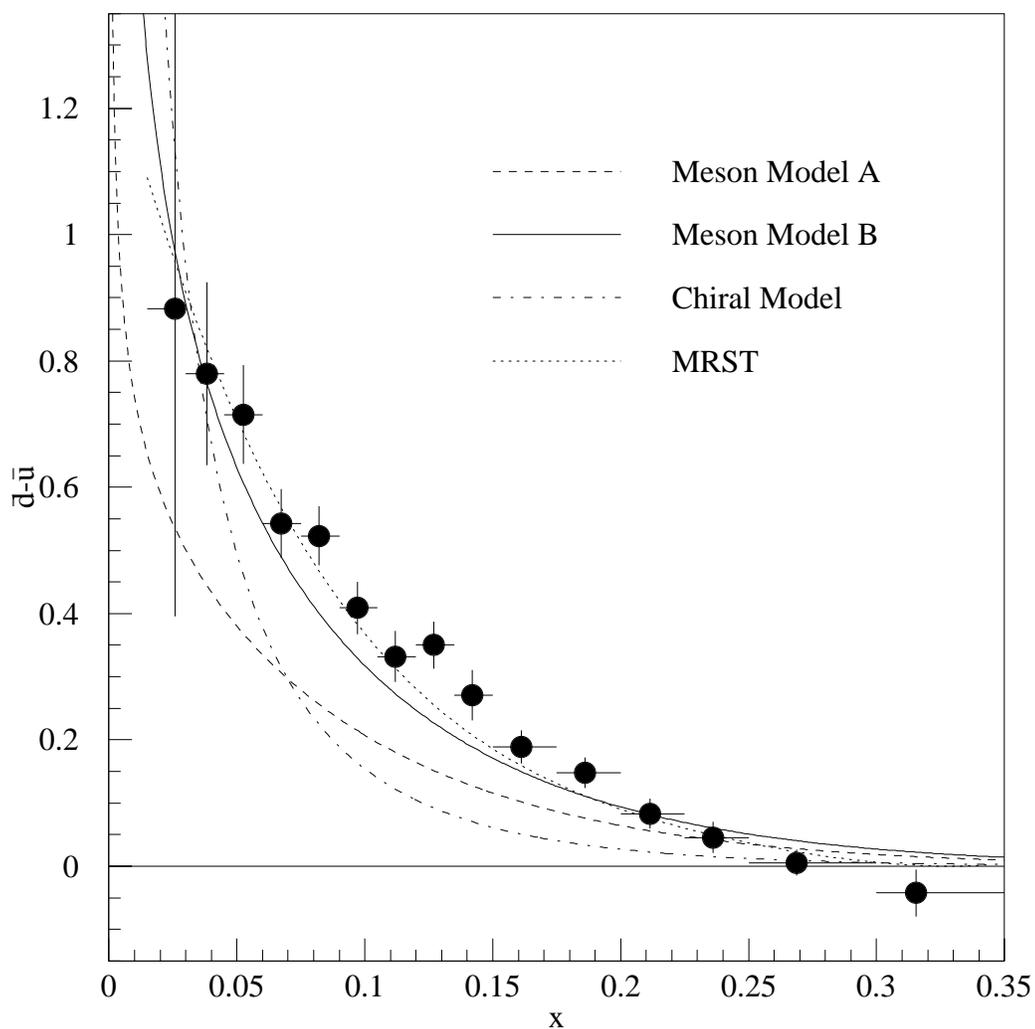}}                           
    \vspace*{-0.25in}                                                      
  \end{center}
  \caption{$\bar d - \bar u$ as a function of $x$.  The E866 results 
		are shown along with predictions of meson and chiral 
		models.  Both are shown for $Q = 6.38$ GeV.}
  \label{fig:dmupion}
\end{figure}

Figure \ref{fig:dmupion} compares $\bar d(x) - \bar u(x)$
from E866 with a meson cloud model calculation.  These calculations
are based on the procedure described in reference~\cite{kumano}.
The SMRS(P2)~\cite{smrs} parameterization was used for the pion structure
functions at a $Q$ value of 6.38~GeV since the E866 results are 
shown at this value of $Q$.  The curve labeled ``Meson Model A'' uses
a dipole form with $\Lambda = 1.0$~GeV for the $\pi NN$ and $\pi N \Delta$
form factors. It has been suggested~\cite{garvey} that $\Delta$ production
experiments~\cite{delta} indicate that the $\pi N \Delta$
form factor is softer than the $\pi NN$ form factor. 
This observation has motivated the calculation for the 
curve in Fig.~\ref{fig:dmupion} labeled "Meson Model B" which
uses a reduced value for the $\pi N \Delta$ form factor of
$\Lambda = 0.8$~GeV.  This calculation fits the E866 measurement
better than Meson Model A.

\begin{figure}
  \begin{center}
    \mbox{\epsffile{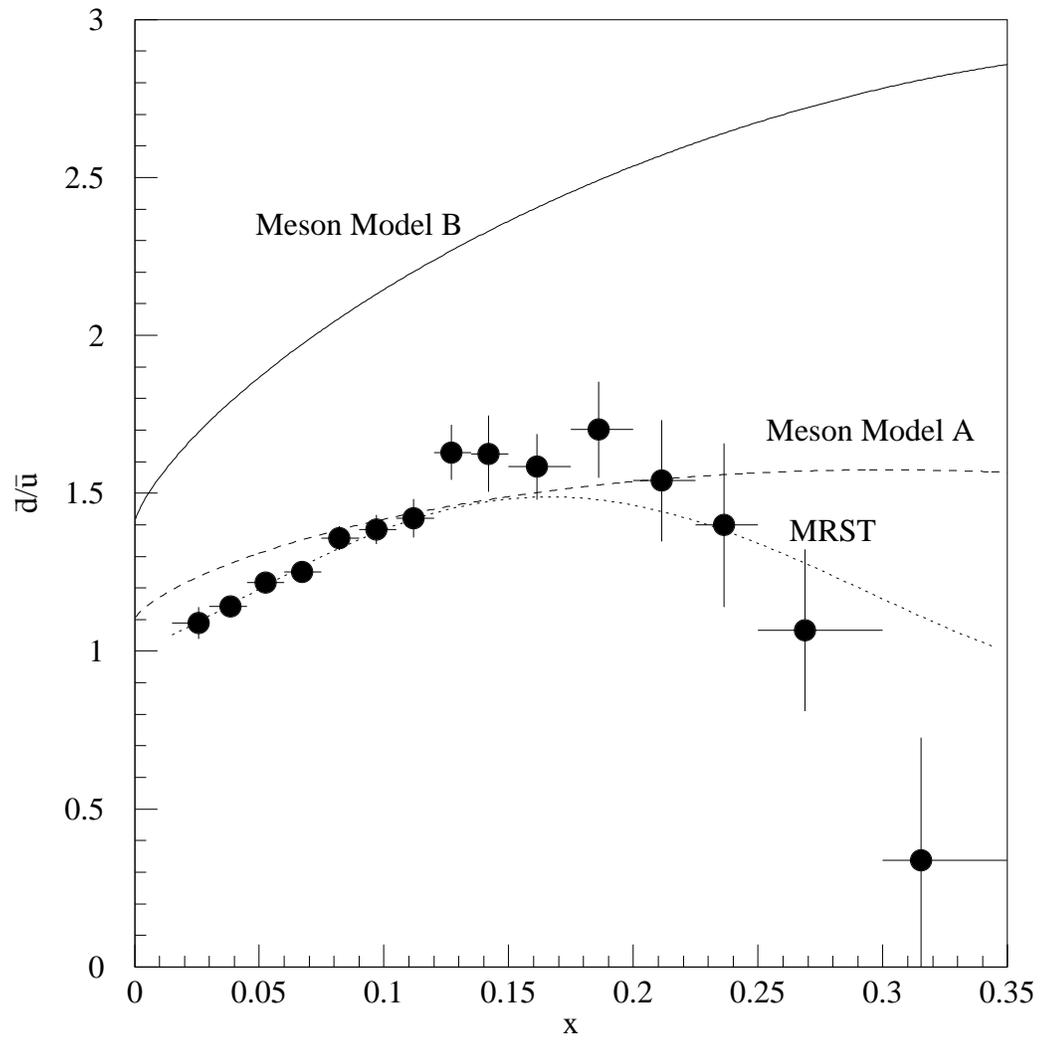}}                           
    \vspace*{-0.25in}                                                      
  \end{center}
  \caption{$\bar d/\bar u$ as a function of $x$.  The E866 results 
		are shown along with predictions of meson models.  
                Both are shown for $Q = 6.38$ GeV.}
  \label{fig:dupion}
\end{figure}

It is also useful to compare the model predictions of $\bar d(x)/\bar u(x)$ 
with the results from E866.  
When making this comparison it is important to remember
that the pion models do not include the perturbative 
production process.  This did not affect the comparison
of the meson models with $\bar d(x) - \bar u(x)$ because
$\bar d(x) - \bar u(x)$ is only sensitive to the flavor 
asymmetry.  However, $\bar d(x)/\bar u(x)$ is sensitive 
to both the asymmetric and the symmetric sea.
Therefore, it is not surprising to see in Fig.~\ref{fig:dupion} 
that the two meson models do not reproduce the E866 results.
This comparison does show that meson model A can not 
accommodate an additional contribution from the perturbative
production process for $0.1<x<0.2$.  Figure~\ref{fig:dupion}
also provides information about the relative importance
of the perturbative versus the non-perturbative sea.

Another class of models which produces a flavor asymmetry in the nucleon
via the inclusion of virtual mesons are the chiral quark 
models~\cite{quigg,szczurek}.  The chiral quark model is similar
to the meson model, except the virtual meson couples directly to
a quark, not the nucleon as a whole.  So antiquarks are produced in
processes such as $u \rightarrow d\pi^+$ and $d \rightarrow u\pi^-$.  
The excess of $\bar d$ over $\bar u$ in the proton is due to the additional
up valence quark in the proton.

Figure \ref{fig:dmupion} shows the chiral model
prediction for $\bar d(x) - \bar u(x)$.  This prediction
was calculated following the formulation in reference~\cite{szczurek}.
The quantity $\bar d(x) - \bar u(x)$ was calculated at $Q = 0.5$~GeV
and then evolved to $Q = 6.38$~GeV.  As seen in Fig.~\ref{fig:dmupion},
this model predicts a much smaller mean-$x$ for $\bar d(x) - \bar u(x)$
than that predicted by the meson models.  This is a result of the mesons
in the chiral model coupling to the valence quarks which carry on average $1/3$ 
of the nucleon momentum.  The $x$ dependence of the E866 data favor 
the meson model over the chiral model.  

The last nonperturbative flavor asymmetry producing
mechanism to be mentioned here is the coupling of instantons to 
the valence quarks.  This model~\cite{instanton} does not 
seem to agree with any of the quantities measured by E866.  
Specifically this model predicts that the cross section ratio 
would increase at high $p_T$, which is in total disagreement
with Fig.~\ref{fig:ratiopt}.
It is not known if better agreement with the data can be obtained
by adjusting the parameters within the model.

\section{Future Experiment}

The sharp drop towards unity in $\bar d(x)/\bar u(x)$ above $x = 0.2$ was 
unexpected.  This has prompted interest~\cite{thomas} in extending the measurement
of $\sigma^{pd}/2\sigma^{pp}$ to higher values of $x$.  
A new experiment is currently being designed and proposed~\cite{p906} to make
this measurement using the 120 GeV proton beam from the new Main Injector at Fermilab.

The Drell-Yan cross section is inversely proportional to the incident
beam energy.  The Main Injector can provide protons at 120~GeV, which
means that the Drell-Yan cross section will be about a factor of seven 
higher compared to the 800~GeV beam used for this measurement.  The higher
cross section means that the measurement of $\sigma^{pd}/2\sigma^{pp}$
could be extended to near $x = 0.5$.  

\section{Conclusion}

FNAL E866/NuSea was an extremely successful experiment.  The stated goal of 
measuring $\sigma^{pd}/2\sigma^{pp}$ with a systematic uncertainty of 1.5\% 
over the kinematic range of $0.04\le x \le 0.3$ was exceeded.  By keeping 
the systematic uncertainty under 1\% while measuring $\sigma^{pd}/2\sigma^{pp}$
from $x=0.015$ to $x=0.35$ allowed for the extraction of $\bar d(x)/\bar u(x)$
over this wide kinematic range.  The surprising result of $\bar d(x)/\bar u(x)$
decreasing above $x=0.2$ has instigated several new global fits. 
Finally, as discussed in the previous section, this 
measurement has motivated plans for a further experiment to continue the
study of this interaction.  These results clearly demonstrate the importance
of this measurement, which went well beyond initial expectations for this 
experiment.


\begin{vita}
Rusty Shane Towell was born in Indio, California on April 1, 1969, 
the son of Delbert Clayton Towell and Helen Fay Towell. 
After graduating from Abilene High School in 1986, 
he entered Abilene Christian University in Abilene, Texas.  
He received the degree
of Bachelor of Science from Abilene Christian University in May 1990.
During the following years he was an officer in the United States 
Navy where he taught at the Naval Nuclear Power School in 
Orlando, Florida.  In September 1994 he entered the Graduate 
School of The University of Texas at Austin.
\end{vita}

\end{document}